\documentclass[twocolumn]{aastex631}
\usepackage{amsmath}
\usepackage{colortbl}

\newcommand{\sph}{S_\mathrm{ph}}
\newcommand{\feh}{\left[\mbox{Fe/H}\right]}
\newcommand{\teff}{T_\mathrm{eff}}
\newcommand{\prot}{P_\mathrm{rot}}

\shorttitle{Photometric magnetic activity and age relation}
\shortauthors{Mathur et al.}
\graphicspath{{./}{figures/}}

\begin{document}

\title{Magnetic activity evolution of solar-like stars: \\
I. $S_{\rm ph}$-Age relation derived from {\it Kepler} observations}

\correspondingauthor{Savita Mathur}
\email{smathur@iac.es}

\author[0000-0002-0129-0316]{Savita Mathur}
\affil{Instituto de Astrof\'isica de Canarias (IAC), E-38205 La Laguna, Tenerife, Spain}
\affil{Universidad de La Laguna (ULL), Departamento de Astrof\'isica, E-38206 La Laguna, Tenerife, Spain}

\author[0000-0002-9879-3904]{Zachary R. Claytor}
\affiliation{Department of Astronomy, University of Florida, 211 Bryant Space Science Center, Gainesville, FL 32611, USA}
\affiliation{Institute for Astronomy, University of Hawai‘i at Mānoa, 2680 Woodlawn Drive, Honolulu, HI 96822, USA}

\author[0000-0001-7195-6542]{\^Angela R. G. Santos}
\affil{Instituto de Astrof\'isica e Ci\^encias do Espa\c{c}o, Universidade do Porto, CAUP, Rua das Estrelas, PT4150-762 Porto, Portugal}

\author[0000-0002-8854-3776]{Rafael A. Garc\'{i}a}
\affil{Universit\'e Paris-Saclay, Universit\'e Paris Cit\'e, CEA, CNRS, AIM, 91191, Gif-sur-Yvette, France}

\author{Louis Amard}
\affiliation{AIM, CEA, CNRS, Universit\'e Paris-Saclay, Universit\'e de Paris, Sorbonne Paris Cit\'e, F-91191 Gif-sur-Yvette, France}

\author[0000-0003-0142-4000
]{Lisa Bugnet}
\affiliation{Institute of Science and Technology Austria, Klosterneuburg, Austria}

\author[0000-0001-8835-2075]{Enrico Corsaro}
\affiliation{INAF -- Osservatorio Astrofisico di Catania, Via S. Sofia 78, I-95123, Italy}

\author{Alfio Bonanno}
\affiliation{INAF -- Osservatorio Astrofisico di Catania, Via S. Sofia 78, I-95123, Italy}

\author[0000-0003-0377-0740]{Sylvain N. Breton}
\affil{INAF -- Osservatorio Astrofisico di Catania, Via S. Sofia 78, I-95123, Italy}
\affil{Universit\'e Paris-Cit\'e, Universit\'e Paris-Saclay, CEA, CNRS, AIM, F-91191, Gif-sur-Yvette, France }

\author[0000-0003-4556-1277]{Diego Godoy-Rivera}
\affil{Instituto de Astrof\'isica de Canarias (IAC), E-38205 La Laguna, Tenerife, Spain}
\affil{Universidad de La Laguna (ULL), Departamento de Astrof\'isica, E-38206 La Laguna, Tenerife, Spain}

\author{Marc H. Pinsonneault}
\affiliation{Ohio State University, McPherson Laboratory, 140 W 18th Ave, Columbus, OH 43210, USA}

\author[0000-0002-4284-8638]{Jennifer van Saders}
\affiliation{Institute for Astronomy, University of Hawai‘i at Mānoa, 2680 Woodlawn Drive, Honolulu, HI 96822, USA}




\begin{abstract}

The ages of solar-like stars have been at the center of many studies such as exoplanet characterization or {  Galactic}-archaeology. While ages are usually computed from stellar evolution models, relations linking ages to other stellar properties, such as rotation and magnetic activity, have been investigated. With the large catalog of 55,232 rotation periods, $\prot$, and photometric magnetic activity index, $\sph$ from {\it Kepler} data, we have the opportunity to look for such magneto-gyro-chronology relations. Stellar ages are obtained with two stellar evolution codes that include treatment of angular momentum evolution, hence using $\prot$ as input in addition to classical atmospheric parameters. We explore two different ways of predicting stellar ages on {  three subsamples with spectroscopic observations: solar analogs, late-F and G dwarfs, and K dwarfs}. We first perform a Bayesian analysis to derive relations between $\sph$ and ages between 1 and 5\,Gyr, and other stellar properties. For late-F and G dwarfs, and K dwarfs, the multivariate regression favors the model with $\prot$ and $\sph$ with median differences of 0.1\%.and 0.2\% respectively. We also apply Machine Learning techniques with a Random Forest algorithm to predict ages up to 14\,Gyr with the same set of input parameters. For late-F, G and K dwarfs together, predicted ages are on average within 5.3\% of the model ages and improve to 3.1\% when including $\prot$.
These are very promising results for a quick age estimation for solar-like stars with photometric observations, especially with current and future space missions.


\end{abstract}



\section{Introduction} \label{sec:intro}

The importance of stellar ages is undeniable for different fields of astronomy whether it is to study galactic evolution, planetary systems, and of course stellar physics. Stellar ages are model-dependent by their nature \citep{2010ARA&A..48..581S}, and as a consequence a variety of approaches for estimating ages can be found in the literature. These range from ones that rely heavily on theoretical models to those that are primarily empirical in nature. Stars spin down as they age, which makes rotation a potentially powerful chronometer. However, only a few stellar evolution codes include internal angular momentum transport and magnetized winds \citep{2008Ap&SS.316...43E,2008Ap&SS.316...31D,2013ApJS..208....4P,2016A&A...587A.105A}. A major reason why rotation is not more firmly embedded in stellar models is that there is currently no consensus agreement on the most important mechanisms for transport. All published models disagree with some aspects of observed internal rotation profiles  \citep[e.g.][]{2013A&A...555A..54C,2019ARA&A..57...35A}. Earlier codes focused on hydrodynamic angular momentum transport. Additional transport mechanisms such as internal gravity waves \citep{2005A&A...440..981T,2014ApJ...796...17F,2017A&A...605A..31P} and magnetic field stresses \citep{2002A&A...381..923S,2010A&A...517A..58D,2019MNRAS.485.3661F,2021A&A...650A..53B,2021A&A...647A.122M} are very likely to be important, but are usually not included because of the difficulty to properly implement these multidimensional processes, and because of the lack of observational constraints on internal properties (e.g. magnetic field). 
However, these models still represent a good reference as they reproduce the Sun and they are based on the best physics known so far.

For a long time and for large number of stars, isochrone fitting of observables such as colors, magnitudes, or atmospheric parameters (effective temperature, $\teff$, and surface gravity, $\log g$) have been used to derive stellar parameters including ages. When done for field stars, this can lead to uncertainties on ages larger than 50\% \citep[e.g.][]{2014A&A...569A..21L}. Isochrone ages become particularly uncertain on the lower main sequence, where stars experience little nuclear evolution over the age of the universe. The age estimate can be improved for cluster stars as they share a common origin as well as other properties like composition \citep[e.g.][]{2019A&A...623A.108B,2021ApJS..257...46G}.

Stellar oscillations are powerful diagnostics of the global properties of stars. The study of these oscillations, asteroseismology, can provide additional constraints to the stellar models. Indeed \citet{2014A&A...569A..21L} showed how the age precision increases when spectroscopic and asteroseismic data are combined. Asteroseismic analyses have been performed in hundreds of solar-like stars \citep[e.g.][]{2008A&A...488..705A,2011Sci...332..213C,2016PASP..128l4204L,2019LRSP...16....4G,2021ApJ...922..229C,2022A&A...657A..31M} mostly with data from space missions such as Convection, Rotation, and Transits \citep[CoRoT;][]{2006cosp...36.3749B}, {\it Kepler} \citep{2010Sci...327..977B}, K2 \citep{2014PASP..126..398H}, and the Transiting Exoplanet Survey Satellite \citep[TESS;][]{2015JATIS...1a4003R}, allowing the determination of precise seismic ages for around a hundred stars with well characterized individual frequencies \citep[e.g.][]{2015MNRAS.452.2127S,2017A&A...601A..67C,2017ApJ...835..173S}. Unfortunately such seismic analysis has been done on only a small sample of solar-like stars that the aforementioned space missions observed, due to the short-cadence requirement and the small amplitude of the modes that, for instance, can be suppressed by surface magnetic activity \citep[e.g.][]{2010Sci...329.1032G,2011ApJ...732L...5C,2014A&A...571A..35B,2018ApJS..237...17S,2019FrASS...6...46M}.

Another way of estimating stellar ages has been developed and used in the last decades and is based on  semi-empirical methods. Indeed, it has been observed that rotation rate and surface magnetic activity decay when a solar-like star gets older \citep{1972ApJ...171..565S}. These observations are explained by the loss of angular momentum via magnetized stellar winds \citep[e.g.][]{1988ApJ...333..236K,1989ApJ...338..424P}. The rotation-age relationship has been investigated using clusters, where the most reliable ages were available, leading to the dawn of {\it gyrochronology} \citep[e.g.][]{2003ApJ...586..464B,2007ApJ...669.1167B}. To derive such relationship, we need calibrators with independent precise ages (clusters, binaries, or seismic targets).

Similarly, magnetochronology has been developed where the observable is a magnetic activity index such as the one from Ca H \& K lines \citep[e.g.][]{1978ApJ...226..379W,1995ApJ...438..269B}, Zeeman Doppler imaging \citep[e.g.][]{2014MNRAS.444.3517M}, X-ray luminosity \citep[e.g.][]{2018MNRAS.479.2351W, 2021A&A...649A..96J}, or photometry \citep[e.g.][]{2013ApJ...769...37B,2014JSWSC...4A..15M}. 

A few activity-age relations have been derived by calibrating activity as a function of stellar observables, with ages computed mostly from isochrone fitting,  \citep{2008ApJ...687.1264M,2013A&A...551L...8P, 2014MNRAS.441.2361V,2016A&A...594L...3L,2018A&A...619A..73L}.
Usually, those relations rely on magnetic activity proxies that require X-ray, UV, or high-resolution spectroscopic observations that are not trivial to obtain in particular for a large sample of stars. Thankfully, in the era of high-precision photometric space-based missions, measurements of the photometric magnetic activity became more easily available for large samples. In particular, {\it Kepler} allowed the detection of starspot modulation for several tens of thousands of solar-like stars. With the TESS mission already observing millions of stars and the upcoming PLAnetary Transits and Oscillations \citep[PLATO;][]{2014ExA....38..249R} mission that will observe several hundreds of thousands of stars, photometric magnetic activity will be accessible for much larger samples. Therefore, having a magnetic-age relation based on photometric observations would be useful to constrain stellar ages. With this in mind, in this work we study the possibility of using the photometric magnetic activity proxy, $S_{\rm ph}$ \citep{2010Sci...329.1032G, 2014A&A...562A.124M}, computed with {\it Kepler} observations to estimate ages of stars.

In Section~2, we briefly review how the rotation periods and magnetic activity proxies have been computed for the {\it Kepler} targets. We describe our method for inferring stellar ages in Section~3 and we compare them with other published values (Section~4). Different relations are then derived between the photometric magnetic activity proxy and age, along with other stellar parameters by means of a Bayesian approach (Section~5). In Section~6, machine learning tools are investigated to predict the ages of main-sequence solar-like stars. The results and limitations of the different ways of computing or predicting stellar ages are discussed in Section~7. Finally,  in Section~8 we provide the conclusions of this work.

\section{Magnetic activity proxy and rotation period}

Rotation period measurements are possible thanks to the presence of active regions on the stellar surface that create a brightness modulation whose periodicity is related to the surface rotation. With four years of quasi-continuous observations of $\sim$\,160,000 solar-like stars on the main sequence and on the subgiant branch \citep{2017ApJS..229...30M}, the \textit{Kepler} mission provided the longest and best-quality light curves to study the rotation and magnetic activity of solar-like stars. 
The most recent catalog of surface rotation periods inferred from {\it Kepler} observations \citep[][hereafter S19 and S21]{2019ApJS..244...21S,2021ApJS..255...17S} yielded a sample of 55,232 solar-like stars on the main sequence as well as subgiants with rotation measurements. 

The S19 and S21 rotation periods, $P_{\rm rot}$, were obtained by analyzing four different sets of {\it Kepler} light curves \citep{2010ApJ...713L..87J, 2011MNRAS.414L...6G} with a rotation pipeline that uses three different methods: a time-frequency analysis with wavelets tools \citep{1998BAMS...79...61T,liu2007,2010A&A...511A..46M}, an auto-correlation function \citep{2014A&A...572A..34G,2014ApJS..211...24M}, and the composite spectrum, a combination of the first two methods \citep{2017A&A...605A.111C}. Finally, the machine learning algorithm ROOSTER \citep{2021A&A...647A.125B} was applied to select the most likely rotation periods among all the results. 

With up to 4 years of time series data and given that a detection of rotational modulation implies that the star is active, we can characterize the typical variability of our targets precisely. To do so, we define a magnetic proxy that we call $S_{\rm ph}$ \citep{2014A&A...562A.124M}. It is computed as the standard deviation of the light curve by taking subseries of length 5\,$\times P_{\rm rot}$ as explained in \citet{2014JSWSC...4A..15M}. Finally, the average $S_{\rm ph}$ value is adopted, as computed by S19 and S21. {  $\sph$ values are computed in parts per million (ppm).} This index has been shown to be a good proxy for magnetic activity as tested for the Sun \citep{2017A&A...608A..87S} and solar-analogs \citep{2016A&A...596A..31S, 2018ApJ...852...46K}.

\section{Age inference}\label{sec:ageYREC}

We used stellar evolution models that include angular momentum evolution to estimate ages of the stars in our sample. These models were produced using the Yale Rotating Evolution Code \citep[YREC;][]{1989ApJ...338..424P,2001ApJ...555..990B}, and the global stellar properties were then used to infer angular momentum evolution scenarios. 

The models are identical to those presented by \citet{Claytor2020}, including the fast-launch initial rotation conditions of \citet{2013ApJ...776...67V} ($P_\mathrm{init} = 8.1$ d), except that the models undergo weakened magnetic braking as described by \citet{2016Natur.529..181V}. Following this prescription, the braking weakens when the star reaches a stellar Rossby number (ratio between the rotation period and the convective overturn time at one pressure scale height above the base of the convective zone) of 2.16. Further rotational evolution is purely from the change in the stellar moment of inertia. Stellar evolution models including weakened braking better reproduce the observed rotation periods and seismic ages of solar-like stars observed by the {\it Kepler} mission \citep{2016Natur.529..181V, vanSaders2019,2020ApJ...900..154M,Hall2021}.

We fit stellar evolution models to observational data using the interpolation and Markov-Chain Monte Carlo (MCMC) tools in \texttt{kiauhoku} \citep{Claytor2020, kiauhoku}. For the MCMC we used a $\chi^2$ log-likelihood of the form:
\begin{equation}
    \mathcal{L} = -\frac{1}{2} \sum_{i} \frac{\left(x_i - x_i'\right)^2}{\sigma_{x_i}^2}, \nonumber
\end{equation}
where $x_i$ and $\sigma_{x_i}$ are the observational input parameters and uncertainties, respectively, $x_i'$ is the computed value from the model, and $i$ iterates over the input parameters.

\begin{figure*}[ht]
    \centering
    \includegraphics[width=15cm]{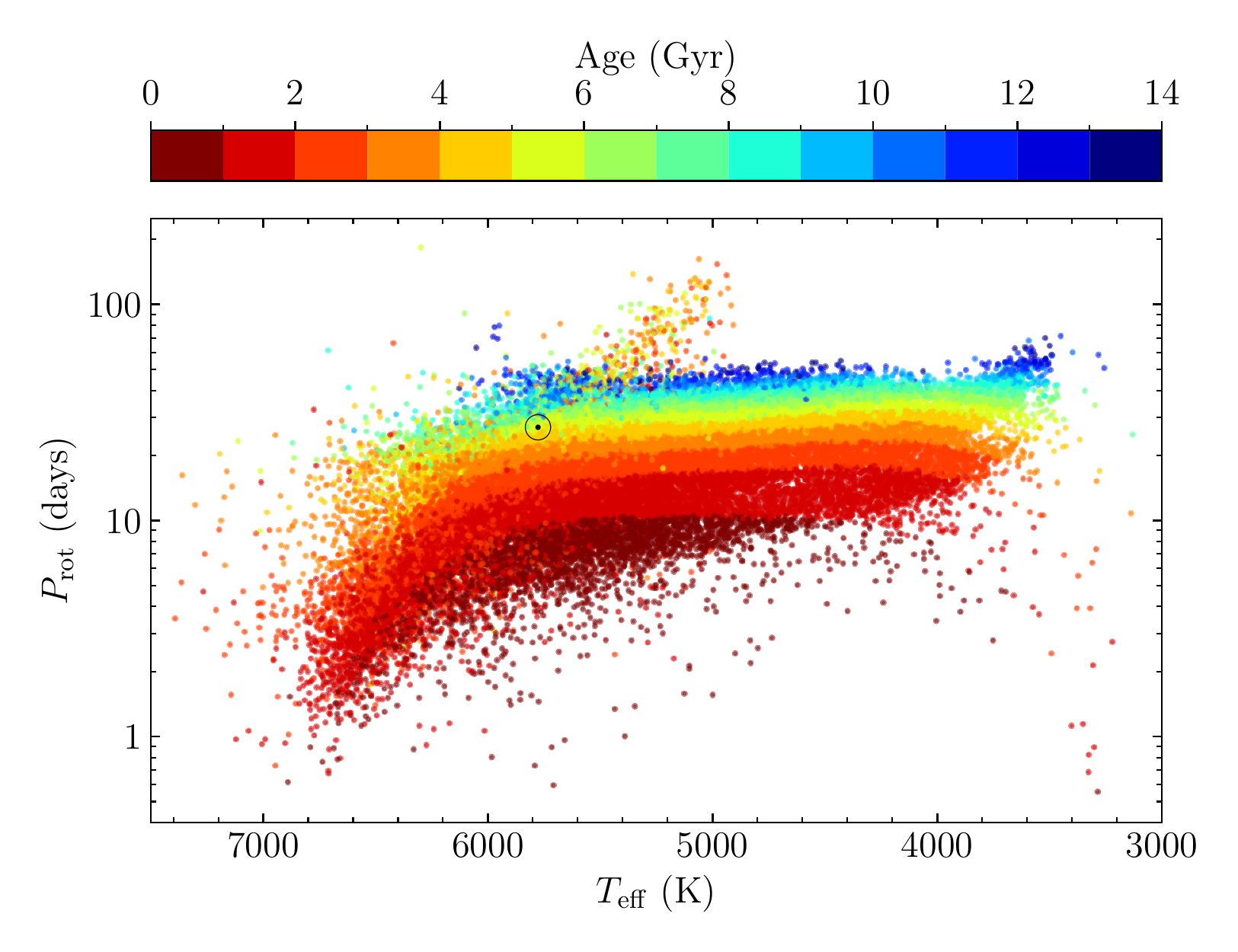}
    \caption{$P_\text{rot}$ vs. $T_\text{eff}$ color-coded by ages computed with \texttt{kiauhoku} for 41,440 stars. The Sun is represented with its usual symbol in black with a rotation period of 26.43\,d \citep{2023A&A...672A..56S}.}
    \label{Prot_Teff_ages}
\end{figure*}

The observational data used as input include the effective temperature, metallicity, luminosity, and rotation period. These stellar parameters were obtained from different catalogs that we describe here. Regarding the atmospheric parameters, we took in priority values from three spectroscopic surveys: the {\it Kepler} Community Follow-up Observation Program \citep[CFOP,][]{2018ApJ...861..149F} with mid- and high-resolution spectroscopic observations of {\it Kepler Object of Interest} and seismic targets, the Data Release 16 (DR16) of the Apache Point Observatory for Galactic Evolution Experiment survey \citep[APOGEE,][]{2020ApJS..249....3A} and the DR7 of the Large Sky Area Multi-ObjectFiber Spectroscopic Telescope \citep[LAMOST,][]{2012RAA....12..723Z,2020ApJS..251...15Z}. We then completed the atmospheric parameters for the stars without spectroscopic observations with first the {\it Gaia-Kepler} stellar properties catalog \citep{2020AJ....159..280B} and then with the DR25 {\it Kepler} stellar properties catalog \citep{2017ApJS..229...30M}. We then used {\it Gaia} DR2 luminosities from \citet{Berger2020} and rotation periods from \citet{2019ApJS..244...21S, 2021ApJS..255...17S}.

For the full sample of S19 and S21 of more than 55,000 stars, we performed two MCMC runs using slightly different sets of input parameters. To derive age and other fundamental stellar parameters, we used temperature, metallicity, luminosity, and rotation period. For 3,167 targets ($\sim$\,6\% of the full sample) with missing luminosities, we used the spectroscopic surface gravity instead. However, occasionally we were unable to fit a single model to both the luminosity (or gravity) and rotation period. We flagged those stars (flag set to 1) if that was the case and they were removed from the analysis below. {  However, these atypical stars can reflect that the observations do not agree among themselves or that the stars went through an unusual evolution (such as binarity) and they can be interesting for further analysis. They represent $\sim$\,1\% of the full sample so removing them from our analysis to have a clean sample with an usual evolution will not bias our results.} We also note that the Rossby number is another important quantity to consider in terms of magnetic activity. However the study with Rossby number is out of the scope of this paper and will be presented in a different study (Mathur et al. in prep.).

\section{Age comparison with other models}

We computed ages using the method described in the previous section for 55,232 {\it Kepler} main-sequence and subgiant stars. We describe the results in terms of statistics in order to study general evolution of photometric magnetic activity. We also compare the ages from different stellar evolution models using the same or different input parameters to estimate systematics and possible biases.

\subsection{Ages from YREC models combined with \texttt{kiauhoku}}


\begin{figure*}[ht]
    \centering
    \includegraphics[width=15cm]{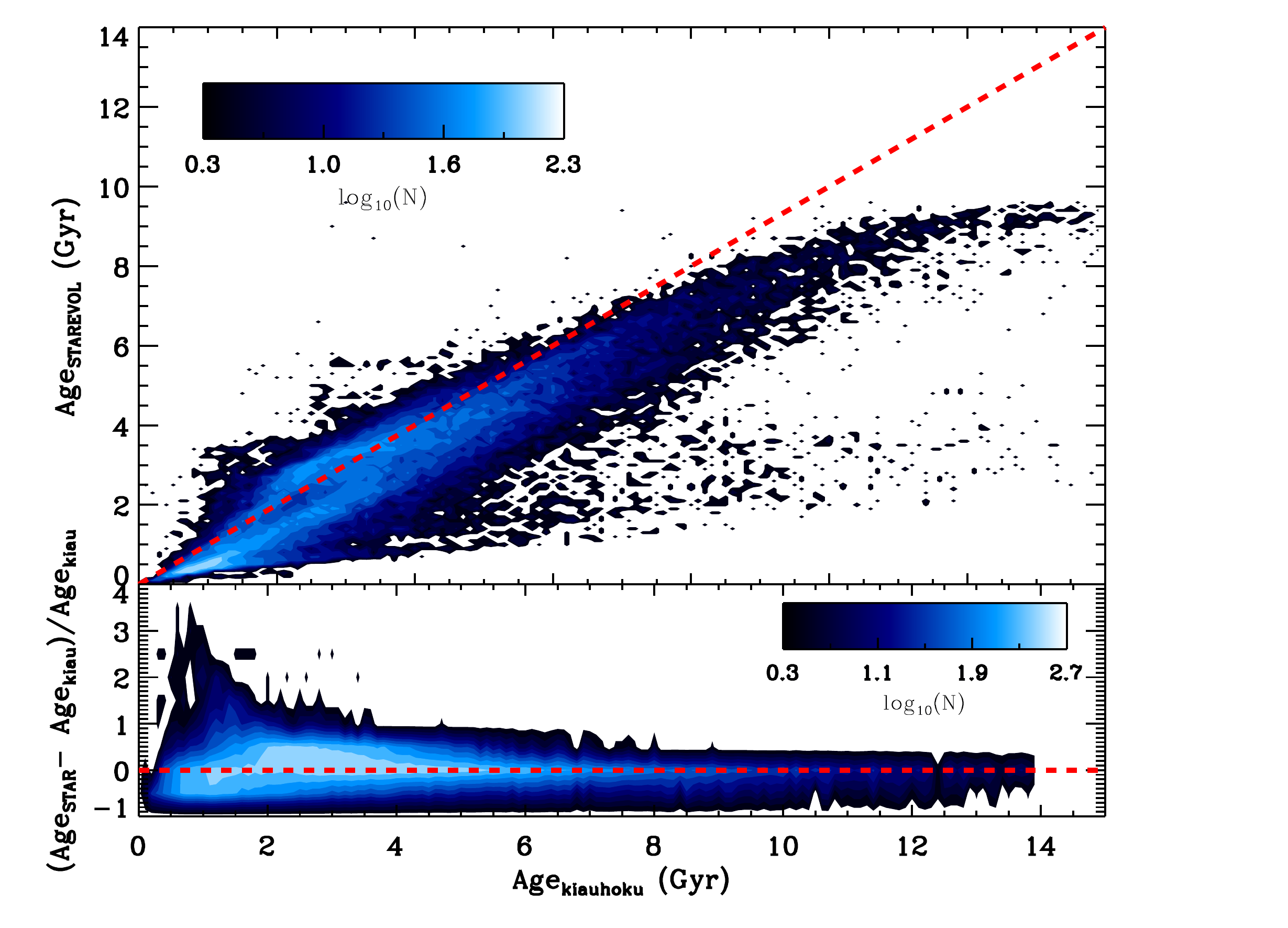}
    \caption{ {\it Top:} Comparison of ages computed by \texttt{kiauhoku} and STAREVOL color-coded by density number of stars. The dashed line corresponds to the 1:1 relation. {\it Bottom:} Ratio of age differences over $\sigma$, the square root of the sum of the quadratic uncertainties from each method. The dotted-dashed lines represent $\pm$\,1\,$\sigma$.}
    \label{ages_YREC_STAR_comp}
\end{figure*}

From the \texttt{kiauhoku} analysis, ages were obtained for the full sample of stars. However there are some caveats to consider in order to select the most reliable ages. For this analysis, we wanted to remove as much as possible potential binary stars. Hence, we discarded stars for which {\it Gaia} Renormalized Unit Weight Error (RUWE) is larger than 1.2, as well as stars that are part of the {\it Gaia} DR3 non-single star catalog \citep{2022arXiv220605439H}, which are either SB1 or eclipsing binaries. Finally we removed the stars where the {\it CPCB1} flag (for close binary candidates based on the light curves, see S19 for more details) was set to 1. As aforementioned, for approximately 3000 stars, no luminosity from {\it Gaia} was available and we used surface gravity. This also means that the Renormalized Unit Weight Error (RUWE) was not defined for those stars. We thus discarded them as well. From the modeling results we removed stars with age older than 14 Gyr, stars whose posterior age uncertainty was greater than 60\% (which is the optimal cut to remove stars whose age posteriors were poorly converged), and stars whose posterior luminosity mean and median were not mutually consistent (this occurred when the sampler favored two very different model fits equally, producing a bimodal posterior). If any other flag was set to 1 (see Section 3), we also discarded the stars from our analysis. This leads to a sample of 41,440 stars with reliable ages according to our criteria. The ages for these stars are provided in Table~\ref{tab:models}. In the remainder of this paper, we will refer to these ages as Age$_{\rm kiauhoku}$.

In Figure~\ref{Prot_Teff_ages}, we plot the rotation period from S19 and S21 as a function of effective temperature and color-coded with Age$_{\rm kiauhoku}$. We clearly see the boundaries of the gyrochrones. The range 4-5\,Gyrs is shown in yellow. When going into more details, we find some fast old rotators with an age\,$>$\,5\,Gyr and $P_{\rm rot} <$\,20\,days mostly in the range of effective temperature 5,000-6,000\,K. 
The stars with $P_{\rm rot}$ longer than 50\,days with $T_{\rm eff}$ between 5,000 and 6,000\,K correspond to subgiant stars that have already slowed down, due to radius expansion.

\begin{longrotatetable}
\begin{deluxetable*}{cccccccccccccc}
\tablewidth{700pt}
\tabletypesize{\scriptsize}
\tablecaption{Input parameters for modeling ($\teff$, $L$, [Fe/H]) and resulting stellar fundamental parameters ($\log$\,g, $M$, $R$, Age, EEP) from \texttt{kiauhoku}.}
\startdata
KIC & $T_\text{eff}$ (K) & $\log L/L_{\odot}$ & [Fe/H] (dex)& $\log\,g$ (dex)&  $M$ (M$_\odot$)& $R$ (R$_\odot$)& Age (Gyr)& EEP & RUWE & flag$_{\rm spec}$ & flag$_{\rm KOI}$ & flag$_{\rm bin}$ & flag\\
\hline
\hline
757099 &  5364\,$\pm$\,  93&   -0.066\,$\pm$\,   0.028& 0.080\,$\pm$\,0.140& 
4.497$^{+0.016}_{-0.016}$& 1.10$^{+0.03}_{-0.03}$& 0.98$^{+0.03}_{-0.03}$& 
 0.05$^{+ 0.01}_{- 0.00}$& 203.9&    2.2& 3& -999& 2& 1\\
757450 &  5301\,$\pm$\, 107&   -0.182\,$\pm$\,   0.036& 0.240\,$\pm$\,0.130& 
4.513$^{+0.013}_{-0.014}$& 0.98$^{+0.03}_{-0.02}$& 0.91$^{+0.03}_{-0.02}$& 
 2.61$^{+ 0.33}_{- 0.32}$& 262.5&    1.0& 3&    0& 0& 0\\
891916 &  5650\,$\pm$\, 134&    0.239\,$\pm$\,   0.296& 0.020\,$\pm$\,0.150& 
4.513$^{+0.036}_{-0.862}$& 1.03$^{+0.67}_{-0.06}$& 0.93$^{+2.30}_{-0.07}$& 
 0.46$^{+ 0.69}_{- 0.07}$& 223.5&    8.7& 3& -999& 0& 1\\
892195 &  5333\,$\pm$\,  92&   -0.136\,$\pm$\,   0.028& 0.070\,$\pm$\,0.140& 
4.495$^{+0.015}_{-0.016}$& 0.97$^{+0.03}_{-0.02}$& 0.92$^{+0.03}_{-0.03}$& 
 3.42$^{+ 0.76}_{- 0.67}$& 283.1&    1.1& 3& -999& 0& 0\\
892713 &  6238\,$\pm$\, 126&    1.252\,$\pm$\,   0.030& 0.080\,$\pm$\,0.210& 
3.495$^{+0.069}_{-0.046}$& 1.71$^{+0.12}_{-0.09}$& 3.85$^{+0.24}_{-0.35}$& 
 1.33$^{+ 0.18}_{- 0.17}$& 529.4&    1.0& 3& -999& 0& 0\\
892834 &  4823\,$\pm$\,  86&   -0.584\,$\pm$\,   0.032&-0.060\,$\pm$\,0.120& 
4.625$^{+0.011}_{-0.008}$& 0.78$^{+0.02}_{-0.02}$& 0.71$^{+0.02}_{-0.02}$& 
 1.39$^{+ 0.16}_{- 0.14}$& 219.9&    1.1& 3& -999& 0& 0\\
892882 &  5149\,$\pm$\, 153& -999.000\,$\pm$\,-999.000&-0.180\,$\pm$\,0.300& 
4.573$^{+0.036}_{-0.035}$& 0.86$^{+0.06}_{-0.07}$& 0.79$^{+0.06}_{-0.06}$& 
 3.28$^{+ 0.52}_{- 0.44}$& 247.8& -999.0& 3& -999& 0& 1\\
893033 &  4707\,$\pm$\,  84&   -0.715\,$\pm$\,   0.034&-0.240\,$\pm$\,0.110& 
4.647$^{+0.010}_{-0.012}$& 0.70$^{+0.02}_{-0.02}$& 0.66$^{+0.02}_{-0.02}$& 
 3.92$^{+ 0.52}_{- 0.50}$& 233.5&    1.1& 3& -999& 0& 0\\
893209 &  6051\,$\pm$\, 108&    0.586\,$\pm$\,   0.036& 0.030\,$\pm$\,0.140& 
4.156$^{+0.032}_{-0.029}$& 1.42$^{+0.03}_{-0.05}$& 1.65$^{+0.06}_{-0.07}$& 
 1.82$^{+ 0.30}_{- 0.19}$& 335.3&    1.8& 3& -999& 0& 1\\
893286 &  5297\,$\pm$\, 100&   -0.302\,$\pm$\,   0.039&-0.040\,$\pm$\,0.140& 
4.550$^{+0.014}_{-0.013}$& 0.88$^{+0.02}_{-0.03}$& 0.82$^{+0.02}_{-0.03}$& 
 4.01$^{+ 0.61}_{- 0.57}$& 263.4&    1.0& 3& -999& 0& 0\\
893383 &  5680\,$\pm$\, 103&   -0.152\,$\pm$\,   0.034&-0.170\,$\pm$\,0.130& 
4.518$^{+0.016}_{-0.016}$& 0.91$^{+0.03}_{-0.03}$& 0.87$^{+0.03}_{-0.03}$& 
 3.63$^{+ 0.80}_{- 0.69}$& 274.1&    1.0& 3& -999& 0& 0\\
\enddata
    \label{tab:models}
\tablecomments{RUWE is from {\it Gaia} DR3. flag$_{\rm spec}$ corresponds to the origin of the atmospheric parameters (0:CFOP, 1:APOGEE, 2:LAMOST, 3:B20). flagKOI is the flag for KOI (-999=non-KOIs, 0=confirmed planet hosts, 1=candidate planet-hosts, 2=false-positives). flagbin is the flag for potential binaries (0:not a binary according to other works, 1: CPCB1, 2: binaries as flagged in S19 and S21{  , 3: binaries from {\it Gaia} NSS)}. The column flag is set to 1 if either of the previous flag is positive, when RUWE$>$1.2, when luminosity is not available or when \texttt{kiauhoku} did not converge. The full table is available online in a machine readable format.}
\end{deluxetable*}
\end{longrotatetable}

\subsection{Comparison with ages from the STAREVOL code}\label{sec:YREC_STAREVOL}

To check the reliability of our ages, a second set of ages was also computed with the same input parameters described above but with a different stellar evolution code.
We used the stellar evolution grid of \citet{Amard2019}, which was computed using the STAREVOL code \citep[][]{2000A&A...358..593S,Lagarde2012,2016A&A...587A.105A,Amard2019}. The models include a self-consistent treatment of internal angular momentum transport, and angular momentum extraction by magnetised winds following \cite{Matt2015}, calibrated to reproduce open clusters rotation period distributions \cite[see the grids paper for more details][]{Amard2019}.
We use a MCMC maximum likelihood tool adapted from the SCePtER software by \cite{Valle2014} to interpolate in the grid of evolution models at intermediate rotation rates. The estimated values of age and mass are obtained by averaging the age and the mass of all the models with likelihood greater than $0.99 \times \mathcal{L}_{\textrm{max}}$ similarly to what was done in \cite{Amard2020} to estimate only the stellar masses. The stellar parameters obtained with STAREVOL are given in Appendix~\ref{appendix:age_comp_Teff}.

This allows us to quantitatively assess the impact of the different physics included in the models. The set of stars in common with reliable ages from both stellar evolution codes consists of 40,220 stars. We can see (Figure~\ref{ages_YREC_STAR_comp}) that \texttt{kiauhoku} and STAREVOL ages {   are well correlated in general} with a Spearman correlation coefficient of 0.89. {   However we discuss the differences in more detail below.} In the lower panel, the difference between the two sets of ages is {   centered on 0 with some departure for younger stars. We also found that compared to the reported uncertainties from both stellar modeling approaches, the age differences are} on average of -1.22\,$\sigma$, where $\sigma$ is the sum of the quadratic uncertainties from each method. This suggests that the true uncertainty in the age is dominated by the systematic uncertainty inherent in choosing a specific model.

\begin{figure}[ht]
    \centering
    \includegraphics[width=9cm]{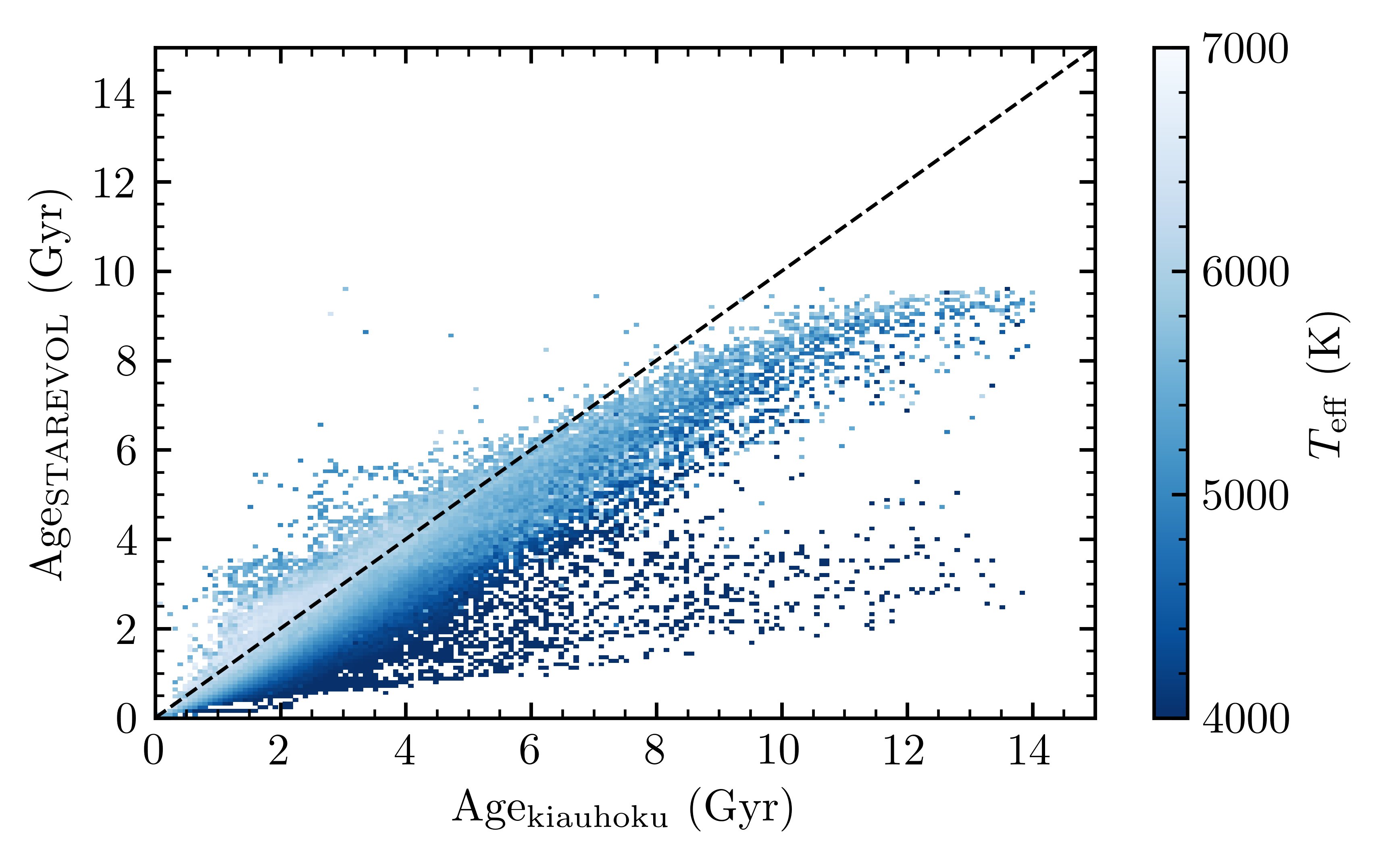}
    \caption{{  Ages from \texttt{kiauhoku} compared to STAREVOL color-coded with $\teff$. }}
    \label{age_comp_Teff}
\end{figure}

 While there is generally good agreement between the \texttt{kiauhoku} and STAREVOL ages, there is somewhat worse agreement in older ($>5$ Gyr) and in cooler ($<4500$ K) stars, with up to 5$\sigma$ {  difference} {  (see Fig.~\ref{age_comp_Teff})}. These discrepancies can be explained by the differences in angular momentum transport in the models underlying the two ages estimates. The most likely source of {  disagreement} is the presence of internal differential rotation in the STAREVOL models, which the YREC-based \texttt{kiauhoku} models lack. Differential rotation implies that the torque is stronger at the age of the Sun \citep[See][]{Amard2016,SomersPinsonneault2016}
 , resulting in faster spin-down and a younger age given a rotation period. The effect is more pronounced at old ages, where the models have had more time to diverge in evolution, and in cool stars, which experience stronger differential rotation.

 Other differences between the models include the weakening of magnetic braking \citep[e.g.,][]{2016Natur.529..181V} in the YREC-based models at a given Rossby number, which the STAREVOL models do not include. The weakened braking allows for older ages at shorter periods, which may contribute to the age discrepancy, especially for the older \texttt{kiauhoku} ages. Additionally, both sets of models are calibrated to solar-mass stars, which could lead to different and discrepant behaviors in the coolest regime \citep[See \textit{e.g.}][]{2020ApJ...889..108A}. 

 We also note that STAREVOL predicts significantly younger stars below 1\,Gyr compared to \texttt{kiauhoku}. The differences at these young ages could be explained by the different initial conditions used. \cite{Claytor2020} begin the evolution with a 0.28\,Myr disc-locking time at a rotation period of 8.1 days, while \cite{Amard2019} start at 4.5\,days after 5\,Myr. By 5\,Myr, the YREC-based models have spun-up under contraction to periods of 1.25 to 1.75\,days. The initial conditions are only expected to affect the age estimates significantly for stars younger than 1 Gyr. Older than this, spin-down ensures that stars ``forget” their initial conditions \citep[e.g.,][]{Claytor2020}.

 Finally, stellar metallicity may also contribute to the age discrepancies, as it has a non-negligible impact on the wind torque \citep{2020ApJ...889..108A}. Recently \citet{2022ApJ...939L..26B} showed that metallicity has a significant impact on the rotational evolution of active stars in general.

\subsection{Comparison with published ages of the Kepler sample}


Several catalogs of ages of the {\it Kepler} field stars have been compiled and we now compare them with our derived ages. The most classical method consists of isochrone fittings in a Color-Magnitude Diagram (CMD) or in a Kiel diagram where atmospheric parameters are fitted. This was done in a homogeneous way for the full or almost full {\it Kepler} sample in several {\it Kepler} star properties catalogs \citep{2011AJ....142..112B,2014ApJS..211....2H,2017ApJS..229...30M,2020AJ....159..280B}. In the most recent of these catalogs, which we refer to as B20, {\it Gaia} DR2 \citep{2018A&A...616A...1G} luminosity was also used. 

Another homogeneous and large-scale catalog was derived using gyro-kinematics relations from age-velocity dispersion \citep[hereafter L21;][]{2021AJ....161..189L} where they combined gyrochronology (using previous rotation periods measurements from {\it Kepler}) and vertical velocities from {\it Gaia} or LAMOST observations. It has been observed that vertical velocity increases with ages. However, by themselves, vertical velocities can only provide {  statistical} ages. By combining them with gyrochronology, it is then possible to extract individual ages. This work led to a catalog of almost 30,000 stars with gyro-kinematic ages. 


\begin{figure*}[h!]
    \centering
    \includegraphics[width=8.5cm]{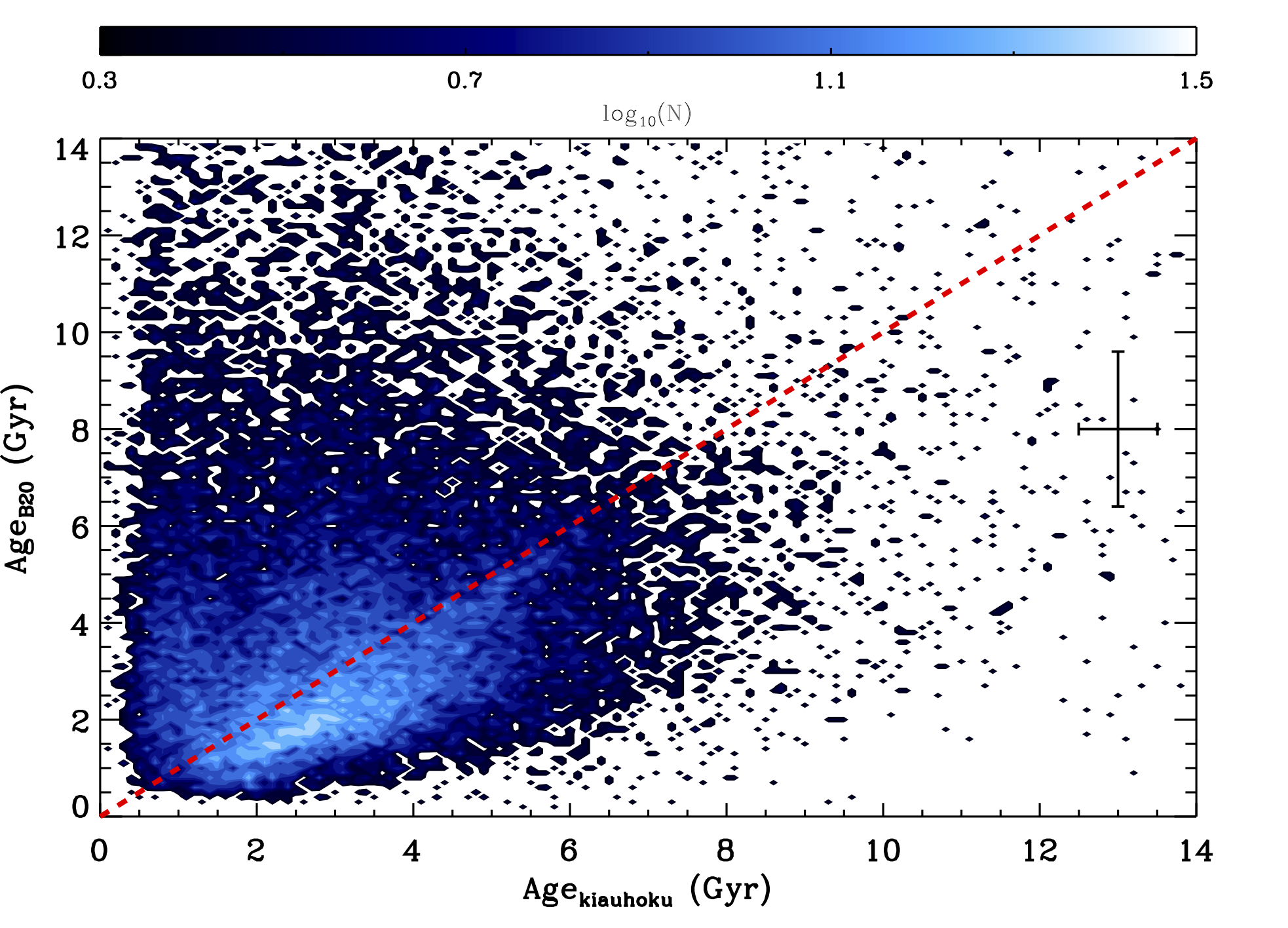}
    \includegraphics[width=8.5cm]{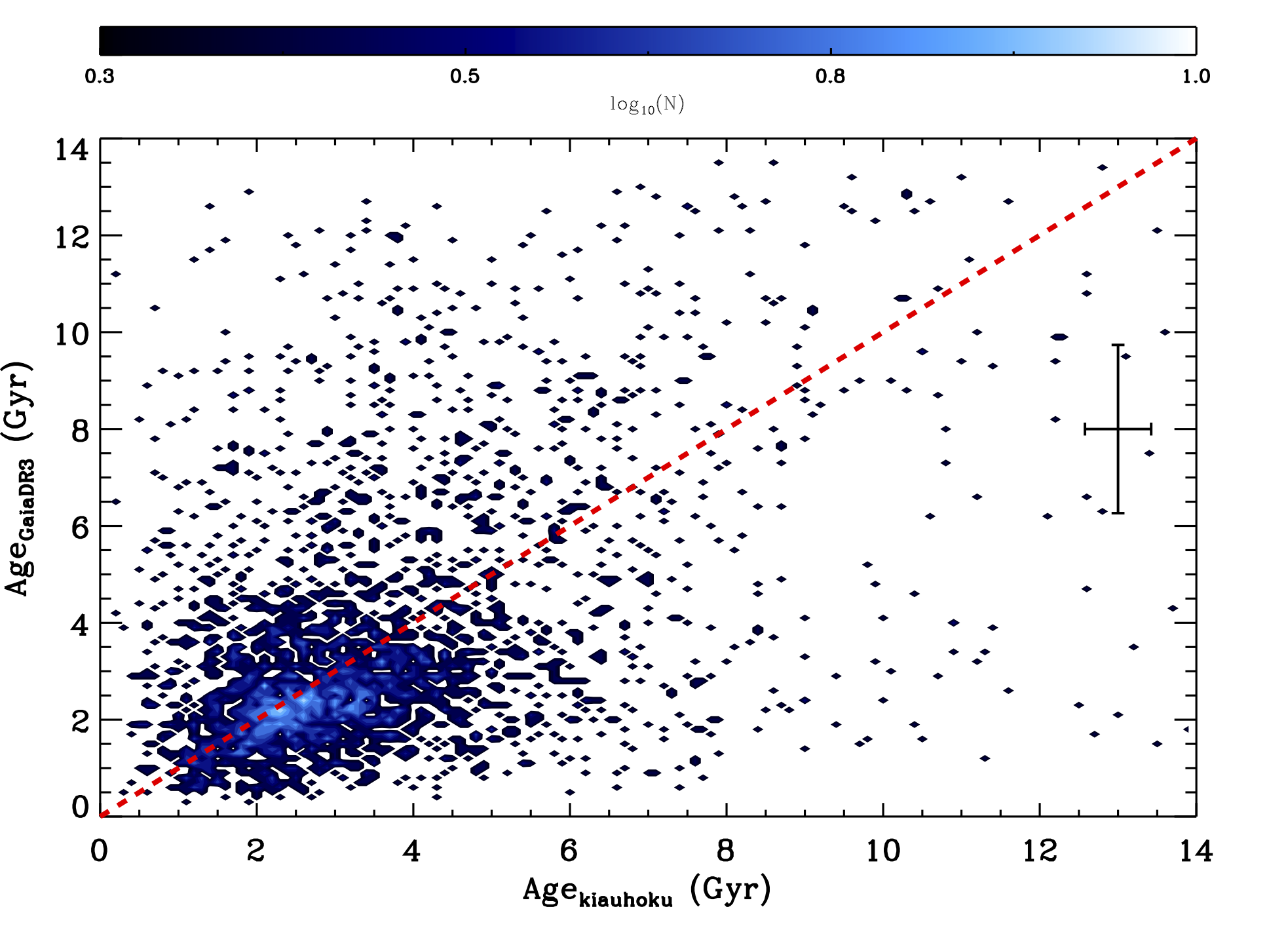}\\
    \includegraphics[width=8.5cm]{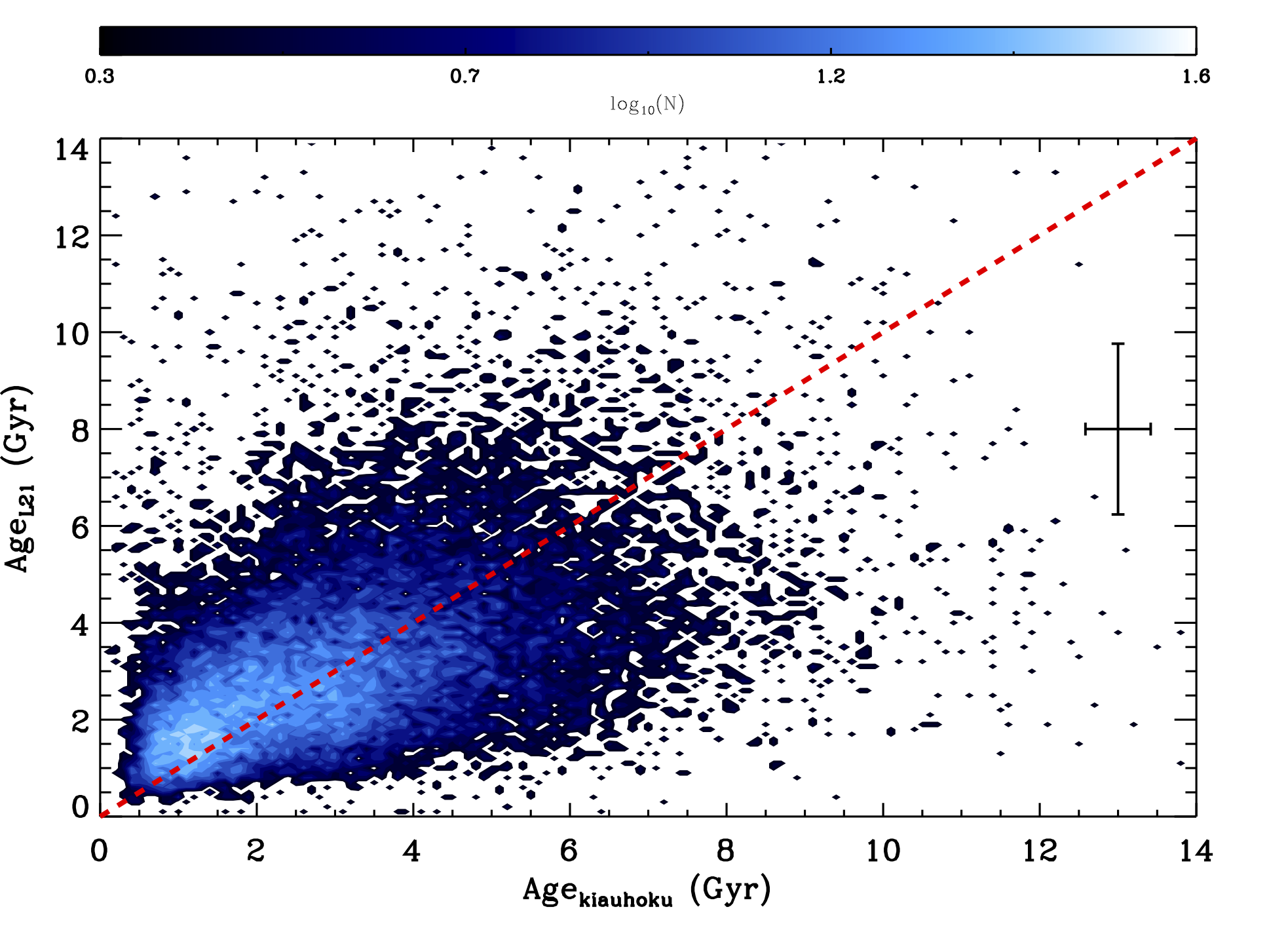}
    \includegraphics[width=8.5cm]{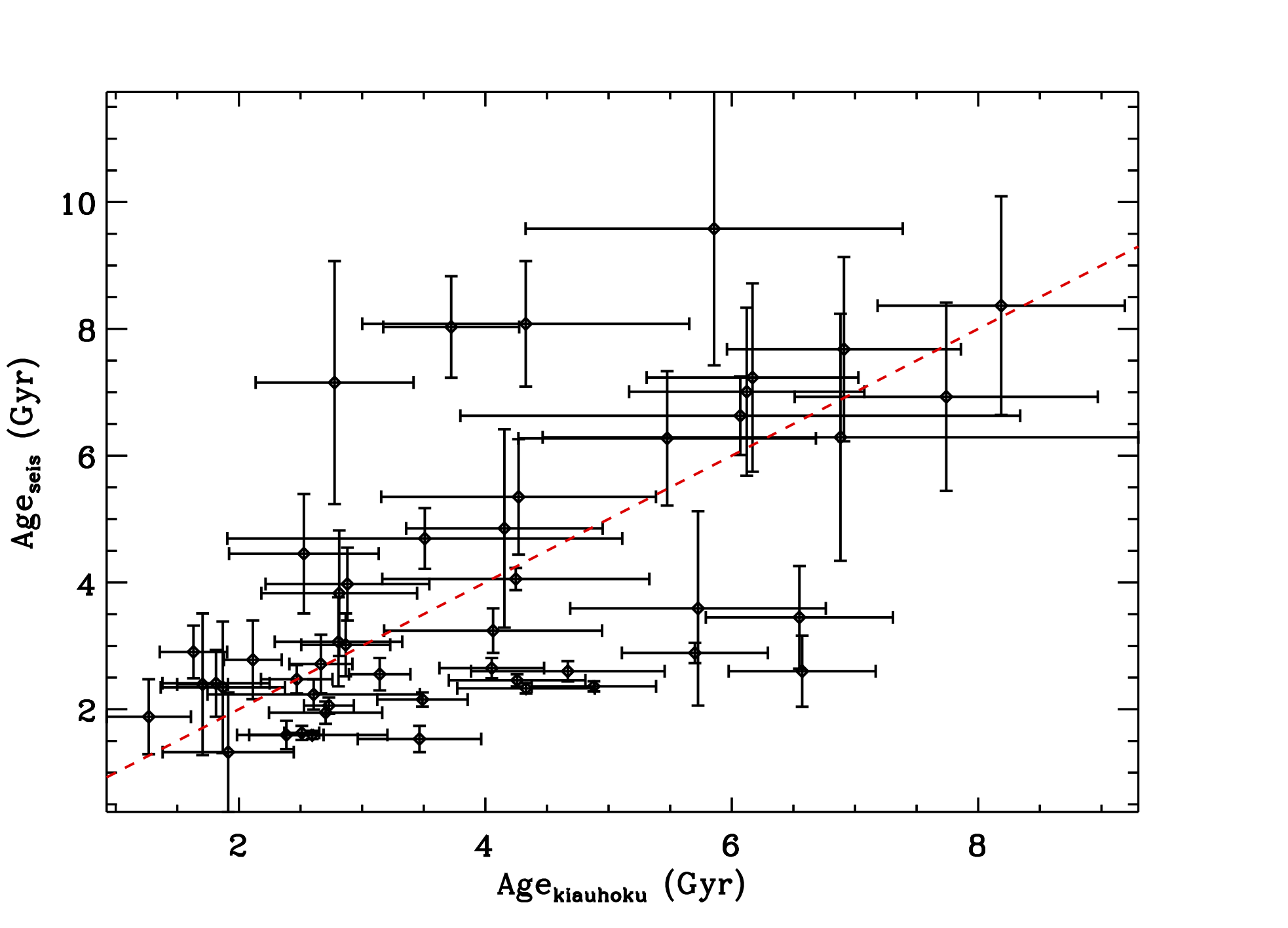}\\
    \caption{Literature ages as a function of the ages obtained in this work (Age$_\texttt{kiauhoku}$). Top left panel:  {\it Gaia-Kepler} catalog (B20). Top right panel: {\it Gaia} DR3 FLAME ages (C22). Bottom left panel: gyro-kinematic ages (L21). Bottom right panel: seismic ages (SA17). The red dashed line corresponds to the 1:1 line. In the cases of the B20, {\it Gaia} DR3, and L21, only the average uncertainties are represented for clarity purposes. }
    \label{published_ages_comp}
\end{figure*}

We also took advantage of the {\it Gaia} DR3 ages obtained with the Final Luminosity Age Mass Estimator \citep[FLAME,][hereafter C22]{2022arXiv220605864C}, that is available for 2945 of our targets. These ages are the most reliable for the {\it Gaia} solar-like stars as they are computed based on spectroscopic parameters as well as astrometry and photometry. 

Finally, on a smaller scale but with a higher level of precision, we consider the catalogs of seismic ages for dwarfs and subgiant stars. These ages are obtained by finding the best-fit models to atmospheric as well as seismic parameters (global or with individual mode frequencies). A catalog based on the global seismic parameters was done for more than 400 stars \citep{2017ApJS..233...23S} while the detailed analysis with individual frequencies was done for a smaller sample of 99 stars \citep[hereafter SA17]{2015MNRAS.452.2127S,2017ApJ...835..173S}. As it was shown that the precision on stellar parameters is improved when using the additional information from the frequencies of the modes \citep{2012ApJ...749..152M}, we make the comparison with the ages from the smaller sample of stars. SA17 provided ages from 6 different modeling pipelines and we compare our ages with the ones from the BAyesian STellar Algorithm pipeline for consistency between the two seismic samples. However we note that in terms of seismic ages, there can be discrepancies of 1\,Gyr up to 4\,Gyr between model pipelines.

The comparison with the B20 isochrone ages (for 26,262 stars in common) shows a weak correlation, with a Spearman correlation coefficient of 0.27, as it can be seen in the top left panel of Figure~\ref{published_ages_comp}. Some disagreement was expected given that isochrones are inaccurate or imprecise for main-sequence stars cooler than the Sun, which make up two-thirds of our sample. The luminosity of such stars does not evolve fast enough to constrain ages precisely, even with {\it Gaia} measurements. For cool stars, rotation period is a more precise tracer of age. The comparison of the gyro-kinematic ages with the isochrone ages also shows similar disagreement (see appendix of L21). 

Compared to the {\it Gaia} DR3 FLAME ages (top right panel of Fig~\ref{published_ages_comp}), the agreement is reasonably good with a Spearman correlation coefficient of 0.38, so a rather weak to moderate correlation.

In the bottom left panel of Figure~\ref{published_ages_comp}, we can see that there is a better agreement between the gyro-kinematic ages from L21 and \texttt{kiauhoku} ages. The Spearman correlation coefficient of 0.53 suggests a moderate correlation.  While gyro-kinematics ages were derived for individual stars, because they are computed for a given vertical velocity in bins of magnitude, temperature, rotation period, and Rossby number, the ages vary smoothly and a star with an unusual evolution or parameter would get averaged out and might have an over- or under-estimated age. We note that for stars older than $\sim$\,5\,Gyr according to our models, the gyro-kinematics ages are smaller, similarly to what was seen in the comparison with STAREVOL ages (Sect.~\ref{sec:YREC_STAREVOL} and Figure~\ref{ages_YREC_STAR_comp}).  This could reflect the effect of smoothing of the parameter space used by L21. We note that some stars with gyro-kinematics ages are older than 14\,Gyr (up to 25\,Gyr) that have been removed. 

\begin{figure*}[ht]
    \centering
    \includegraphics[width=8cm]{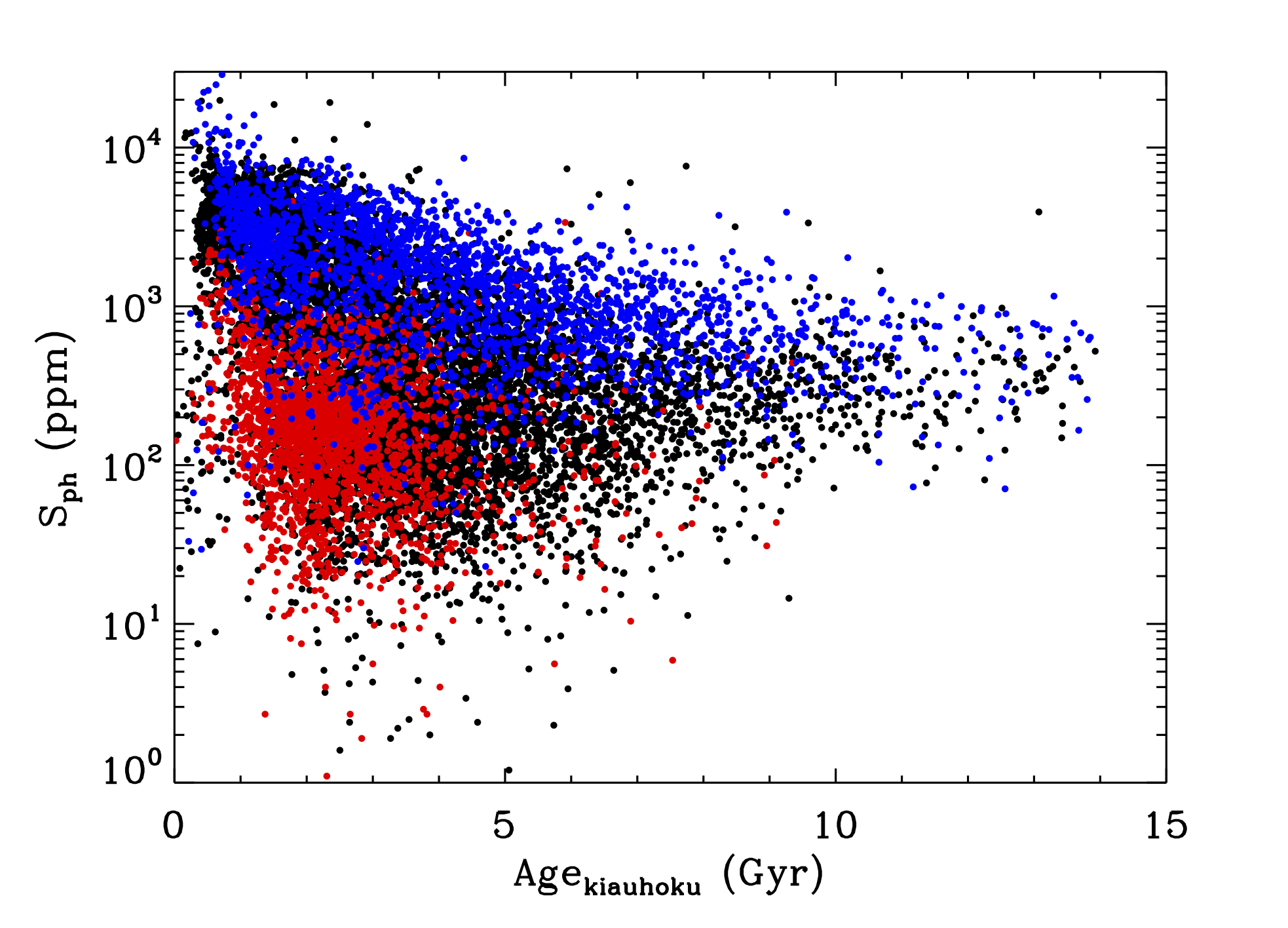}
    \includegraphics[width=8cm]{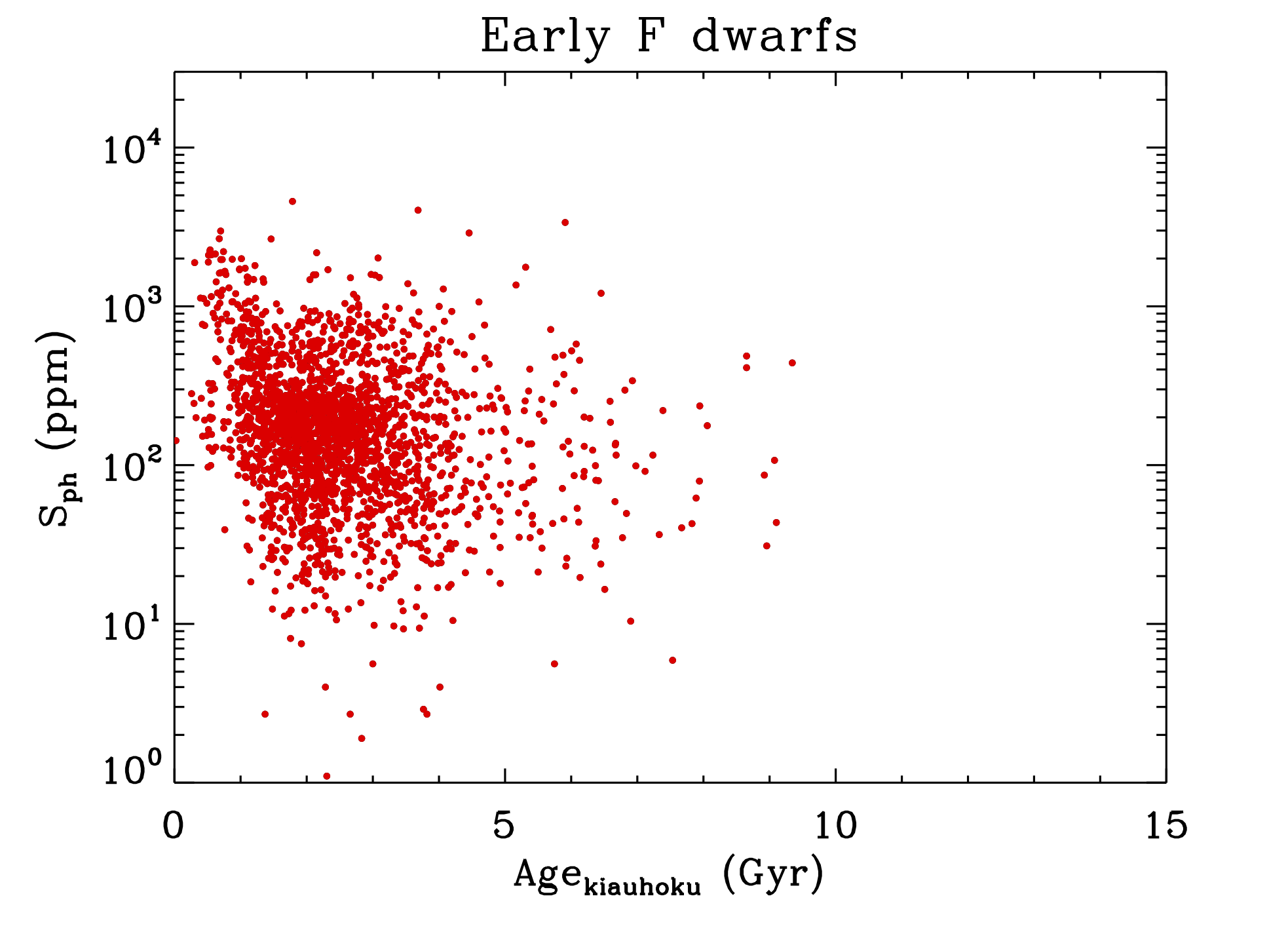}
    \includegraphics[width=8cm]{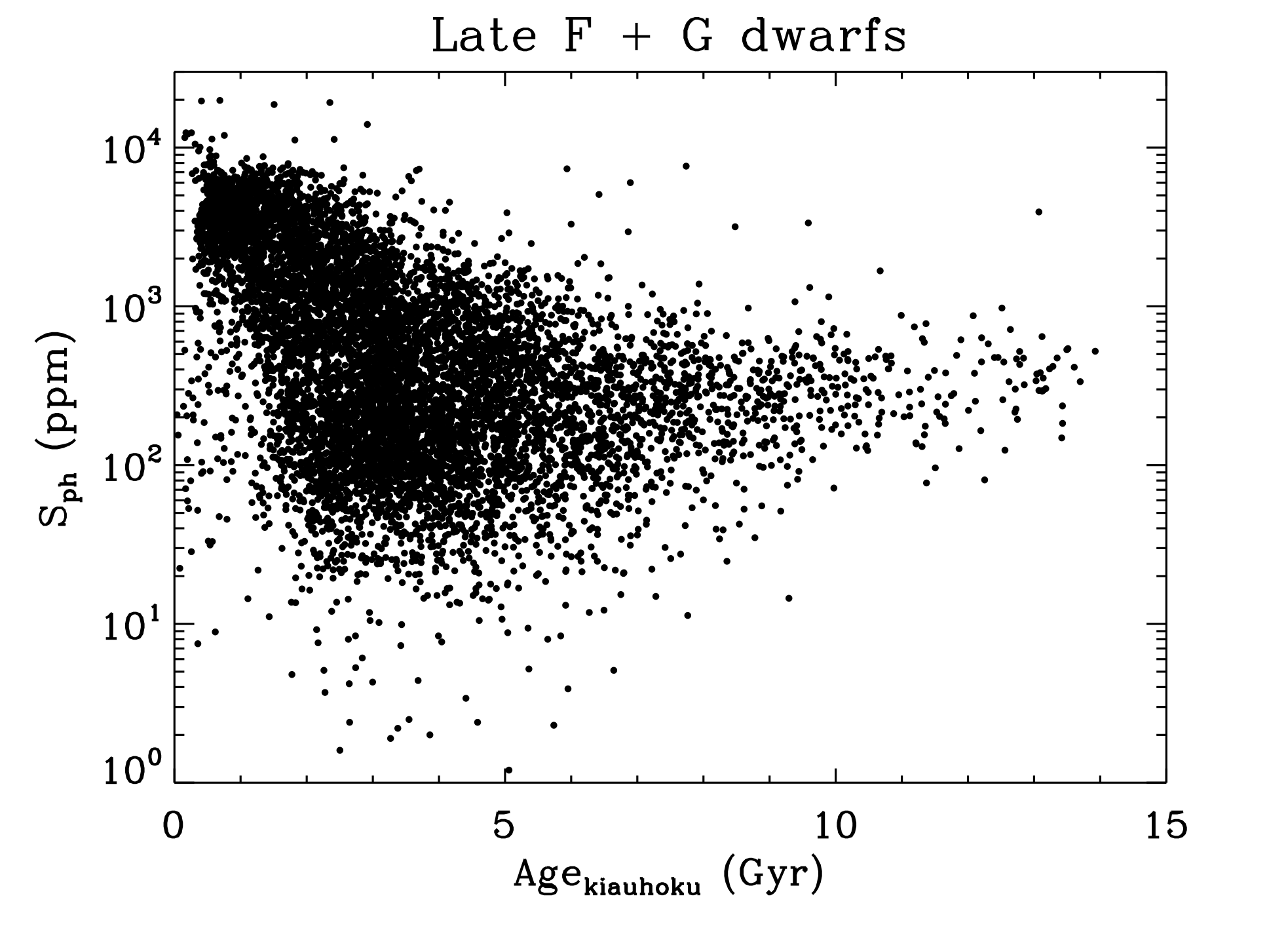}
   \includegraphics[width=8cm]{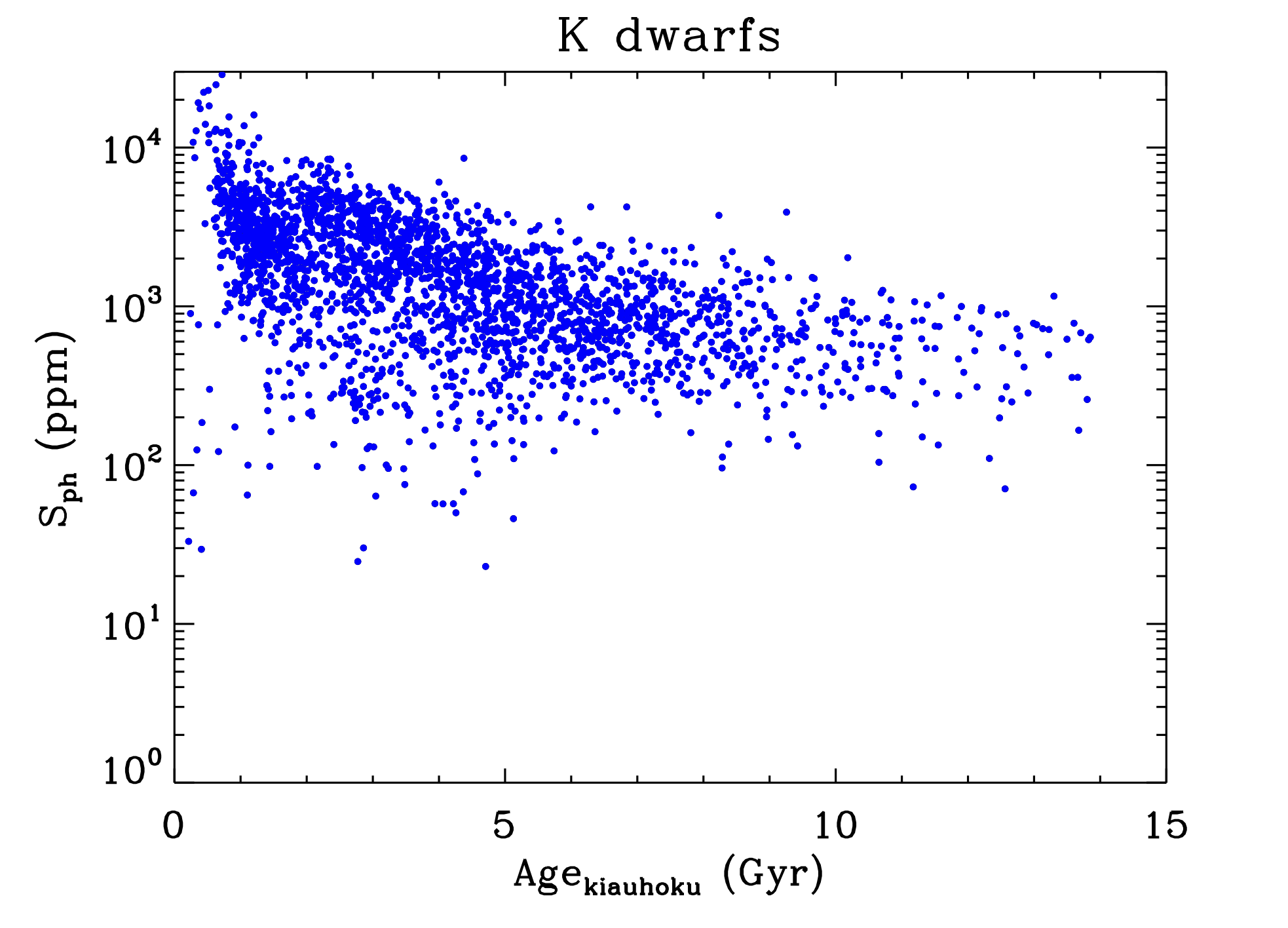}
    \caption{$S_\text{ph}$-age relation for early F (red symbols), late F and G (black symbols), and K (blue symbols) dwarfs all together with spectroscopic data and separated by spectral type. }
    \label{Sph_age_spec}
\end{figure*}

The comparison with the seismic ages in the bottom right panel of Figure~\ref{published_ages_comp} shows a strong correlation with a Spearman correlation coefficient of 0.62. There are some stars with seismic ages between 2 and 4\,Gyr that are older in our models and have small uncertainties (both from seismology and gyrochronology). These stars all have temperatures hotter than 6000 K and are near the Kraft break \citep{Kraft1967}, where the outer convective envelope becomes thin and stars spin down substantially less over time. In this regime, stars' rotation rates are more sensitive to their initial speeds, and the ages are more sensitive to the particular braking prescription used by the models. 

We fit these stars with different spin-down models, varying the models' starting speed and whether they experience weakened magnetic braking halfway through the main sequence. Our default \texttt{kiauhoku}/YREC models include weakened braking and have an initial rotation period of 8.1 days, consistent with observations of other seismic samples and young open clusters. Even using models with no weakened braking and slow launch condition of 13.8 days \citep[consistent with the slow rotators of open clusters;][]{2013ApJ...776...67V}, we recovered the seismic ages for only two of the discrepant stars. 

In case the discrepancy was caused by the rotation period, we also tried relaxing the rotation constraint and fitting a YREC isochrone to the temperature, metallicity, and luminosity of the discrepant seismic stars. This {  reduced the  disagreement} for about half of the stars, although this was at least partly due to wider age posteriors because the models were less constrained.

Since the discrepancy with the seismic stars was not fully solved by varying the braking prescription or relaxing the rotation constraint, it is possible that the offsets arise because we use different models with different input physics compared to SA17. \citet{Tayar2022} showed that fitting different models to the same input parameters can result in large scatter in inferred ages, as high as 50\% near the main sequence. We emphasize that while there may be scatter between different model-dependent ages, our homogeneous modeling procedure produces ages that are internally consistent and highly precise (15\% median relative uncertainty).

\section{Magneto-gyro-chronology}\label{sec:bayesfit}

In this section we investigate how the average magnetic activity, measured through the $S_{\rm ph}$, varies as a function of stellar age, measured using gyrochronology (Sect.~\ref{sec:ageYREC}). To do so, we focus on the main-sequence solar-like stars {  with an Equivalent Evolutionary Phase, EEP \,$<$\,454, which indicates the end of the main sequence, defined here as the point where the core hydrogen fraction falls below $10^{-12}$, consistent with \citet{2016ApJS..222....8D}. We also select stars} with spectroscopic atmospheric parameters, which are the most reliable parameters available for our sample, yielding a sample of 14,637 stars. The goal is to explore the possibility of estimating stellar ages via $S_{\rm ph}$. 

Given that we have very few M dwarfs, in Figure~\ref{Sph_age_spec} we only show the photometric magnetic activity proxy as a function of the age computed with \texttt{kiauhoku} for early F, late-F and G, and K dwarfs.  We can see different behaviours. On the one hand, the K dwarfs are more active than the late-F and G dwarfs and their magnetic activity levels decay more slowly with age. On the other hand, early-F dwarfs {  have a more complex shape that could have two interpretations. Either there is no correlation between $\sph$ and age or when computing the median values on bins of 0.5\,Gyr, we find a slope up to 2\,Gyr. However for the latter, there are fewer points for younger ages that could bias that interpretation.} 
We will thus focus on late F, G, and K dwarfs in the remainder of this paper.

In this section we will perform different fits between the $S_{\rm ph}$ and age by means of the Bayesian inference tool DIAMONDS \citep{2014A&A...571A..71C}. For the fit, we adopt a standard Normal Likelihood function following \citet{2013MNRAS.430.2313C}:

\begin{equation}
{\mathcal{L}} 
\left(\boldsymbol{\theta}\right) = \sum_{i=1}^N \frac{1}{\sqrt{2 \pi} \widetilde{\sigma}_i} \exp \left[ -\frac{1}{2} \left( \frac{\ln \mbox{Age}_i^{\mathrm{obs}} - \ln \mbox{Age}_i^{\mathrm{th}} \left( \boldsymbol{\theta} \right)}{\widetilde{\sigma}_i} \right)^2 \right]
\label{eq:likelihood}
\end{equation}

\noindent where $i$ refers to a single element, $N$ is the total number of observations and uncertainties correspond to the relative uncertainties in Age, i.e. $\widetilde{\sigma}_{\mbox{Age}}$= $\sigma_{\mbox{Age}}/\mbox{Age}$. We also compute the Bayesian Evidence ($\ln \mathcal{E}$) to solve the model comparison problem and identify the best statistical model (as a trade-off between fitting quality and model complexity).

\subsection{Magnetochronology for solar analogs}
\label{sec:linfit}

Studies that are looking for magnetochronology relations usually search for a power-law relation, which translates into a simple linear relation in the log-log space of the magnetic activity proxy as a function of stellar age. Because of the way uncertainties are taken into account in the Bayesian framework, the coefficients found for a given relation do not correspond to the coefficients for the inverse relation. As we want to predict the ages from the magnetic activity proxy, we fitted a power law between the ages determined from the stellar evolution model and $\sph$, which in the log-log space can be expressed as:

\begin{equation}\label{eq:AgeSph}
    \ln ({\rm Age}) = a_0 + a_1 \ln (S_{\rm ph}),
\end{equation}

where Age is in Gyr and $\sph$ in ppm.

We also fitted the inverse relation, i.e. $\sph$ as a function of age:

\begin{equation}\label{eq:SphAge}
    \ln (S_{\rm ph}) = a'_0 + a'_1 \ln ({\rm Age}).
\end{equation}


We will see that the coefficients can be different between the two ways of fitting the relations.

\begin{figure}[ht]
    \centering
    \includegraphics[width=9cm]{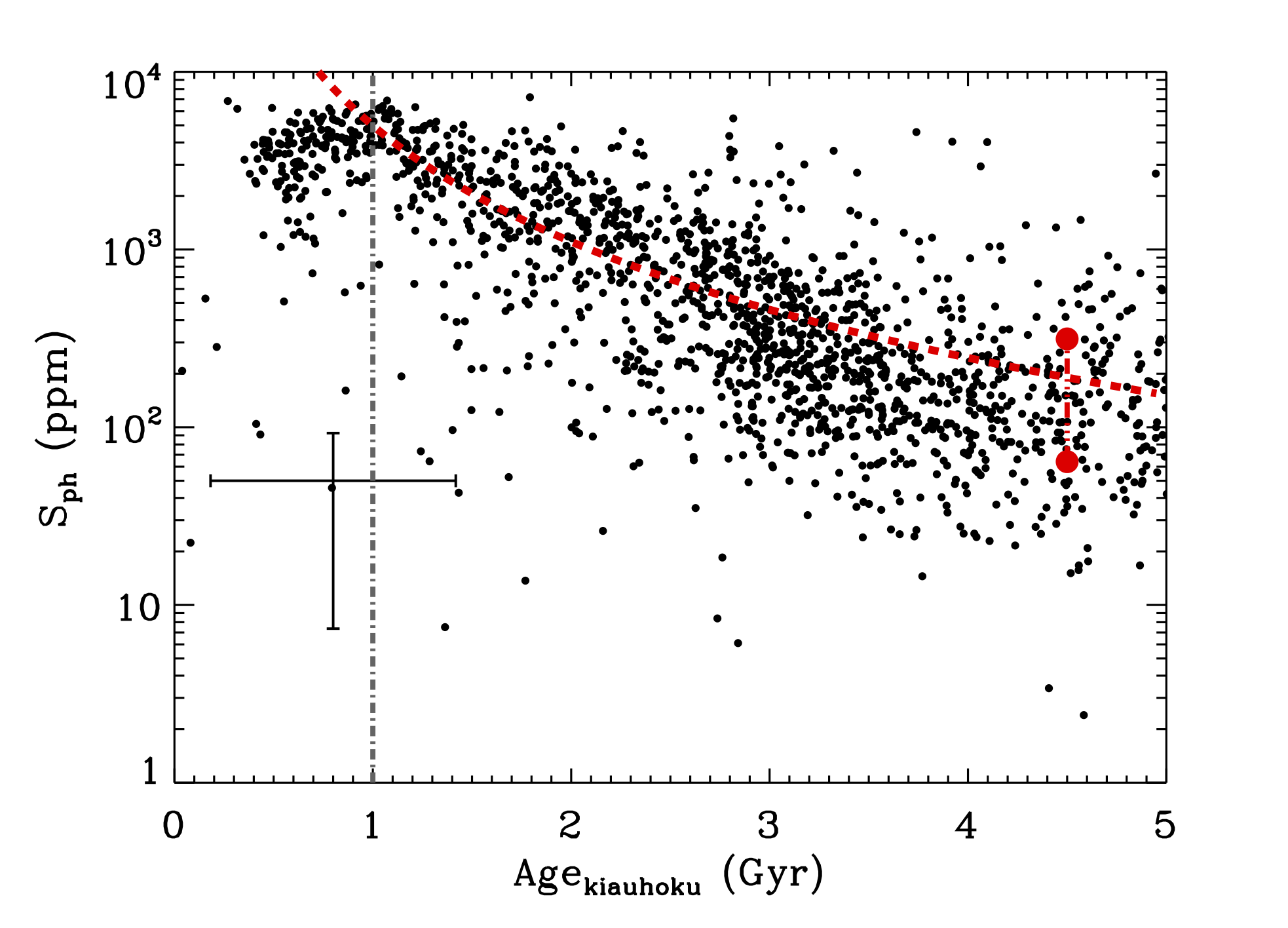}
    \caption{$\sph$ as a function of age for solar analogs where the red dashed line represents the results of the linear fit (Equation~(\ref{eq:SphAge})) done in the log-log space. The fits were performed between 1 and 5\,Gyr and the grey dotted-dashed line shows that lower limit in age. The red large circles represent the $\sph$ values of the Sun between minimum and maximum activity.} 
    \label{Sph_age_SA_fits}
\end{figure}

To have less scatter in our plot and to focus on the most reliable ages, we first selected stars with an effective temperature ranging between 5,700\,K and 5,900\,K (corresponding to 2,064 stars). We will refer to these stars as solar analogs. Figure~\ref{Sph_age_SA_fits} shows how the photometric magnetic activity proxy varies with the stellar age for these solar analogs. We clearly see a trend where the magnetic activity decreases as the star evolves. This behavior is expected and has been seen in other proxies of stellar magnetic activity such as Ca HK or magnetic field \citep[e.g. ][]{2008ApJ...687.1264M,2014MNRAS.441.2361V}. In Figure~\ref{Sph_age_SA_fits}, we observe a {\it plateau} for ages above $\sim$\,5\,Gyr, also seen by \citet{2022ApJ...937...94M} for a small sample of planet-host stars. We can wonder whether the plateau is real or rather a result of detection biases. From that same figure, we clearly see that in our sample of stars with measured rotation periods around 3-4\,Gyr, we have $S_{\rm ph}$ values lower than the {\it plateau}, meaning that our detection limit could be lower than that observed plateau. In Section~\ref{sec:detection_bias}, we discuss more in details the possibility of a detection bias in our sample for low $\sph$ values.

In addition to the {\it plateau} after 5\,Gyr, there is another one below 1\,Gyr. So we performed a first Bayesian fit between 1 and 5\,Gyr for Eq.~(\ref{eq:AgeSph}) and found $a_0$\,=\,$1.937^{+0.022}_{-0.014}$ and $a_1$\,=\,-0.157\,$\pm$\,0.009. 

As mentioned above, we also performed another Bayesian fit, similar to the previous activity index-age relations, with Eq.~(\ref{eq:SphAge}). In that case we found $a'_0$\,=\,$8.513^{+0.004}_{-0.003}$ and $a'_1$\,=\,-2.170\,$\pm$\,0.007 as shown by the dashed red line in Figure~\ref{Sph_age_SA_fits}, which means that $S_{\rm ph} \propto ({\rm Age})^{-2.17}$. This is quite different from what has been found for other proxies of magnetic activity, such as magnetic field or $\log R'_{\rm HK}$, where the exponent varies between -0.5 and -0.6 {  \citep[e.g.][]{1991ApJ...375..722S,2014MNRAS.441.2361V,2018A&A...619A..73L}}. However given that $S_{\rm ph}$ measures a different type of magnetic activity feature in photometric data than the chromospheric one, we do not expect to find the same coefficients. We remind the reader that in the previous works, ages were obtained from simple isochrone fitting with different stellar evolution models. This can also impact the age determination and lead to different coefficients.

We note that the slope from Equation~(\ref{eq:SphAge}) does not correspond to the inverse relation of Equation~(\ref{eq:AgeSph}), which is due to the different variable for which we optimize the inference process. Indeed in the first case of Equation~(\ref{eq:SphAge}), we use errors on $\sph$ while in the other case we use errors on age.

\begin{figure}
    \centering
    \includegraphics[width=9cm]{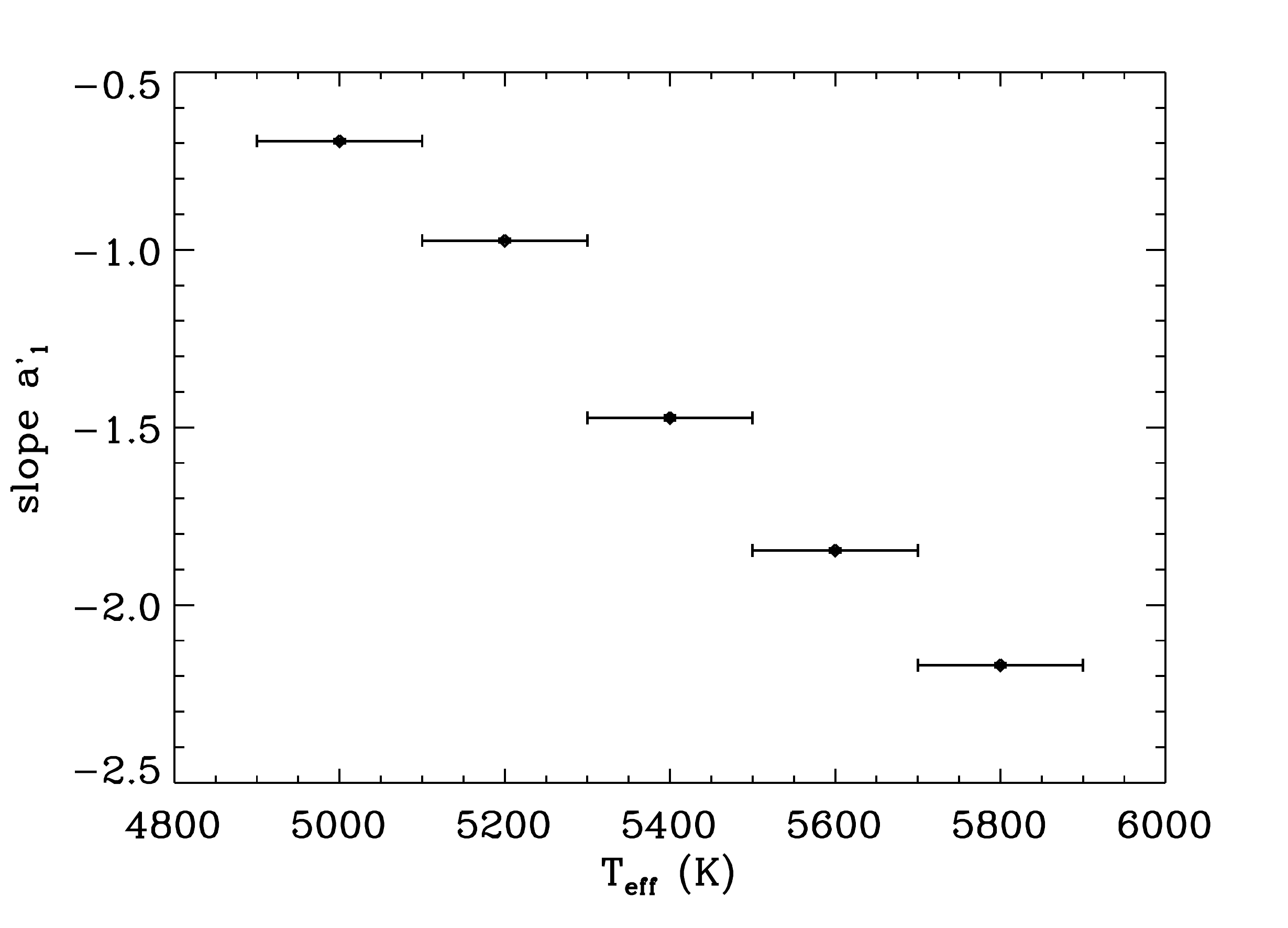}
    \caption{Slope $a'_1$ for Eq.~\ref{eq:SphAge} as a function of effective temperature bins.}
    \label{fig:slope_Teff}
\end{figure}

We then performed the fit by selecting different samples where we changed the range of the effective temperature for both G and K dwarfs and we found that the slope of the $\ln (S_{\rm ph})-\ln (Age)$ relation depends on $T_{\rm eff}$. When $T_{\rm eff}$ decreases, the absolute value of the slope decreases (see Figure~\ref{fig:slope_Teff}), i.e. the relation becomes flatter as seen in Figure~\ref{Sph_age_spec}. This suggests that lower mass stars remain active for longer timescales compared to higher mass stars. The steep $\sph$-Age relation at young ages seen in Figure~\ref{Sph_age_spec} is consistent with these results.

\subsection{Multivariate regression to infer the stellar age}\label{sec:multivar}

As we have seen that the slope in the $S_{\rm ph}$-Age relation varies with effective temperature, we investigate here relationships connecting the Age to multiple observables related to stellar properties. In particular, we consider a multi-power relation of the type:
\begin{equation}
{\rm Age} \left( \alpha_1, \alpha_2, \dots, \alpha_N, \beta \right)  = \beta \prod_{i=1}^{N} O_i^{\alpha_i} \, ,
\end{equation}
\noindent where $O_i$, for $i=1, \dots, N$, are the observables, while $\alpha_i$ are the corresponding exponents that, along with the multiplication factor $\beta$,  have to be estimated from a fit. Following the approach presented by \citet{2013MNRAS.430.2313C} as well as in \citet{2014A&A...571A..35B} and \citet{2017A&A...605A...3C}, we linearize this relation to become:
\begin{equation}
\ln {\rm Age} \left( \alpha_1, \alpha_2, \dots, \alpha_N, \beta \right) = \ln \beta + \sum_{i=1}^N 	\alpha_i \ln O_i \, .
\label{eq:multi-linear}
\end{equation}
In this application, the observables $O_i$ that are taken into account can be: $\sph$, $\teff$,  $L$, $\feh$, $\prot$, thus a total of up to N = 5 components of the multi-linear relation. 

The fit is performed by adopting a specific Likelihood function that is formally equivalent to the standard Normal Likelihood given by Eq.~(\ref{eq:likelihood}), except for the treatment of the uncertainties. Following the work by \citet{2013MNRAS.430.2313C}, the uncertainty on each data point is computed as a total uncertainty arising from that of each of the observables taken into account, including the one on Age. The uncertainty on a single data point $j$ is thus depending on the model parameters $\{ \alpha_i \}$ and it is evaluated as:

\begin{equation}
\widetilde{\sigma_j}^2 \left( \alpha_1, \alpha_2, \dots, \alpha_N \right) = \widetilde{\sigma}_{{\rm Age},j}^2 + \sum_{i=1}^{N} \alpha_i^2 \widetilde{\sigma}_{\mathrm{O},i,j}^2
\label{eq:sig_tilde}
\end{equation}
where $\widetilde{\sigma}$ is the relative uncertainty because we are dealing with logarithmic quantities in the linearized model (e.g. $\widetilde{\sigma}_\mathrm{Age} \equiv \sigma_\mathrm{Age} / \mbox{Age}$).

\subsubsection{Magneto-gyro-chronology for solar analogs}

Focusing on the solar analogs, we tested here a relation between $S_{\rm ph}$, $P_{\rm rot}$, and age. Based on Eq.~(\ref{eq:multi-linear}), this model can be expressed as:

\begin{equation}
\ln \mbox{Age} \left( \alpha_1, \alpha_5, \beta \right) = \ln 	\beta + \alpha_1 \ln \sph + \alpha_5 \ln \prot \,
\label{eq:multi-linear_analogs}
\end{equation}
and with a total uncertainty in a single data point $j$ expressed as in Eq.~(\ref{eq:sig_tilde}) where the observables $O_{i,j}$ belong to \{$\sph$, $P_{\rm rot}$\}.
The estimates of the free parameters are listed in the first line of Table~\ref{tab:multi-linear} for the solar analogs.

The relation with rotation (Equation~(\ref{eq:multi-linear_analogs})) is favored compared to Equation~(\ref{eq:AgeSph}) with a Bayes factor (the difference between the two Evidences) $\ln{B}$ = 165.

\begin{table*}
\small
\centering
 \caption{Fitting coefficients from the multivariate fit for all the models described in Section~\ref{sec:multivar}, predicting the age as a function of other stellar properties (see Equation~(\ref{eq:multivar})).}
\begin{tabular}{lrrrrrcc}
  \hline
  \hline
  \\[-8pt]
 Model & \multicolumn{1}{c}{$\alpha_1$} & \multicolumn{1}{c}{$\alpha_2$} & \multicolumn{1}{c}{$\alpha_3$} & \multicolumn{1}{c}{$\alpha_4$} & \multicolumn{1}{c}{$\alpha_5$} & \multicolumn{1}{c}{$\ln \beta$} & \multicolumn{1}{c} {$\ln\mathcal{E}$}\\[1pt]
 Quantity & \multicolumn{1}{c}{[$\sph$]}& \multicolumn{1}{c}{[$L$]} &\multicolumn{1}{c}{[Fe/H]} & \multicolumn{1}{c}{[$\teff$]} & \multicolumn{1}{c}{[$\prot$]}\\
  \hline
  \\[-8pt]
  Solar Analogs & $-0.122^{+0.017}_{-0.013}$ & -- & -- & -- & $0.387^{+0.015}_{-0.014}$ & $0.728^{+0.049}_{-0.051}$ & --\\[1pt]
 \rowcolor[gray]{0.9}  late-F and G dwarfs & $-0.051^{+0.005}_{-0.005}$ & $0.497^{+0.008}_{-0.008}$ & $-0.618^{+0.026}_{-0.034}$ & $-2.403^{+0.098}_{-0.118}$ & $0.950^{+0.013}_{-0.017}$ & $19.466^{+1.048}_{-0.870}$ & $757$\\[1pt]
  late-F and G dwarfs & $-0.199^{+0.003}_{-0.003}$ &  $0.441^{+0.008}_{-0.008}$ & $-0.774^{+0.022}_{-0.020}$ & $-10.322^{+0.112}_{-0.121}$ & -- & $91.530^{+1.079}_{-0.937}$ & $-3322$\\[1pt]
    late-F and G dwarfs & -- & $0.542^{+0.008}_{-0.009}$ & $-0.661^{+0.033}_{-0.027}$ & $-1.458^{+0.113}_{-0.097}$ & $1.068^{+0.020}_{-0.015}$ & $10.641^{+0.831}_{-0.989}$ & $531$\\[1pt]
 \rowcolor[gray]{0.9}  K dwarfs & $-0.021^{+0.005}_{-0.005}$ & $0.140^{+0.012}_{-0.011}$ & $-0.068^{+0.018}_{-0.021}$ & $-0.373^{+0.100}_{-0.080}$ & $1.384^{+0.013}_{-0.014}$ & $0.326^{+0.709}_{-0.836}$ & $1002$ \\[1pt]
 K dwarfs &  $-0.036^{+0.0110}_{-0.0106}$ & $-0.131^{+0.044}_{-0.047}$ & $-3.114^{+0.066}_{-0.058}$  & $6.244^{+0.392}_{-0.280}$  & -- & $-51.880^{+2.375}_{-3.392}$ & $-2567$ \\[1pt]
  K dwarfs &  -- & $0.149^{+0.012}_{-0.012}$ & $-0.086^{+0.021}_{-0.019}$  & $-0.402^{+0.089}_{-0.092}$  & $1.397^{+0.014}_{-0.014}$ & $0.380^{+0.809}_{-0.764}$ & $998$ \\[1pt]
  \hline
 \end{tabular}
 \flushleft {   Notes.} Median estimators and 68.3\,\% Bayesian credible limits are reported for each free parameter. Models favored according to the Bayesian model comparison (one for late-F and G dwarfs and one for K dwarfs) are highlighted by a gray-shaded background. $\ln\mathcal{E}$ corresponds to the Bayesian Evidence.
\label{tab:multi-linear}
\end{table*}

\subsubsection{Adding atmospheric parameters}


Given the different slopes obtained for different ranges of effective temperature (see Figure~\ref{fig:slope_Teff}), we performed a multivariate regression taking into account the atmospheric parameters of the stars: $T_{\rm eff}$, $L$, and [Fe/H], in addition to $P_{\rm rot}$. This was done with the same Bayesian tool as described before and for stars with only spectroscopic stellar parameters as aforementioned. For that fit, we divided the sample into late F (6000\,K\,$<\teff \le$\,6250\,K), G (5200\,K\,$<\teff \le$\,6000\,K), and K dwarfs (4500\,K\,$<\teff \le$\,5200\,K), allowing to have a broad range of $T_{\rm eff}$ and we used the range 1 to 5\,Gyr. The G and late F dwarfs are put together as they behave similarly in the $\sph$-Age diagram. The relation that is fitted is as follows:

\begin{multline}
\ln \mbox{Age}  = \ln 	\beta + \alpha_1 \ln \sph + \alpha_2 \ln L + \alpha_3 \feh\\
+ \alpha_4 \ln T_\mathrm{eff} + \alpha_5 \ln \prot \,,
\label{eq:multivar}
\end{multline}


with a total relative uncertainty given by
\begin{multline}
\widetilde{\sigma_j}^2  = \widetilde{\sigma}_{\mathrm{Age},j}^2 + \alpha_1^2 \widetilde{\sigma}_{\sph,j}^2 + \alpha_2^2 \widetilde{\sigma}_{L, j}^2 + \alpha_3^2 \sigma_{\mathrm{[Fe/H]}, j}^2\\
+ \alpha_4^2 \widetilde{\sigma}_{T_\mathrm{eff},j}^2 + \alpha_5^2 \widetilde{\sigma}_{\prot,j}^2 \,,
\end{multline}

\noindent where we indicated the uncertainty on $\feh$ as a standard uncertainty (and not relative) because $\feh$ is already a logarithmic quantity. The multivariate regression results for the late-F and G, and K dwarfs are listed separately in Table~\ref{tab:multi-linear}. 

Finally we also tested the statistical significance of the term $\prot$ in the fit, by setting $\alpha_5 = 0$. The results for late-F and G, and K dwarfs are also listed in Table~\ref{tab:multi-linear}.

For both sets of data, late-F and G dwarfs, and K dwarfs, we found that incorporating the rotation term improves the fits ($\alpha_5 \ne 0$). Indeed, the Bayesian model comparison favors the general model given by Eq.~(\ref{eq:multivar}) by a large extent: $\ln B > 4000$ for the late-F and G dwarfs and $\ln B > 3500$ for the K dwarfs. The improvement is also clearly visible from the residuals of the fits, which appear flatter when the rotation term is included. The residuals for the late-F and G dwarfs and the K dwarfs are provided in Appendix~\ref{appendix:residuals}.

To see how the knowledge of the level of magnetic activity with $\sph$ changes the age prediction compared to a pure gyrochronology relation, we also tested the statistical significance of the term $\sph$ by taking $\alpha_1 = 0$ for both sets of late-F and G dwarfs and K dwarfs. The results are given in Table~\ref{tab:multi-linear}. By comparing the evidence values, the relation with $\sph$ is significantly favored with a Bayes factor $\ln B > 200$ for the late-F and G dwarfs. Concerning the K dwarfs, the Bayesian evidence when adding $\sph$ is very similar to the one without including it in the relation. This suggests that for K dwarfs the magnetic activity is not a dominant feature to predict ages. 

As mentioned above, we also fitted the inverse relation to predict $\sph$. The resulting coefficients are given in Appendix~\ref{appendix:Sph_pred}.

\section{Estimating ages with Machine Learning}

\subsection{Description}
We then took advantage of the new artificial intelligence tools that are commonly available and developed by the community, to estimate ages of stars from the same parameters used in the Bayesian fits earlier. This analysis was done for late-F, G, and K dwarfs together. We used Random Forest (RF) algorithms \citep{2001MachL..45....5B}, as they have proved to be very useful in estimating stellar physical parameter from multi-parameters laws \citep[e.g.][]{2014AAS...22312501M, 2018A&A...620A..38B,2021A&A...647A.125B}. Random Forest regressors are based on the aggregation of a large number of random decision trees that are constructed from a training data set and internally validated to give a prediction based on the predictor for future observations. The RF method not only allows the use of a large number of parameters but also estimates their individual impact on the regression (see Section~\ref{importances}). Here we aimed at showing that it is possible to automatically estimate the age, based on the set of physical parameters [$S_{\rm ph}$, $P_{\rm rot}$, $T_{\rm eff}$, $\log (L/L_\odot)$, [Fe/H]].  \\

 We split the sample of stars into two sets. First, a training set was used to train the algorithm to estimate the age from the physical parameters. Then, a test set composed of the remaining of stars was used to test and demonstrate the robustness of the method.

We tried different percentages of training sets from 20\% to 80\% and found very similar results for the accuracy (difference between the predicted age and the model one) in general with some improvements in the scatter. For the rest of the paper, we chose a training set of 50\% of the stars that was a good trade-off to have a reduced scatter but still a large number of stars for the test set (more than 5,000 of them).

In order to take into account uncertainties on physical parameters [$\delta S_{\rm ph}$, $\delta P_{\rm rot}$, $\delta$[Fe/H], $\delta$log$L/L_\odot$], and to improve the training set, we randomly drew, for each of the stars ($\star$) in the training samples, $100$ artificial stars ($\star^\mathcal{A}$) by selecting their physical parameters as follow:
\begin{equation}
\textnormal{X}_{0\le i \le 100}=\textnormal{X}+\delta \textnormal{X} \times \mathcal{G}_{0\le i \le 100} \,
\label{variable}
\end{equation}
where $\textnormal{X}\in[S_{\rm ph}$, $P_{\rm rot}$, $T_{\rm eff}$,  $\log L/L_\odot$, [Fe/H]], $\delta\textnormal{X}\in[\delta S_{\rm ph}$, $\delta P_{\rm rot}$, $\delta T_{\rm eff}$, $\delta \log L/L_\odot$, $\delta$[Fe/H] ] and $\mathcal{G}_{0\le i \le 100}$ random values following a standard normal distribution.  


We also made two other runs with slightly different sets of parameters: one without $S_{\rm ph}$ that is similar to gyrochronology and one without $P_{\rm rot}$ that is similar to magnetochronology. 

\subsection{Relative importance of each parameter}
\label{importances}

\begin{figure}[h!]
    \centering
        \includegraphics[width=9cm]{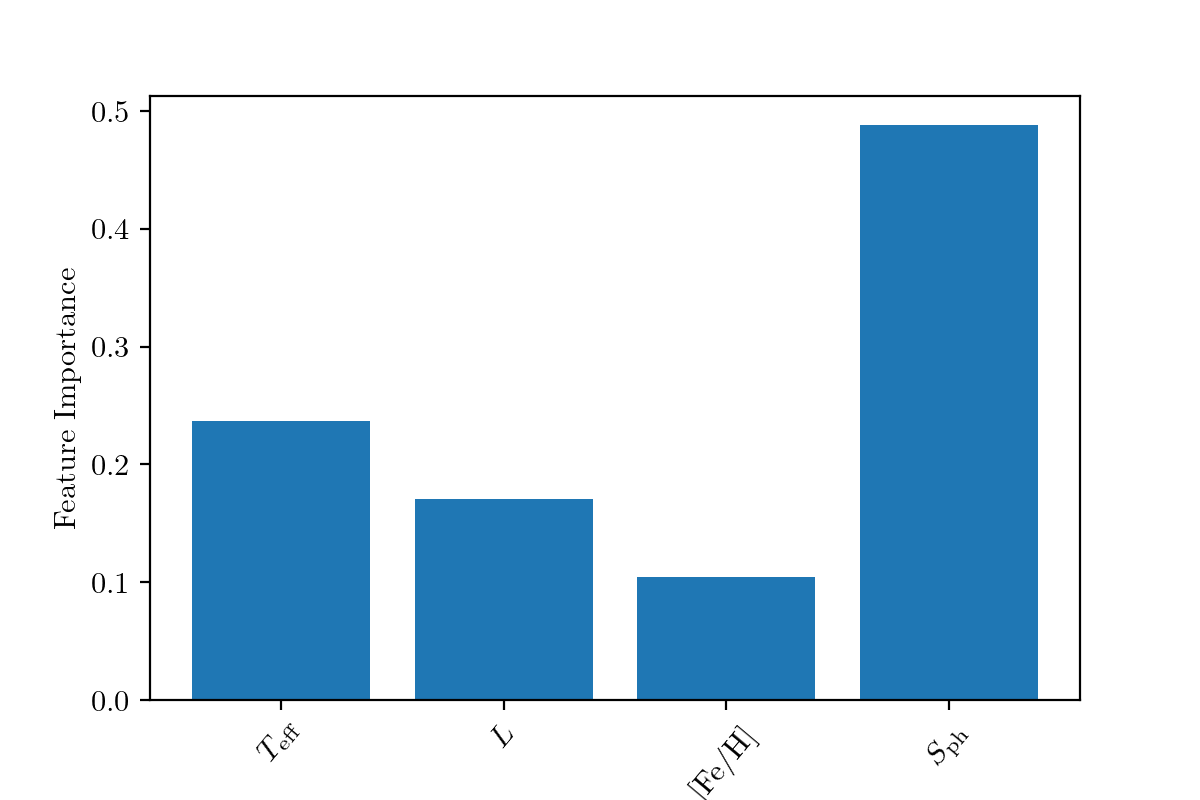}\\
        \includegraphics[width=9cm]{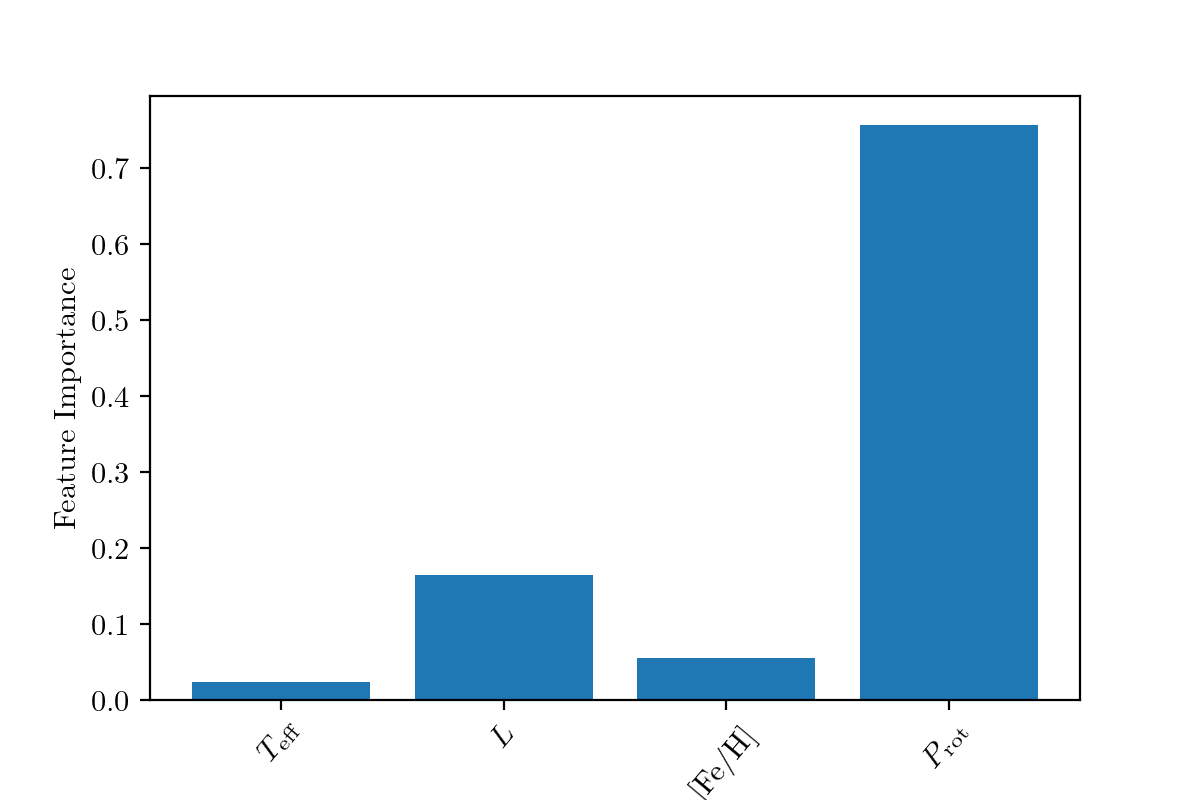}\\
        \includegraphics[width=9cm]{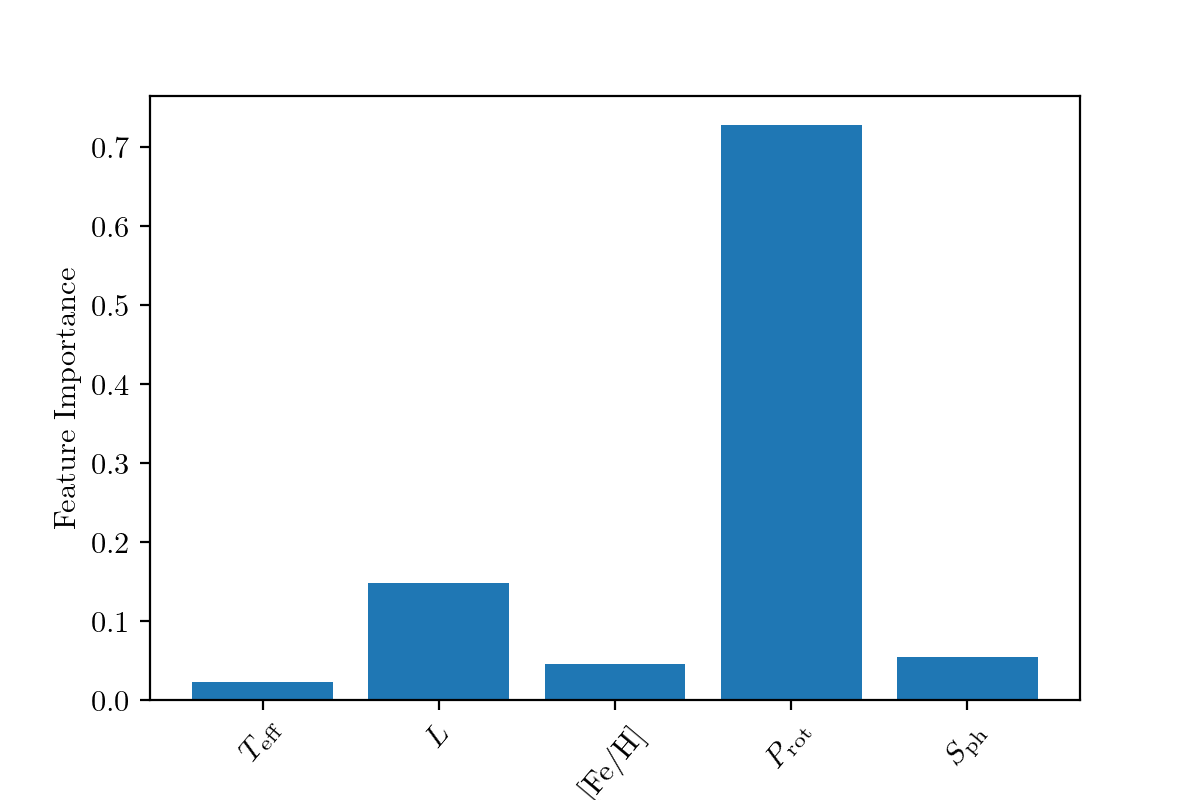}
    \caption{Random Forest Features importance for magnetochronology (top panel), gyrochronology (middle panel), and magneto-gyro-chronology (bottom panel).}
    \label{RF_features}
\end{figure}

In Figure~\ref{RF_features}, we can see the importance of the different input parameters in the training of the RF algorithm.  The so-called {\it gini} \citep{1912vamu.book.....G} importance is a measure of how each parameter contributes to the homogeneity of the nodes and leaves in the trained random forest. The higher the importance value, the higher impact the parameter has during the training process. In the case of not including the rotation period in the RF (i.e. RF magnetochronology, top panel of Fig.~\ref{RF_features}), $S_{\rm ph}$ has the highest weight or importance, followed by the effective temperature, luminosity, and metallicity. This is in agreement with the results of the Bayesian fits done in Section~\ref{sec:bayesfit} where $S_{\rm ph}$ has the largest coefficient. 

In the case of RF gyrochonology (middle panel, Fig.~\ref{RF_features}), the highest importance is rotation period and then luminosity, while temperature and metallicity have very little impact.

Finally, for RF magneto-gyro-chronology, the picture changes drastically where the main feature that impacts the age estimate is still the rotation period followed by luminosity. The photometric magnetic activity proxy has very little impact, as much as [Fe/H]. This emphasizes that for our sample of late-F, G, and K dwarfs, rotation is the dominant factor for age determination and that the magnetic activity proxy is redundant compared to rotation period given the strong correlation between the 2 quantities (see Figure~7 of S21). 



\subsection{Uncertainties from RF}

To obtain uncertainties on the ages predicted by the RF, we randomly draw, for each $\star$ in the test samples, $100$ $\star^\mathcal{A}$ by selecting their physical parameters following Eq.~(\ref{variable}). The standard deviation of the distribution of the age predictions by the RF on the 100 $\star^{\mathcal{A}}$ is then measured for each $\star$ in the test sample.

\begin{figure}[h!]
    \centering
    \includegraphics[width=9cm]{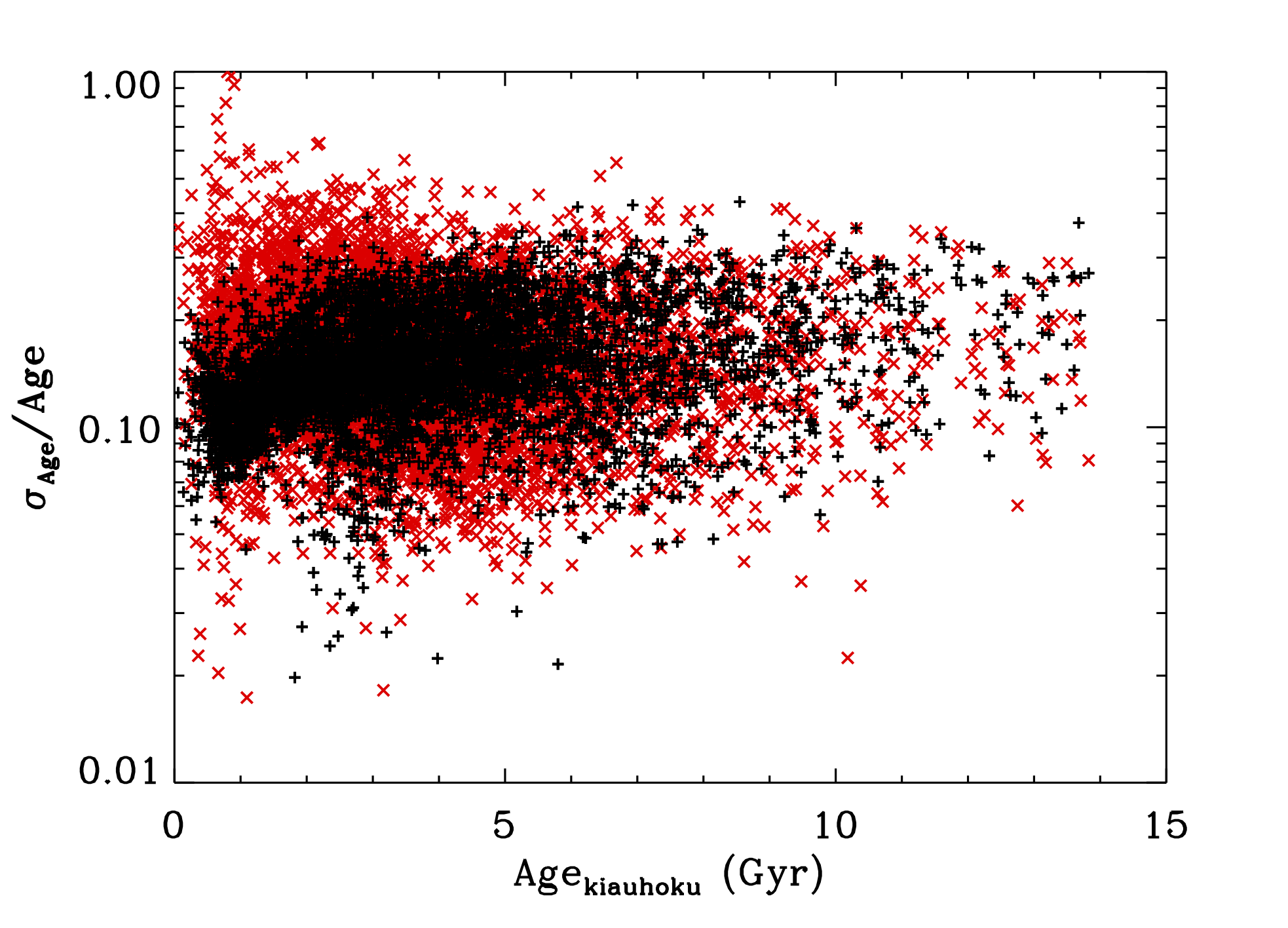}
    \caption{Relative errors from \texttt{kiauhoku} (black crosses) and from the magnetochronology RF algorithm (red crosses) as a function of age. }
    \label{comp}
\end{figure}

In Figure~\ref{comp} are reported such RF relative uncertainties in the case of magnetochronology (red crosses). For most stars in the test sample, the relative uncertainty is comparable to the original uncertainty of the age measurement (black crosses). The median uncertainties for the \texttt{kiauhoku} ages is 14.2\% compared to 15\% for the RF uncertainties. This result is also valid for gyro- and magnetochronology.

\section{Discussion}

Given the different relations found between the magnetic proxy and the ages of the stars, we can evaluate the agreement between the ages predicted  either by the Bayesian fits or the RF and the ones from the models. We remind the reader that the ages used for deriving the relationships are model-dependent. However they still provide good estimates on the stellar ages, in particular given that in our procedure we took into account the most precise spectroscopic observations, the {\it Gaia} luminosity as well as the observed rotation periods as constraints.

\subsection{Ages from magneto-(gyro-)chronology relations}

\begin{figure*}[h!]
    \centering
    \includegraphics[width=8cm]{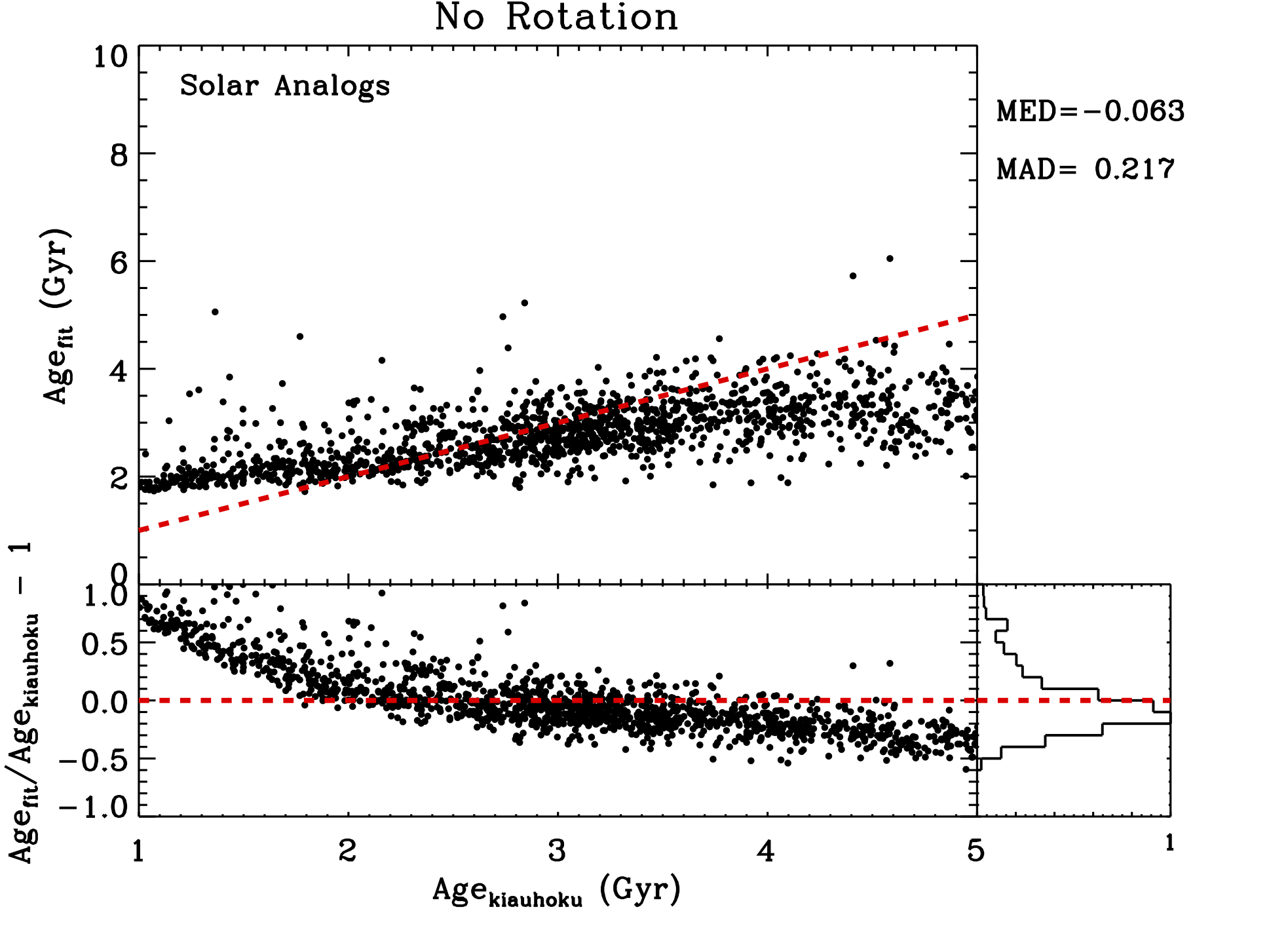}
    \includegraphics[width=8cm]{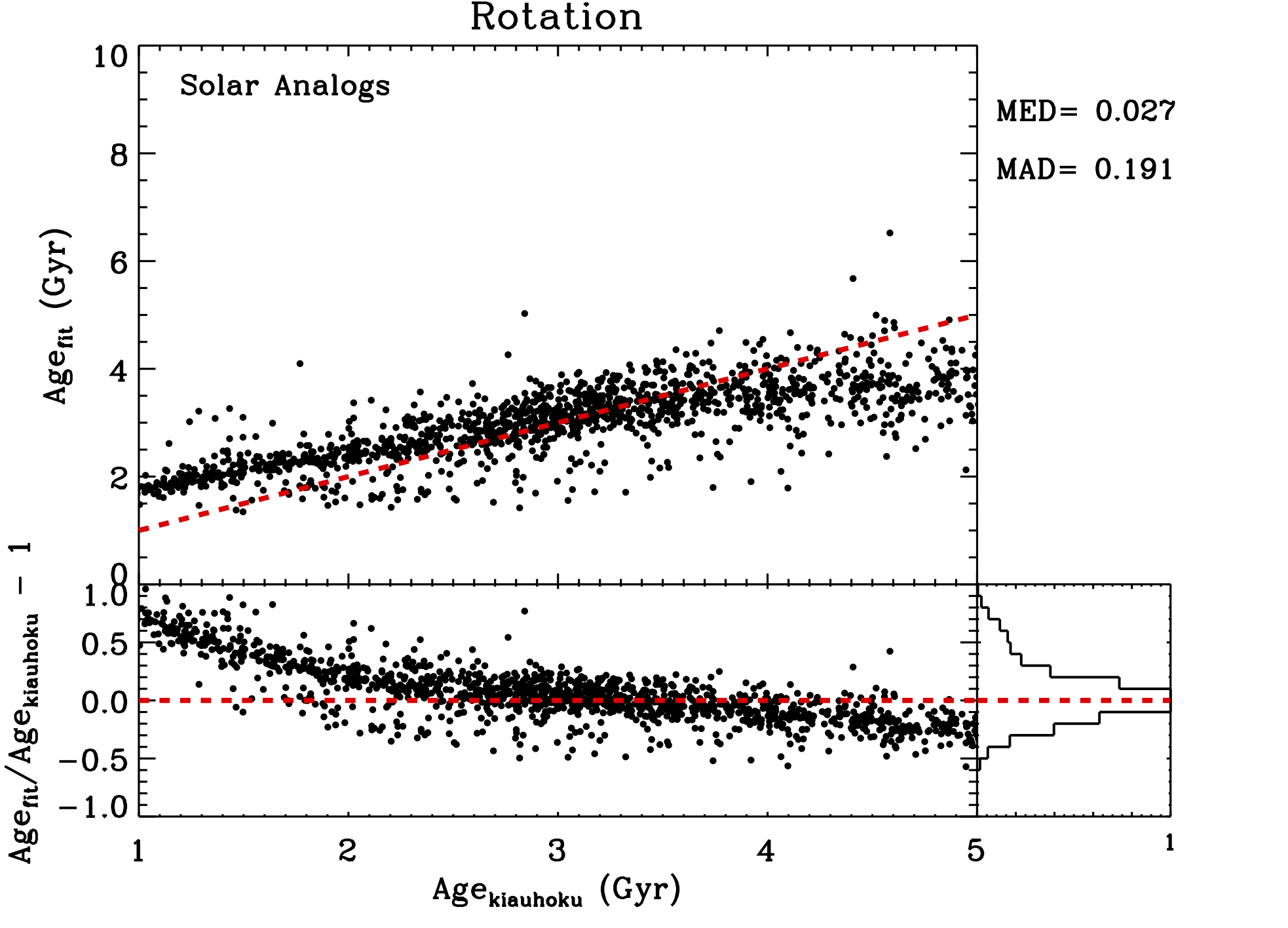}
     \includegraphics[width=8cm]{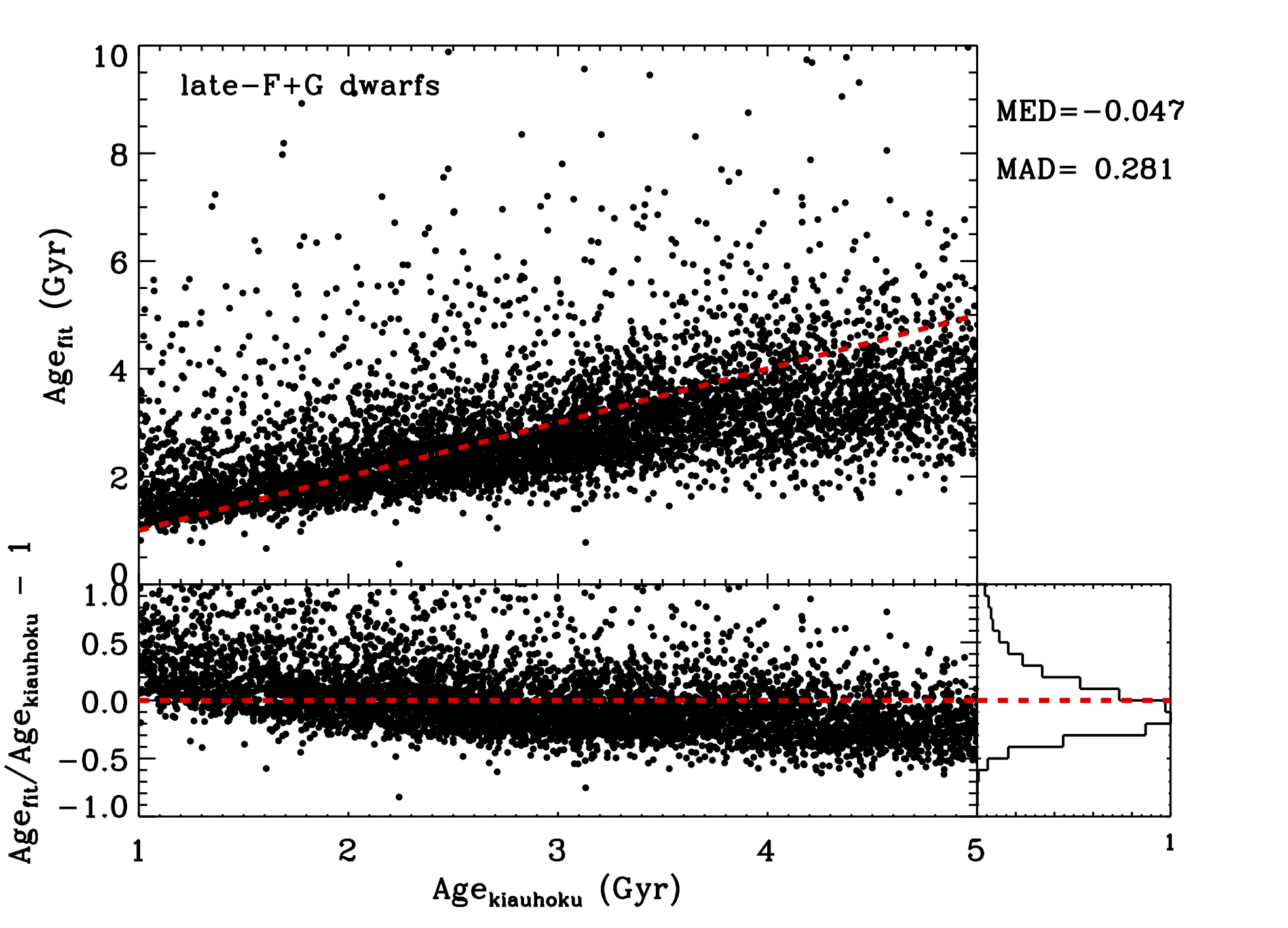}
    \includegraphics[width=8cm]{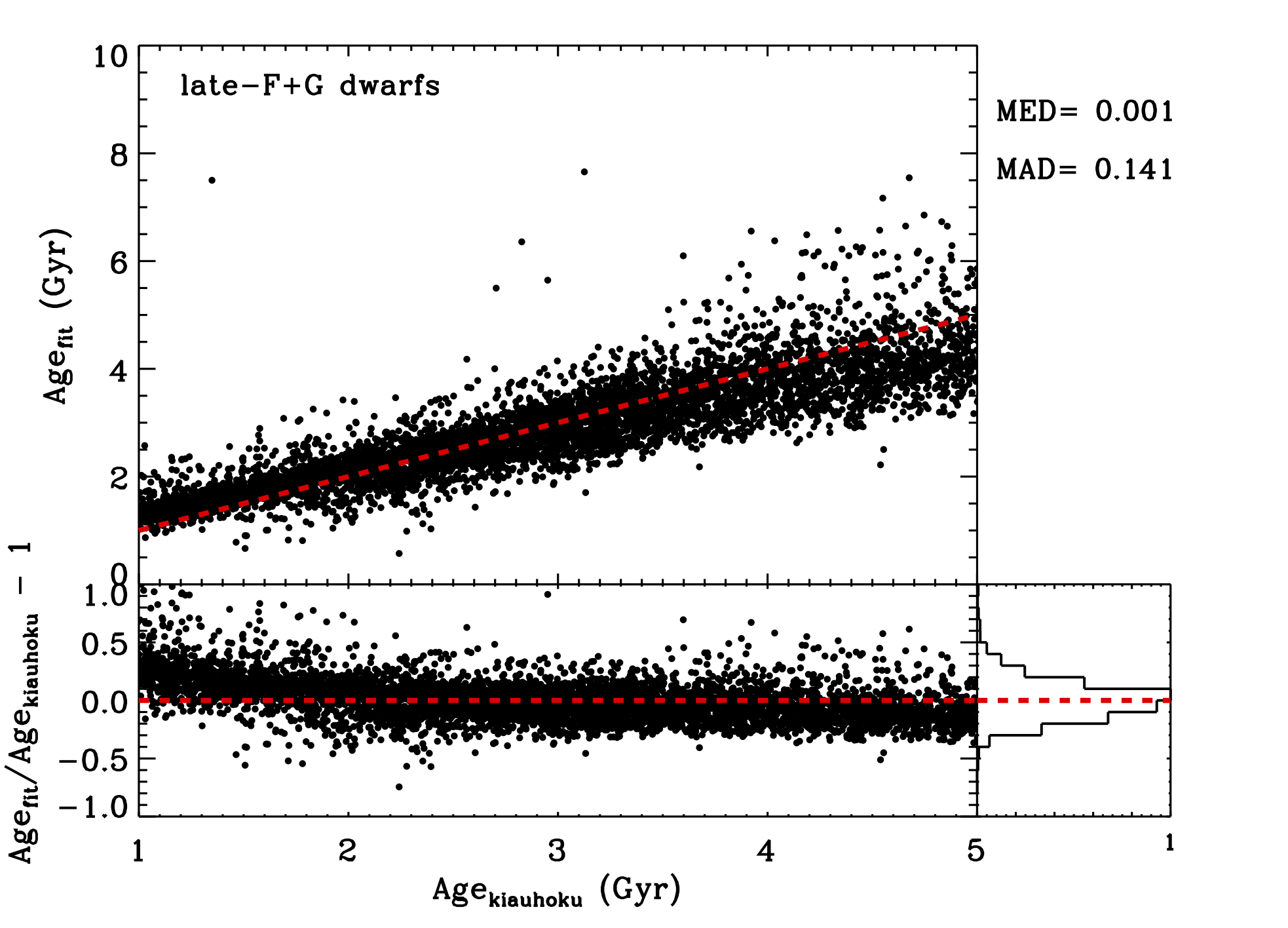}
    \includegraphics[width=8cm]{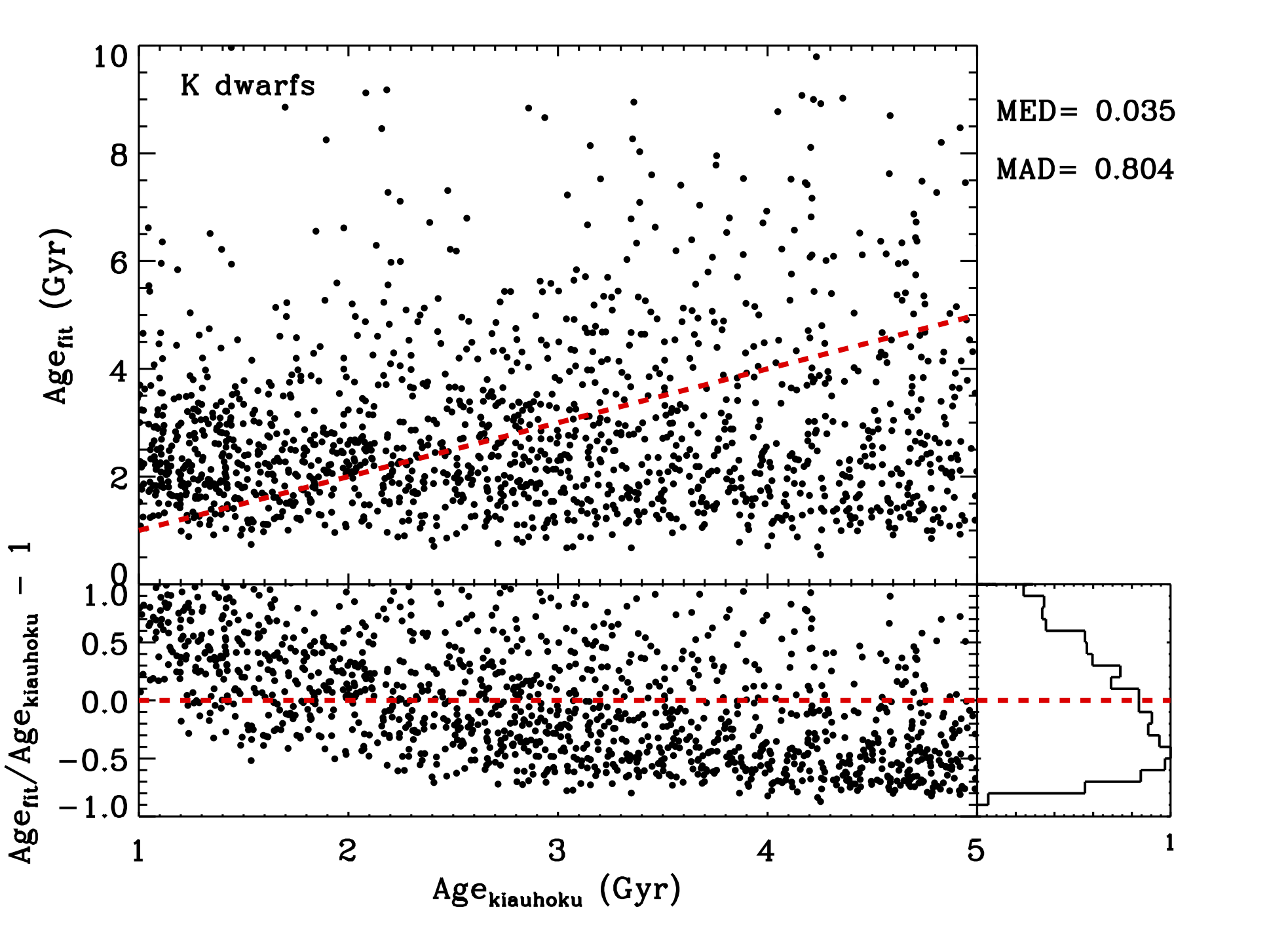}
    \includegraphics[width=8cm]{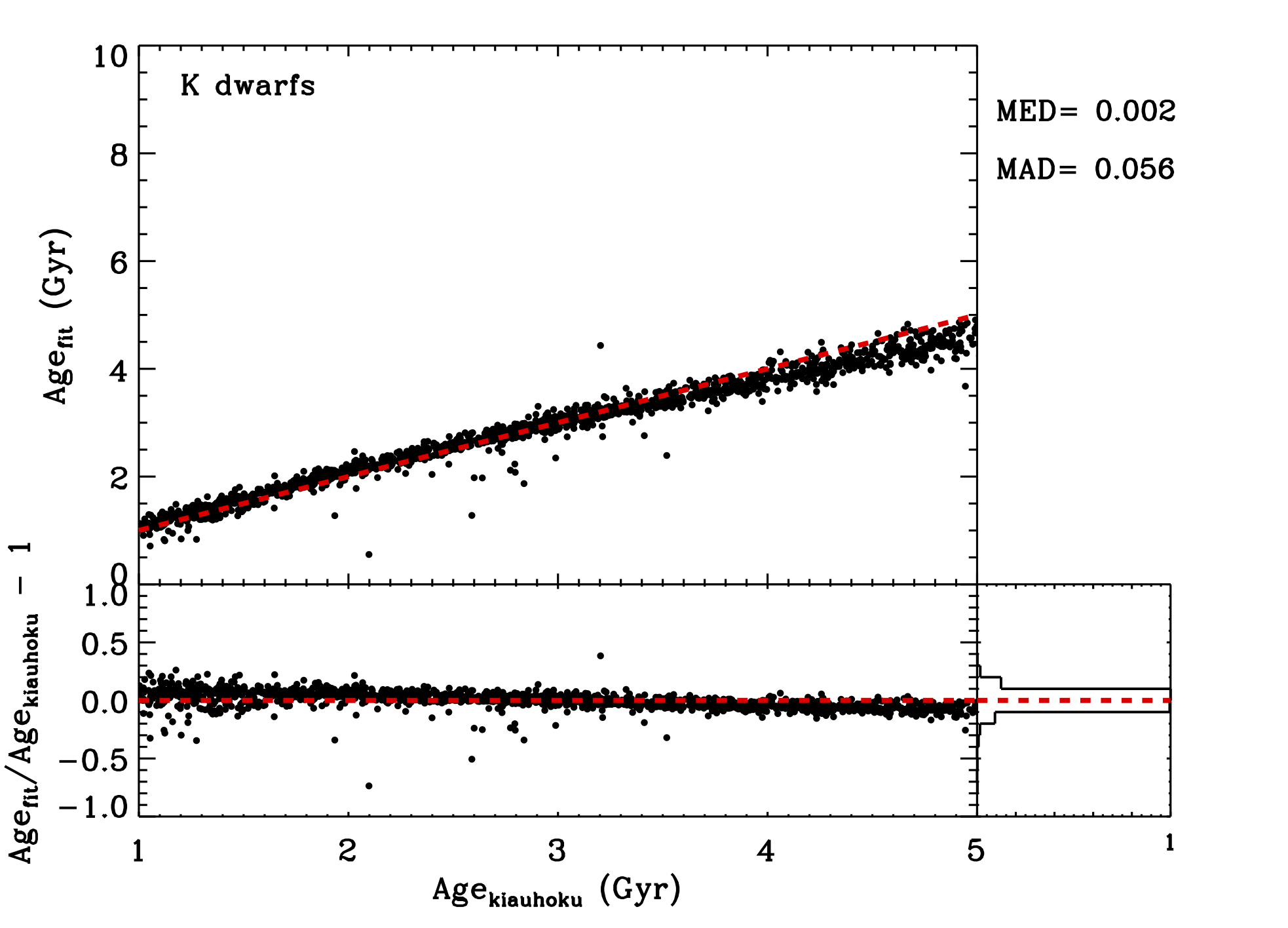}
    \caption{Magnetochronology (without rotation, left column) and magneto-gyro-chronology (with rotation, right column) results from the Bayesian fits. Top row: linear fit for solar analogs. Middle row: multivariate fit for late F and G dwarfs. Bottom row: multivariate fit for K dwarfs. In each figure, the top panel shows the ages computed from the multivariate fit compared to the model ages and the dash red line corresponds to the 1:1 line. The bottom panels of each figure represent the ratio of the ages centered on 0 and the corresponding histogram (as a fraction). The median (MED) and Median Absolute Deviation (MAD) are given for each fit in the top right of each figure.}
    \label{comp_age_fit_norot_rot}
\end{figure*}

From the different relationships derived in Section~\ref{sec:bayesfit}, we computed the predicted ages of the stars based on their stellar parameters. We then evaluated the accuracy of those ages compared to the modeled ones. This was done for the linear and multivariate fits for both magnetochronology and magneto-gyro-chronology relations. In the left panels of Figure~\ref{comp_age_fit_norot_rot}, we can see the comparison between the ages predicted from relations presented in Section~\ref{sec:bayesfit} and calibrated through a Bayesian approach and those coming from the stellar models in the case of magnetochronology (so no $P_{\rm rot}$ was used as input). 

The top left panel shows the results for the linear fit (Eq.~\ref{eq:AgeSph}) for the solar analogs. We can see that there is a trend with an overestimation of the ages below 2\,Gyr and an underestimation for stars older than 4\,Gyr. In particular, this reflects that the analytical relation is not adequate enough to reproduce the age of the young stars. The median of the distribution of the differences between the predicted and modeled ages (bottom right panel) is of -6.3\% with a Median Absolute Deviation (MAD) of 21.7\%. We also notice that the distribution is skewed towards negative values. This means that for the solar analogs, a power law model (Eq.~\ref{eq:AgeSph}) between age and $\sph$, as studied in the past for other magnetic activity proxies, underestimates on average the ages with large uncertainties.

By adding the information on the rotation period to the linear fit (Eq.~\ref{eq:multi-linear_analogs}), the median difference between the model and predicted ages for the solar analogs is of 2.7\%, with a smaller scatter of 19.1\% (top right panel of Figure~\ref{comp_age_fit_norot_rot}). Young ages below 2\,Gyr are still not well retrieved. However, this analytical relation overestimates the ages on average with smaller differences in absolute value compared to the case without rotation. This clearly shows the importance of the knowledge of the rotation period to predict ages of solar analogs where no other stellar parameters are used in the analytical relation.

 For the multivariate relation without rotation period ($\alpha_5$\,=\,0), we show the same comparison between the predicted and observed ages but for late-F and G dwarfs (middle left panel of Figure~\ref{comp_age_fit_norot_rot}) and K dwarfs (bottom left panel of Figure~\ref{comp_age_fit_norot_rot}). While the absolute values of the median differences with the stellar model ages are similar for both sets (-4.7\% for the late-F and G dwarfs and 3.5\% for the K dwarfs), the MAD is much larger for the K dwarfs (80.4\% compared to 28.1\%).  

If we add the information on rotation period ($\alpha_5 \neq$\,0), we clearly see an improvement on the ages prediction for both the late-F and G dwarfs (middle right panel of Figure~\ref{comp_age_fit_norot_rot}) and K dwarfs (bottom right panel of Figure~\ref{comp_age_fit_norot_rot}). For the former, the distribution of the differences is centered close to 0 (median of 0.1\%) with a MAD of 14.1\%. There is a slight bias of overestimated ages below 1.5\,Gyr but on average the ages are well recovered. For the K dwarfs, the improvement is even more striking with a median difference of 0.2\% and a MAD of 5.6\% better than the late-F and G dwarfs. Such an improvement can be explained by the fact that rotation period is a stronger constraint of the ages of K dwarfs. Indeed rotation period is a stronger constraint on stellar age than the classical stellar parameters.

As mentioned in Section~\ref{sec:bayesfit}, we also fitted the relations without $\sph$ to assess the importance of the magnetic activity proxy in the age prediction. The comparison between the predicted and \texttt{kiauhoku} ages is shown in Appendix~\ref{appendix:Bayes_gyro}. While the results are similar to the case where we include $\sph$, the median differences and MAD values are slightly improved when the magnetic activity is taken into account.

\subsection{Age estimates from RF algorithm}

\begin{figure*}[ht]
    \centering
    \includegraphics[width=10cm]{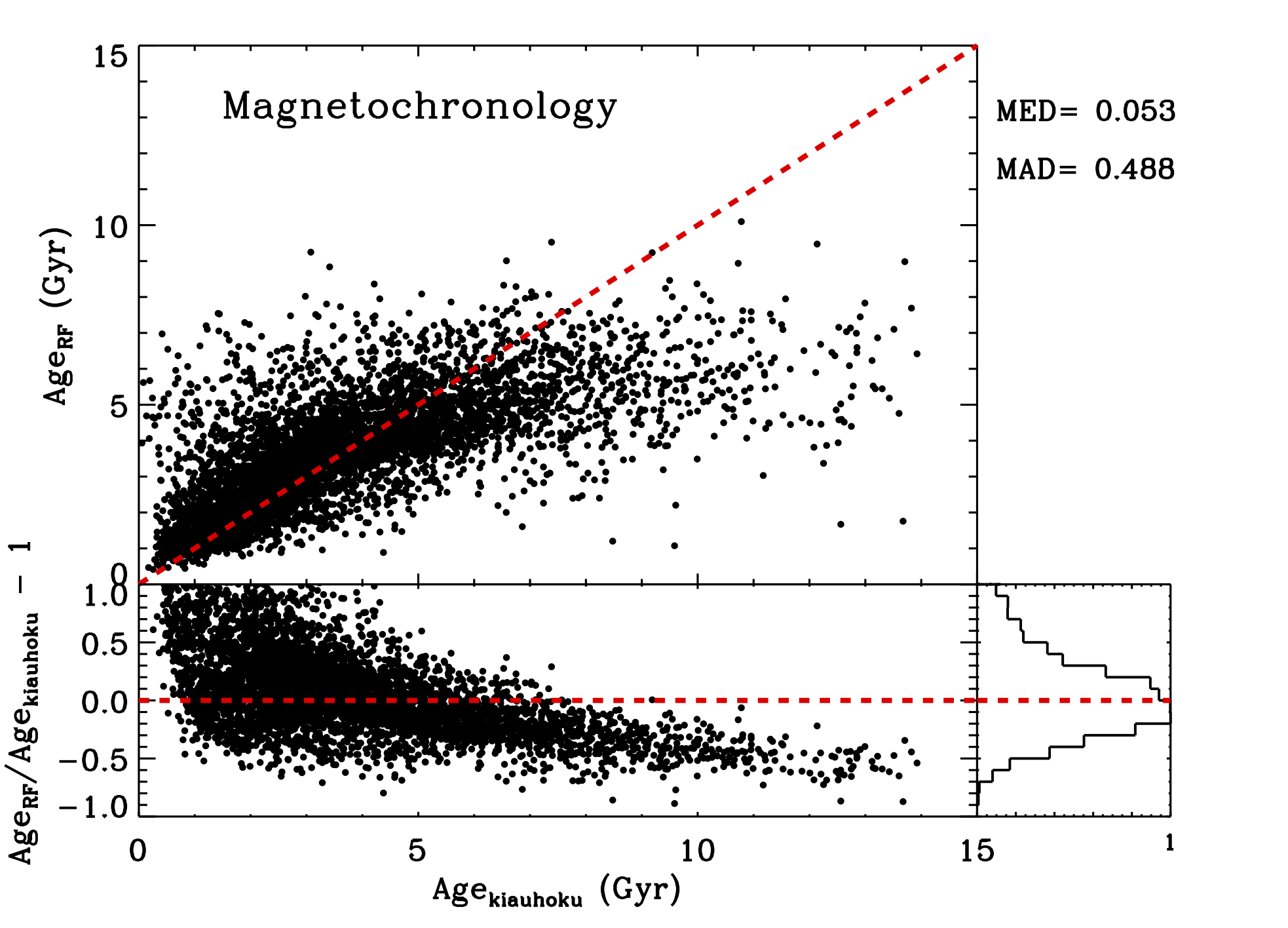}
    \includegraphics[width=10cm]{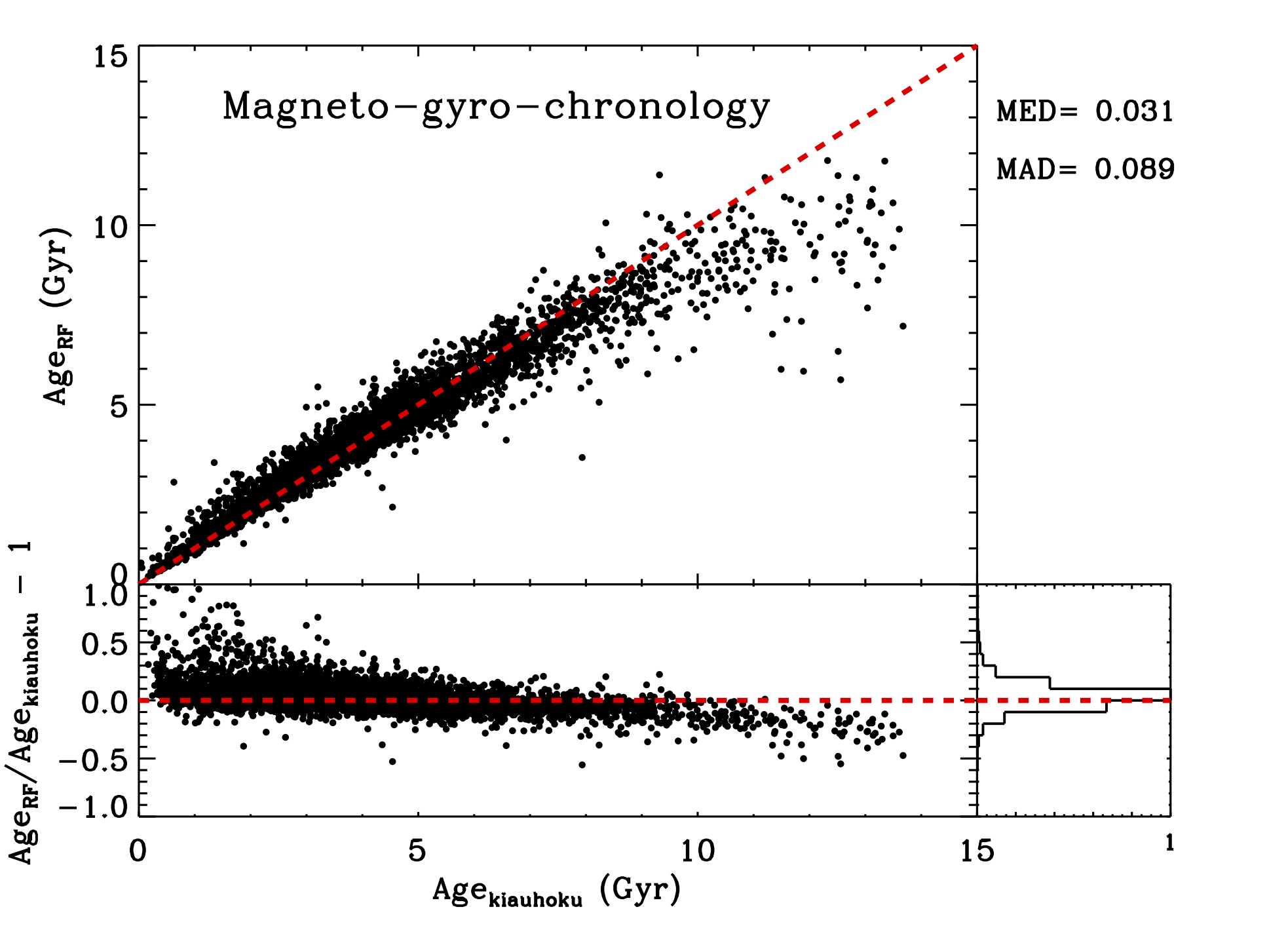}
    \includegraphics[width=10cm]{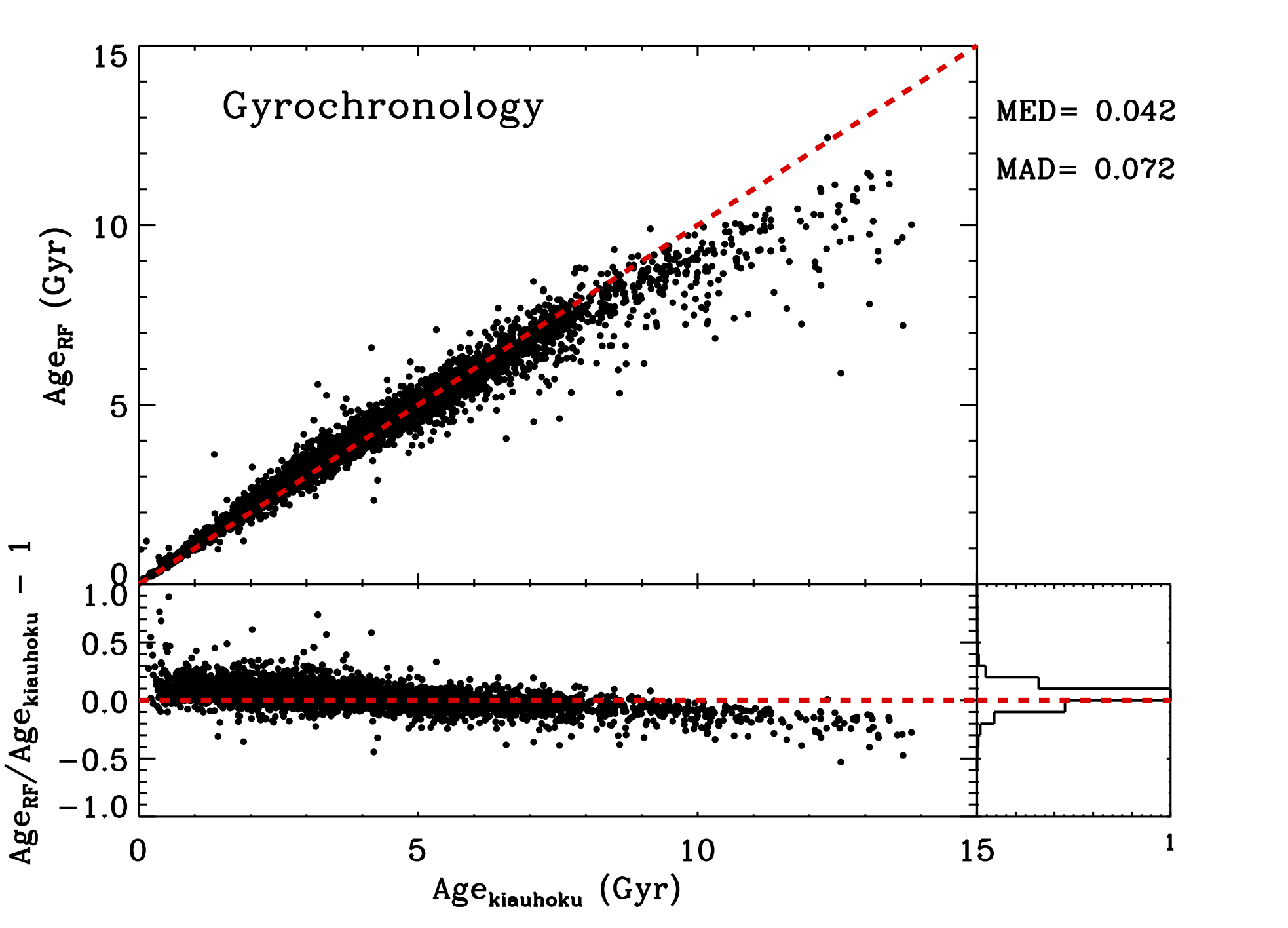}
    \caption{Comparison of ages predicted from the Random Forest with the model ages\ for late F, G, and K dwarfs. Top panel: magnetochronology. Middle panel: magneto-gyro-chronology. Bottom panel: gyrochronology. Same legend as in Figure~\ref{comp_age_fit_norot_rot}.}
    \label{comp_age_RF}
\end{figure*}

We show in the top panel of Figure~\ref{comp_age_RF} the comparison of the predicted ages from the RF analysis without using the rotation period as input. This analysis was done on all late-F, G, and K dwarfs where the test set consisted of 5,452 stars. The predicted ages from the RF go close to 14\,Gyr as the algorithm was trained on the full sample of late-F, G, and K dwarfs. We see a clear trend with increasing ages. Above 5\,Gyr, the RF ages are biased towards smaller ages compared to \texttt{kiauhoku} ones whereas below 5\,Gyr, the RF tends to overestimate the ages with a median difference of 15.9\%. In particular for the old stars, the disagreement can be quite high due to the {\it plateau} that we observed above 5\,Gyr, leading to some degeneracy. If we take into account the full range of model ages, the median difference goes down to 5.3\% with a MAD of 48.8\%.


When adding the rotation period as input (middle panel of Figure~\ref{comp_age_RF}) there is a very good agreement up to ages of $\sim$\,10\,Gyr with less scatter compared to the previous RF run. For the full set, we find a median difference between the RF ages and the \texttt{kiauhoku} ones of 3.1\% with a MAD of 8.9\%. This is around half of the median uncertainties from the RF. 

Finally, we also tested the RF algorithm for pure \emph{gyrochronology} (bottom panel of Figure~\ref{comp_age_RF}), where we used $P_{\rm rot}$ without $S_{\rm ph}$ along with the atmospheric parameters as inputs. The median difference between the predicted ages and the model ones is of 4.2\% with a dispersion of 7.2\%, smaller than the RF uncertainties. By taking into account the magnetic activity proxy, the median difference is improved but the dispersion increases. The use of both inputs $P_{\rm rot}$ and $S_{\rm ph}$ improves in average the age estimates with the RF. 


\subsection{On the limitation of ages computed in this work}

We note that the ages computed in this work depend on models. They rely on a long list of choices of input physics and magnetic braking prescription, as well as calibrators needed to produce expected behavior. While we consider that all our choices are well-motivated, they are inevitably based on an incomplete knowledge of stellar physics and rotation. 

One limitation of gyrochronological ages is the range of available calibrators. Spin-down relations are calibrated using stars for which precise ages and periods are known independently, which limits us to the Sun, a handful open clusters, and more recently a few tens of asteroseismic F, G, and K dwarfs \citep[e.g.,][]{Barnes2010, Angus2015, 2016Natur.529..181V, Curtis2020,2021ApJS..257...46G}. The open clusters are all young with the oldest one of 4\,Gyr, M67, being studied recently \citep{2022ApJ...938..118D}; thus, until recently the Sun was our oldest anchor for gyrochronology at 4.57 Gyr. The few old, seismic dwarfs hint that braking weakens at ages older than the Sun \citep{2016Natur.529..181V, Hall2021}, but the physical processes that weaken the braking are still poorly understood. Furthermore, the benchmark clusters are all near solar composition. Any gyrochronology predictions for chemically different stars are then extrapolations into regimes where there are no calibrators. Spin-down is less constrained in these regimes.


Spin-down relations are also less reliable in regimes where stars lose their convective envelopes. With no surface convection, there is no dynamo to generate a magnetic field, so no braking occurs. In this regime \citep[toward and above $\sim$6300 K, the Kraft break after][]{Kraft1967}, rotation is {   not} meaningful as an age indicator. We note, however, that our use of luminosity in addition to rotation to constrain the age allows for luminosity to take over when rotation becomes less constraining.

Besides the evolution models themselves, the parameters we use as input to the MCMC fitting are also model dependent. Effective temperatures and metallicities are determined by fitting atmospheric models to spectra, which have different scales than those used in the evolution models. Furthermore, the luminosities are determined by fitting yet another set of models to \textit{Gaia} parallaxes and 2MASS colors \citep{Berger2020}. Some authors calibrate their stellar parameter estimates to a more standard scale \citep[e.g., APOGEE;][]{Holtzman2015}, but using different models with different parameter scales necessarily introduces uncertainty into the model estimates \citep{Tayar2022}.

When fitting evolution models to stars, we assume solar abundances of $\alpha$ elements. $\alpha$-abundances influence spin-down independently from metallicity alone, so this assumption adds uncertainty to the estimates as well. For example, \citet{Claytor2020} showed that assuming solar $\alpha$-abundance for a main-sequence star that actually has [$\alpha$/M] = 0.4 results in underestimating the age by as much as 20\%.

Finally, our angular momentum evolution model in YREC assumes that stars spin down as rigid bodies, which may be a poor assumption at some stages of stellar evolution. Open cluster stars exhibit a temporary epoch of stalled spin-down in K dwarfs at 1 Gyr \citep{Curtis2019} which could be explained by a two-zone angular momentum model such as that of \citet{Spada2020}. The departure from expected behavior means that there are physical mechanisms our gyrochronology model does not capture.

Some of these limitations have the effect of stretching or compressing our age scale, while others result in a systematic offset. Taken together, the effects mean that while our ages are internally consistent and precise, any individual age taken from this data set should be vetted cautiously before being used to support new claims.

\subsection{Detection bias: looking at stars without rotation-period estimates}
\label{sec:detection_bias}

We investigate whether the {\it plateau} above 5\,Gyr seen in Fig.~\ref{Sph_age_SA_fits} is real. To evaluate what is the level of bias on the stars where we detected rotation periods in S19 and S21, we study here a sample of more than 100,000 solar-like stars observed by the {\it Kepler} mission for which no rotation periods could be reliably measured for different reasons (instrumental artifacts, low-amplitude signal, low inclination angle, or no stellar magnetic activity during the {\it Kepler} observations). 

By definition $S_{\rm ph}$ is a proxy for magnetic activity measured from the rotational modulation in the light curves. This means that when it is not possible to retrieve a reliable $P_{\rm rot}$, we cannot derive $S_{\rm ph}$. In this section, in order to investigate the detection limit for these measurements, we define a variability index $V_{\rm ph}$, which can be obtained for stars regardless a $P_{\rm rot}$ was measured. We split the light curve in subseries of length 30 days to compute the standard deviation and we take the mean value of all subseries. Similarly to the $S_{\rm ph}$, we subtract the photon noise obtained from the formula by \citet{2010ApJ...713L.120J}. We call this quantity the variability index, $V_{\rm ph}$, as the measured quantity may not be purely related to the magnetic activity of the star, in contrast to $\sph$.



\begin{figure*}[ht]
    \centering
    \includegraphics[width=12cm]{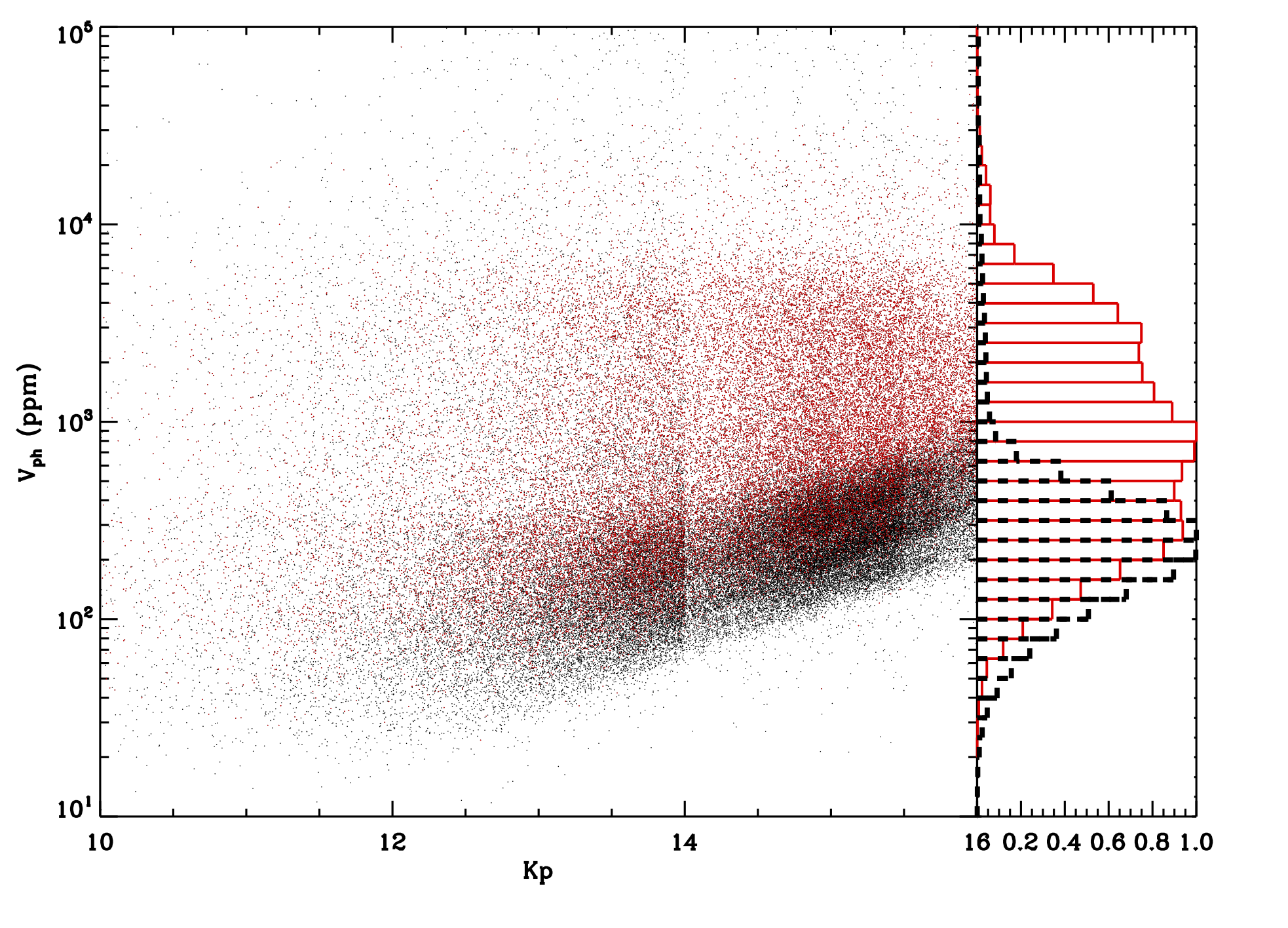}
    \caption{Left panel: Comparison of $V_{\rm ph}$ for stars without $\prot$ (black points) and stars with $\prot$ (red points) as a function of the Kp magnitude. Right panel: Distribution of $V_{\rm ph}$ for the stars with rotation periods (black) and without rotation periods (red). }
    \label{Sph_rot_norot}
\end{figure*}

In the comparison that follows, we also compute $V_{\rm ph}$ for stars with $P_{\rm rot}$ estimates. For these stars, $V_{\rm ph}$ is still related with magnetic activity as we find that $S_{\rm ph}$ and $V_{\rm ph}$ are very similar (see Appendix~\ref{appendix:Vph}).  The comparison between $V_{\rm ph}$ for the {\it rotation} sample and the {\it no rotation} sample in Figure~\ref{Sph_rot_norot} shows that the stars with detected rotational modulation have larger $V_{\rm ph}$ in general than the stars without detection. This can also be seen in the histogram in the right panel of Figure~\ref{Sph_rot_norot}. For active stars, indeed it is easier to observe rotational modulation in epochs of larger magnetic activity. Still, some stars without $P_{\rm rot}$ have large $V_{\rm ph}$ values similar to those of the stars with $P_{\rm rot}$. This could be related to instrumental artifacts, which might still be present in the light curves. {   A lack of modulation detection could also be due to a fully spotted star, a star observed pole-on with active latitudes close to the equator, or with high active latitudes and long-lived spots.}

The rotation analysis done by S19 and S21 extended the rotation periods catalog towards slower rotators, which usually have low $\sph$ or $V_{\rm ph}$ values. Hence in this work, we derived ages older than the age of the Sun for stars with $\prot$ showing that we can still measure rotation periods for these old stars. This is slightly in contrast to the findings of \citet{2022ApJ...937...94M}. Indeed the author studied a sample of 278 planet-host stars selected as having ``robust'' rotation periods from \citet{2015ApJ...801....3M} and computed ages with isochrone fittings based on the models of \citet{2015MNRAS.450.1787A} that also take into account surface rotation periods. He finds that there is a bias in the rotation detection and stars older than the Sun do not have a rotation periods measured. However, as pointed out by \citet{2022ApJ...937...94M} for the S21 sample used in our work, the situation might be different given the larger number of stars with longer rotation periods.

\begin{figure}[ht]
    \centering
    \includegraphics[width=10cm]{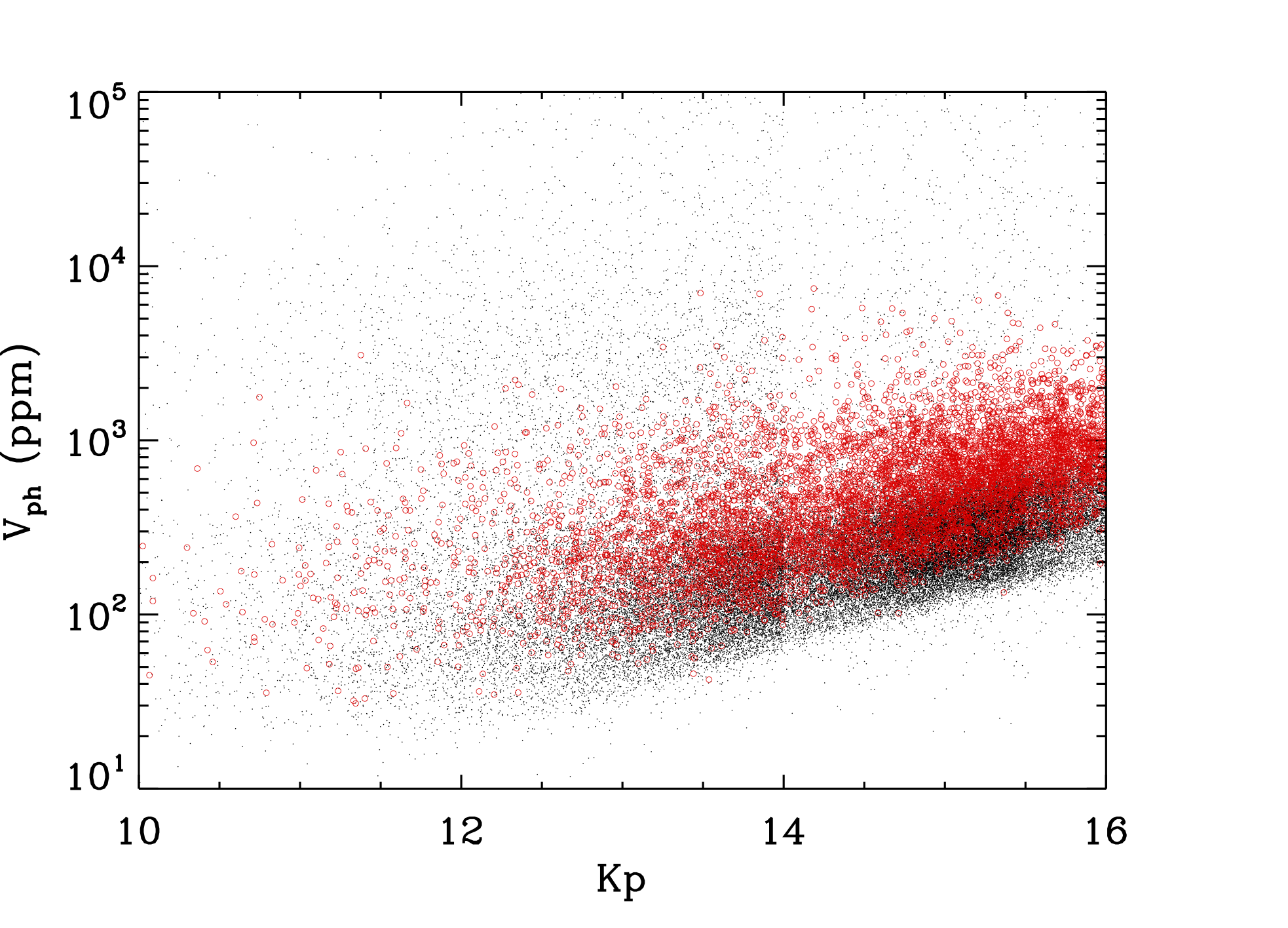}
    \caption{Variability index $V_{\rm ph}$ as a function of the Kp magnitude for stars without rotation periods (black dots) and stars with rotation periods and ages derived in this work above 5\,Gyr (red circles).}
    \label{Vph_Kp_old}
\end{figure}

We looked at stars older than 5\,Gyr in the $S_{\rm ph}$-age diagram to locate them in Figure~\ref{Sph_rot_norot}. We found that most of the stars are above the lower edge, i.e. above the limit of the detection (see Figure~\ref{Vph_Kp_old}). This points in the direction that the level of magnetic activity drops past a given age with a certain flat evolution afterwards. This does not discard the fact that old stars could be even less active.

From previous works, the magnetic activity of the Sun between the minimum and maximum of the solar cycle varies between 67.4 to 314.5\,ppm \citep{2019FrASS...6...46M}. We added those data points in the $\sph$-age diagram for solar analogs (see Figure~\ref{Sph_age_SA_fits}). We can see that the magnetic activity level of the Sun agrees well with its counterpart for a similar age. Hence, it seems that the Sun behaves like Sun-like stars selected from spectroscopic parameters. This also agrees with the analysis of the variability of $\sph$ for solar analogs done by \citet{2023A&A...672A..56S}.

\section{Conclusions}


The measurement of surface rotation and magnetic activity level of solar-like stars observed by the {\it Kepler} mission gives us the opportunity to look for activity-rotation-age relations. 

We computed ages for 55,232 stars with \texttt{kiauhoku} associated with the YREC stellar evolution code that includes treatment of angular momentum evolution as well as weakened magnetic braking by fitting isochrones on $\teff$, L, [Fe/H], and $\prot$. The comparison with ages computed with the STAREVOL code shows a strong correlation with a Spearman correlation coefficient of 0.89 as well as a good agreement with kinematic and seismic ages.

We then investigated two different approaches to infer stellar ages using the photometric activity proxy, $\sph$. 

The first one is based on deriving relations between age and $\sph$ from Bayesian linear and multivariate fits. With that first analysis we obtained the following results:

\begin{itemize}
    \item The solar analogs analysis shows that including the rotation period improves the predicted ages with a median difference of 2.7\% and a MAD of 19.1\%. However there is a clear bias for young and old stars that is not captured by the analytical relations that we fitted. 
    \item For the late-F and G-dwarf sample, the multivariate regression with rotation period is also favored, yielding a median difference between predicted and \texttt{kiauhoku} model ages of 0.1\% and a MAD of 14.1\%. 
    \item For K dwarfs, the use of rotation periods in the relation is necessary to lead to a good agreement between predicted and the \texttt{kiauhoku} model ages, with a median difference of 0.2\% and a MAD of 5.6\%. This reflects the fact that rotation period is a strong constraint in the evolution of K dwarfs.
\end{itemize}

We note that for the late-F and G dwarfs the relations that include $\sph$ are favored by the Bayesian analysis while for the K dwarfs both relations with or without $\sph$ give very similar results.

Readers that are interested in applying our relations may find them in Table~\ref{tab:multi-linear}.

The second approach consists in applying machine learning based on Random Forest algorithms. By training the algorithm with late-F, G, and K dwarfs together with $\teff$, $L$, [Fe/H], and $\sph$ as input parameters, we compared the predicted ages with the model ones and found that:
 \begin{itemize}  
    \item In the magnetochronology run (where no $\prot$ is included), ages above 5\,Gyr are biased towards younger ages. Up to 5\,Gyr, we obtain a median difference of 5.3\% between predicted and the \texttt{kiauhoku} model ages. The magnetic activity index is the dominant feature to predict ages.
    \item The gyrochronology (no $\sph$ included) and magneto-gyro-chronology ($\prot$ and $\sph$ included) runs improve the results with median differences of 4.2\% and 3.1\%, respectively, with an improvement when adding the information from $\sph$. Nevertheless the scatter is slightly larger in the magneto-gyro-chronology  run that in the pure gyrochronology run. In both cases, $\prot$ is the most important feature. 
\end{itemize}


 One advantage of the RF is the ability to predict ages for stars older than 5\,Gyr. The {\it plateau} observed above that age prevents us from deriving an analytical relation. This also suggests that RF finds more complexity in the relation between the different parameters that could not be captured with the analytical relations we had defined.

These results are of course dependent on the stellar evolution models and the physics included to derive the ages. Caveats should be taken into account on the limitations of those models. 

We also investigated the bias in the sample with detected rotation periods. We find that the variability $V_{\rm ph}$ of stars with rotation periods detected is in general larger than the one for stars without rotation periods, suggesting that there can be some bias against low-activity, in principle old, stars. However, we still find old stars (older than 5\,Gyr) with measured rotation periods that are above the detection limit. While we have fewer older stars, our sample seems to be less biased towards younger stars in comparison with previous catalogs.  

These results are very promising for the RF analysis. Indeed, once the algorithm is trained with a given set of stellar models, it requires very short computing time. Given the improvements aforementioned, it finds more complex dependencies that are not necessarily captured by the Bayesian fits that just follow the relation that is fed in. Such an approach can be easily applied to estimate ages of hundreds of thousands of solar-like stars with photometric observations obtained with missions such as the Transiting Exoplanet Survey Satellite \citep{2015JATIS...1a4003R} and PLAnetary Transits and Oscillations \citep{2014ExA....38..249R}.

\newpage
This paper includes data collected by the \emph{Kepler} mission and obtained from the MAST data archive at the Space Telescope Science Institute (STScI). Funding for the \emph{Kepler} mission is provided by the NASA Science Mission Directorate. STScI is operated by the Association of Universities for Research in Astronomy, Inc., under NASA contract NAS 5-26555. We acknowledge
that this research was supported in part by the National Science
Foundation under grant No. NSF PHY-1748958.
S.M.~acknowledges support from the Spanish Ministry of Science and Innovation (MICINN) with the Ram\'on y Cajal fellowship no.~RYC-2015-17697, the grant no. PID2019-107061GB-C66, and through AEI under the Severo Ochoa Centres of Excellence Programme 2020--2023 (CEX2019-000920-S). S.M. and D.G.R acknowledge support from the Spanish Ministry of Science and Innovation (MICINN) with the grant no.~PID2019-107187GB-I00. Z.R.C. acknowledges support from National Aeronautics and Space Administration via the TESS Guest Investigator Program (grant No. 80NSSC18K18584). The work presented here was partially supported by the NASA grant NNX17AF27G. A.R.G.S. acknowledges the support by FCT through national funds and by FEDER through COMPETE2020 by these grants: UIDB/04434/2020 \& UIDP/04434/2020. A.R.G.S. is supported by FCT through the work contract No. 2020.02480.CEECIND/CP1631/CT0001.  R.A.G.\, L.A.\ and S.N.B.\ acknowledge the support from PLATO and GOLF CNES grants.  S.N.B. acknowledges support from PLATO ASI-INAF agreement n. 2015-019-R.1-2018.

%



\software{astropy \citep{2013A&A...558A..33A,2018AJ....156..123A}, \texttt{kiauhoku} \citep{Claytor2020, kiauhoku}, NumPy \citet{numpy}, SciPy \citet{scipy}, STAREVOL \citep{2000A&A...358..593S, Amard2019}, D\textsc{iamonds} \citep{2014A&A...571A..71C}, matplotlib \citep{matplotlib}, pandas \citep{reback2020pandas,mckinney-proc-scipy-2010_pandas}, KADACS \citep{2011MNRAS.414L...6G, 2015A&A...574A..18P}}



\newpage

\appendix

\section{Age comparison as a function of effective temperature}\label{appendix:age_comp_Teff}

The ages from STAREVOL are given in Table~\ref{tab:STAREVOL}.

\begin{table*}[h]
    \centering
    \begin{tabular}{ccc}
    \hline
    KIC &  $M$ (M$_\odot$)&  Age (Gyr)\\
\hline
\hline
757450 &  1.02\,$\pm$\,0.02& 1.68$^{+ 0.36}_{- 0.32}$\\
892195 &  1.02\,$\pm$\,0.02& 2.77$^{+ 0.63}_{- 1.03}$\\
892713 &  1.46\,$\pm$\,0.00& 2.19$^{+0.00}_{- 0.00}$\\
892834 &  0.81\,$\pm$\,0.02& 0.65$^{+ 0.11}_{- 0.11}$\\
893033 &  0.72\,$\pm$\,0.01& 2.80$^{+ 0.51}_{- 0.51}$\\
893286 &  0.90\,$\pm$\,0.03& 3.21$^{+ 0.73}_{- 0.77}$\\
893383 &  0.93\,$\pm$\,0.02& 3.19$^{+ 0.95}_{- 0.88}$\\
893505 &  1.41\,$\pm$\,0.04& 3.32$^{+ 0.16}_{- 0.14}$\\
893507 &  1.48\,$\pm$\,0.00& 3.04$^{+ 0.02}_{-0.02}$\\
893559 &  0.87\,$\pm$\,0.04& 0.89$^{+ 0.17}_{- 0.16}$\\
1026287 &  0.80\,$\pm$\,0.04& 3.68$^{+ 1.50}_{- 1.45}$\\
...\\
\hline
    \end{tabular}
    \caption{Masses and ages computed with STAREVOL as described in Section~\ref{sec:YREC_STAREVOL}. The full table is available in a machine readable format.}
    \label{tab:STAREVOL}
\end{table*}




\section{Residuals from the Bayesian multivariate regression for G and K dwarfs}\label{appendix:residuals}

\begin{figure}[ht]
    \centering
    \includegraphics[width=5cm]{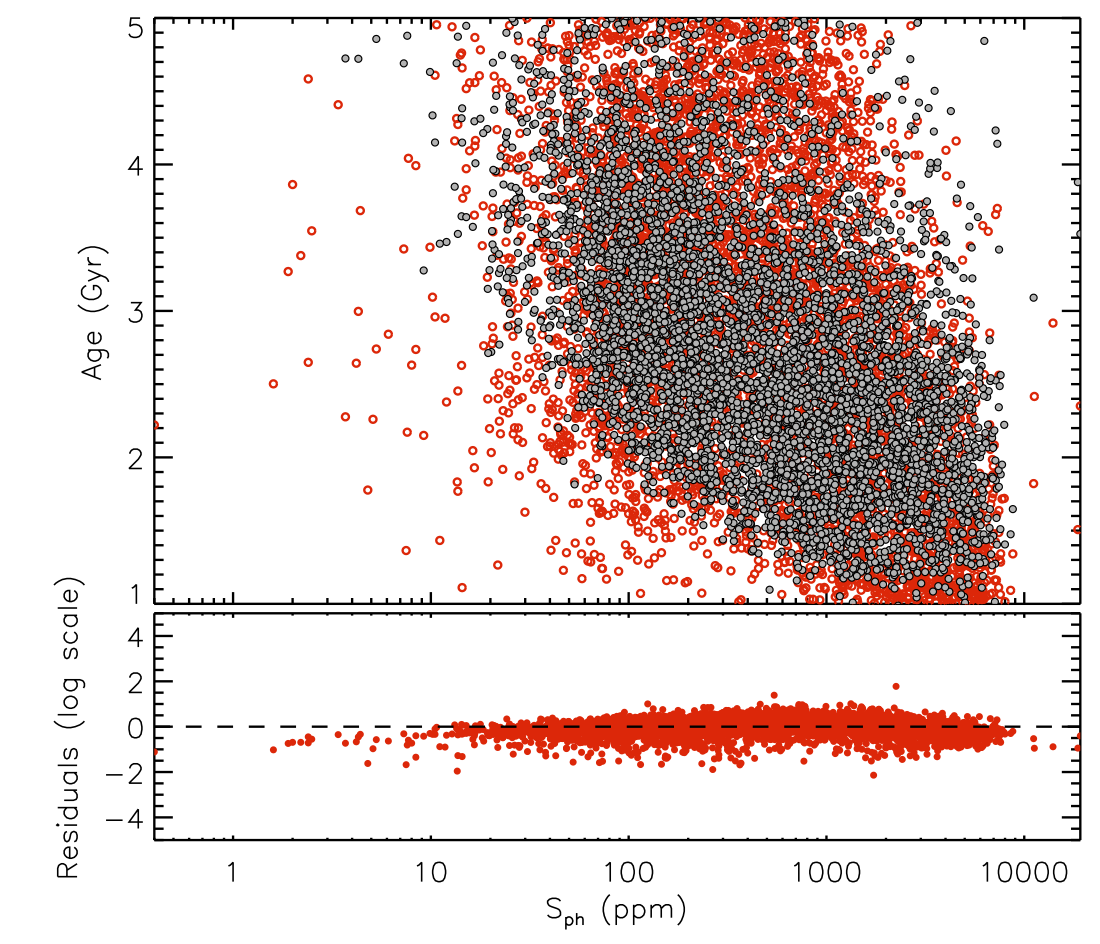}
    \includegraphics[width=5cm]{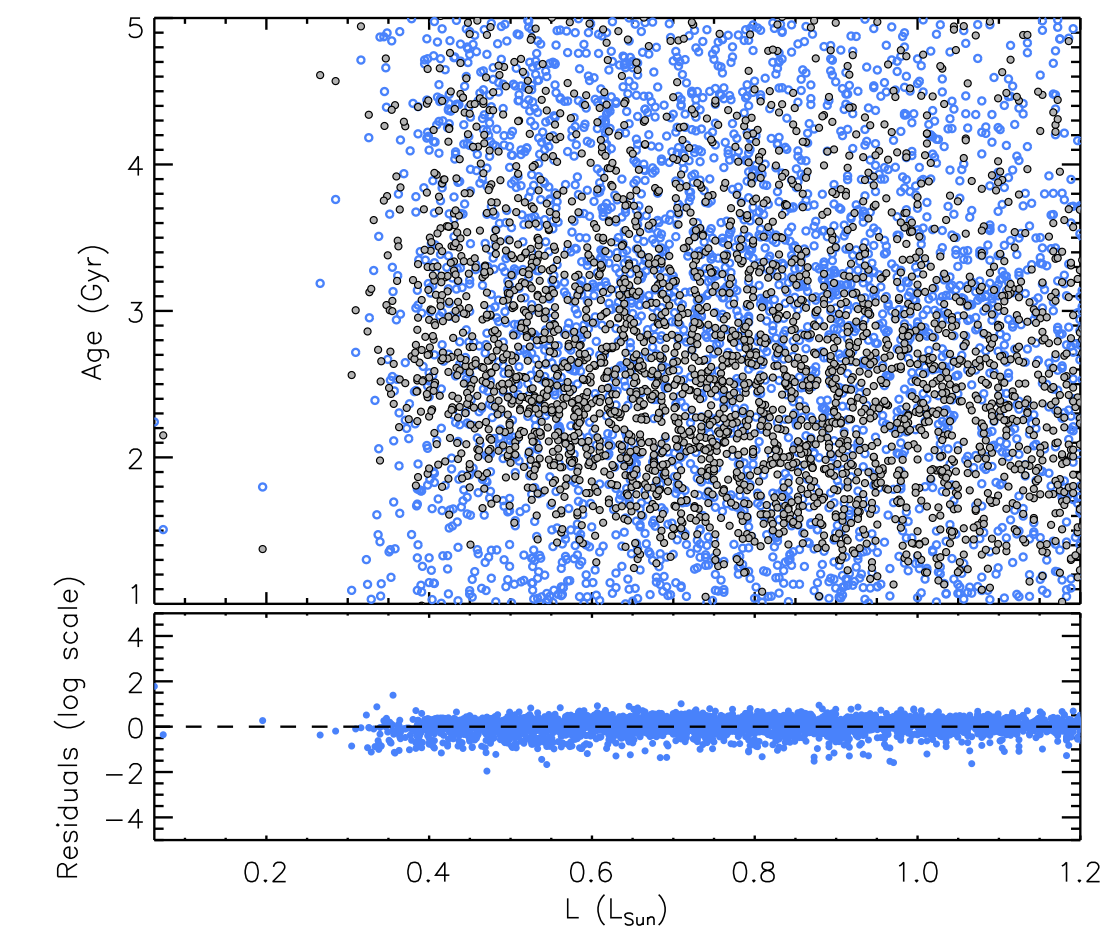}\\
    \includegraphics[width=5cm]{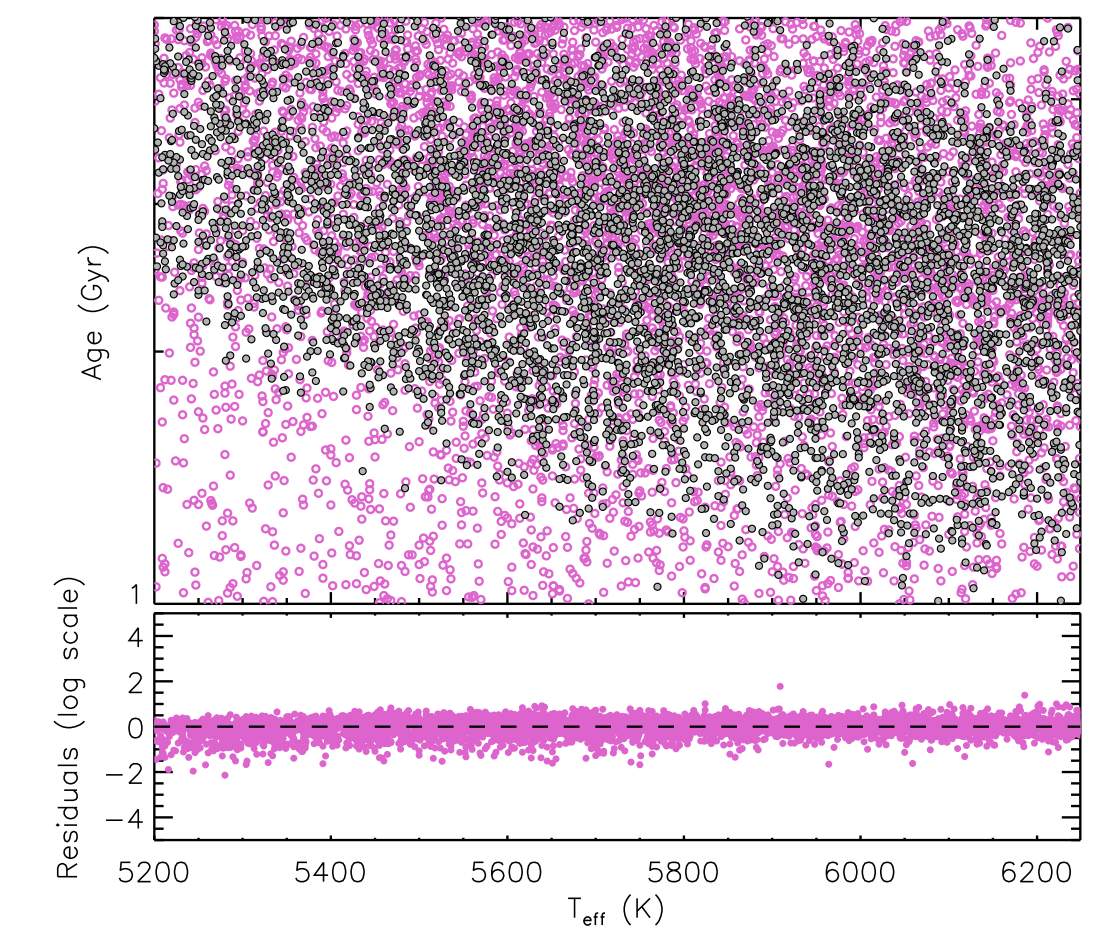}
    \includegraphics[width=5cm]{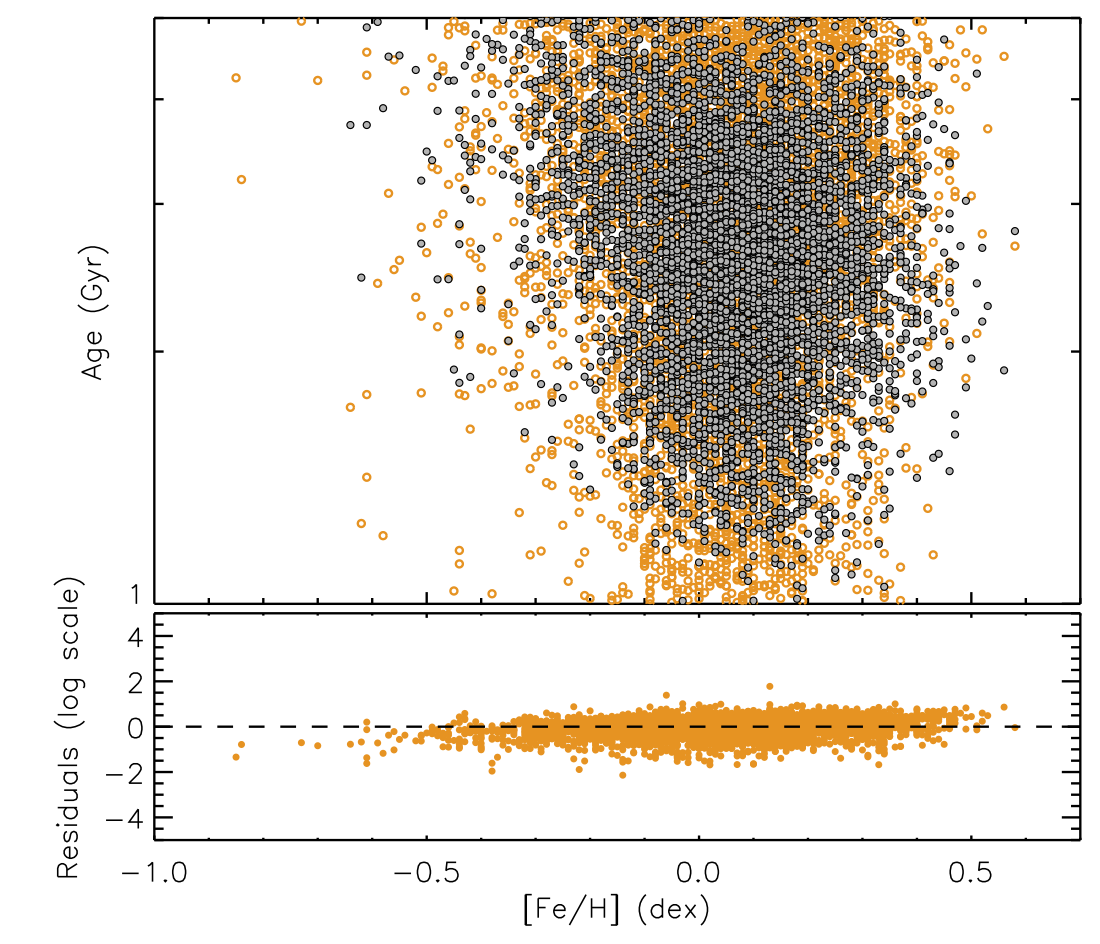}
    \caption{Residuals for late F and G dwarfs for the multivariate regression of Section~\ref{sec:bayesfit} without $\prot$: Age (top left panel), L (top right panel), $\teff$ (bottom left panel), and [Fe/H] (bottom right panel). The colored symbols are the observables and the grey symbols correspond to the predicted values from the Bayesian fits.}
    \label{multi_G_noprot}
\end{figure}
\begin{figure}[ht]
    \centering
    \includegraphics[width=5cm]{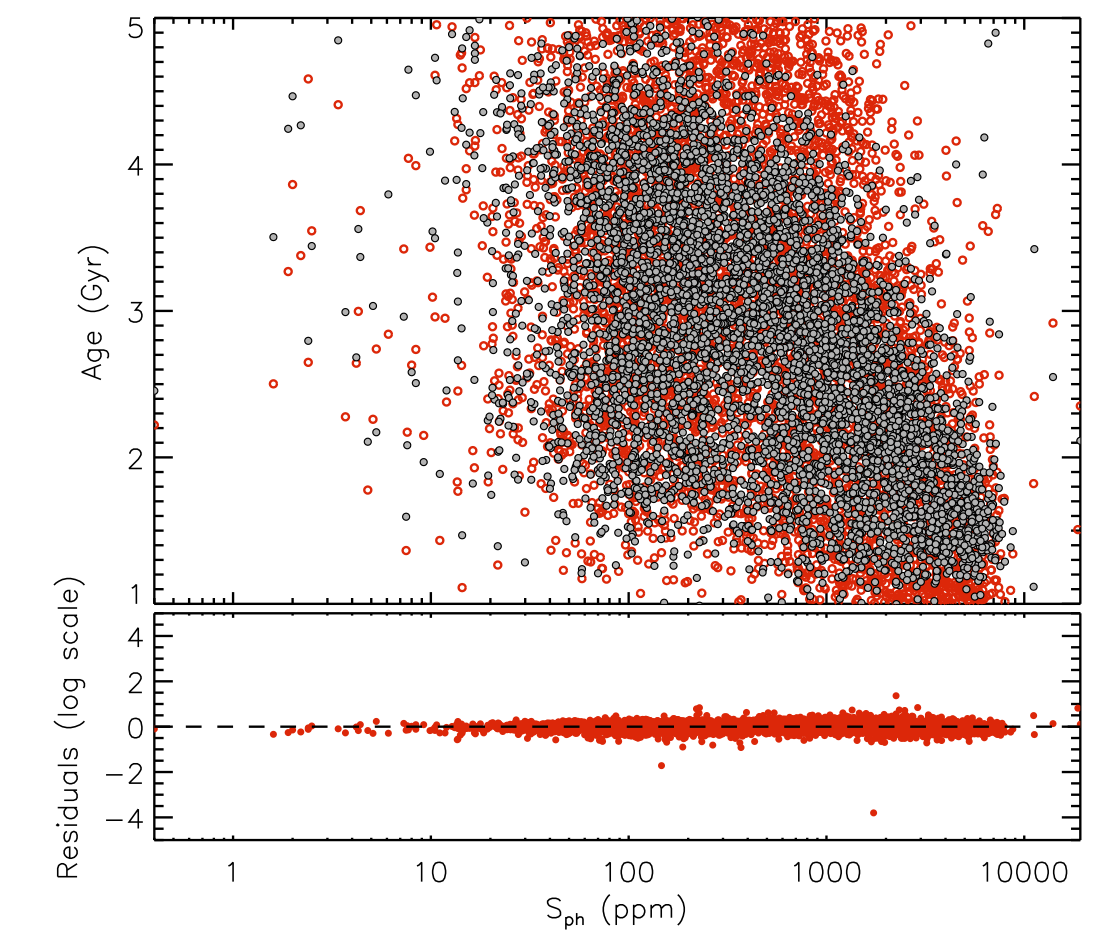}
    \includegraphics[width=5cm]{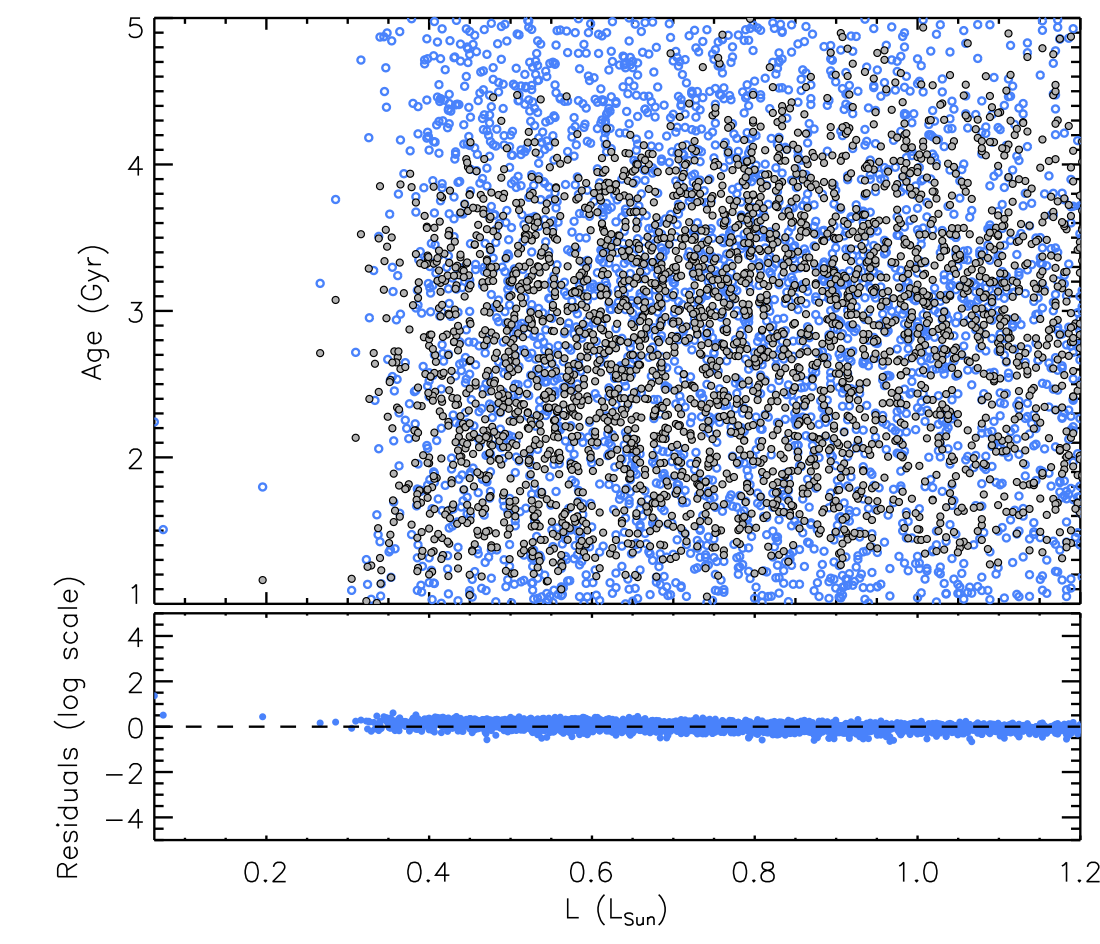}\\
    \includegraphics[width=5cm]{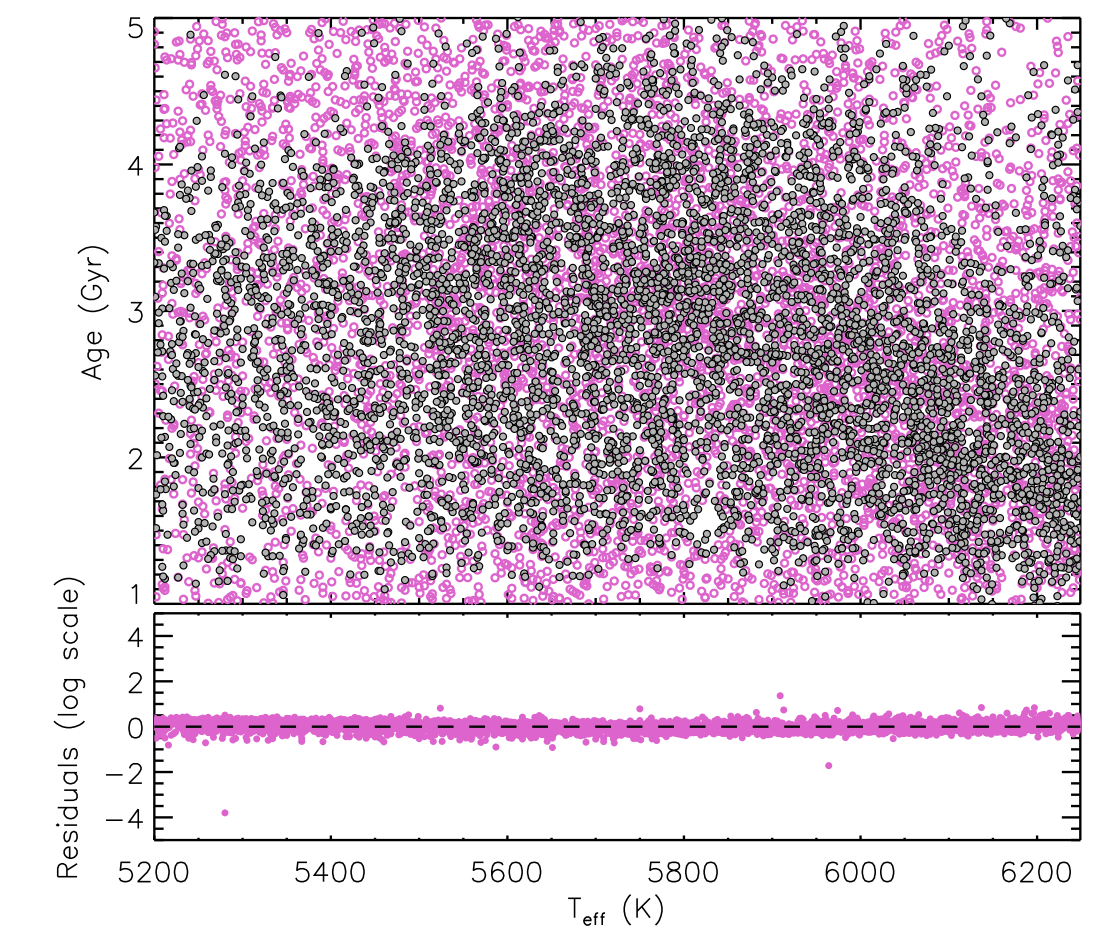}
    \includegraphics[width=5cm]{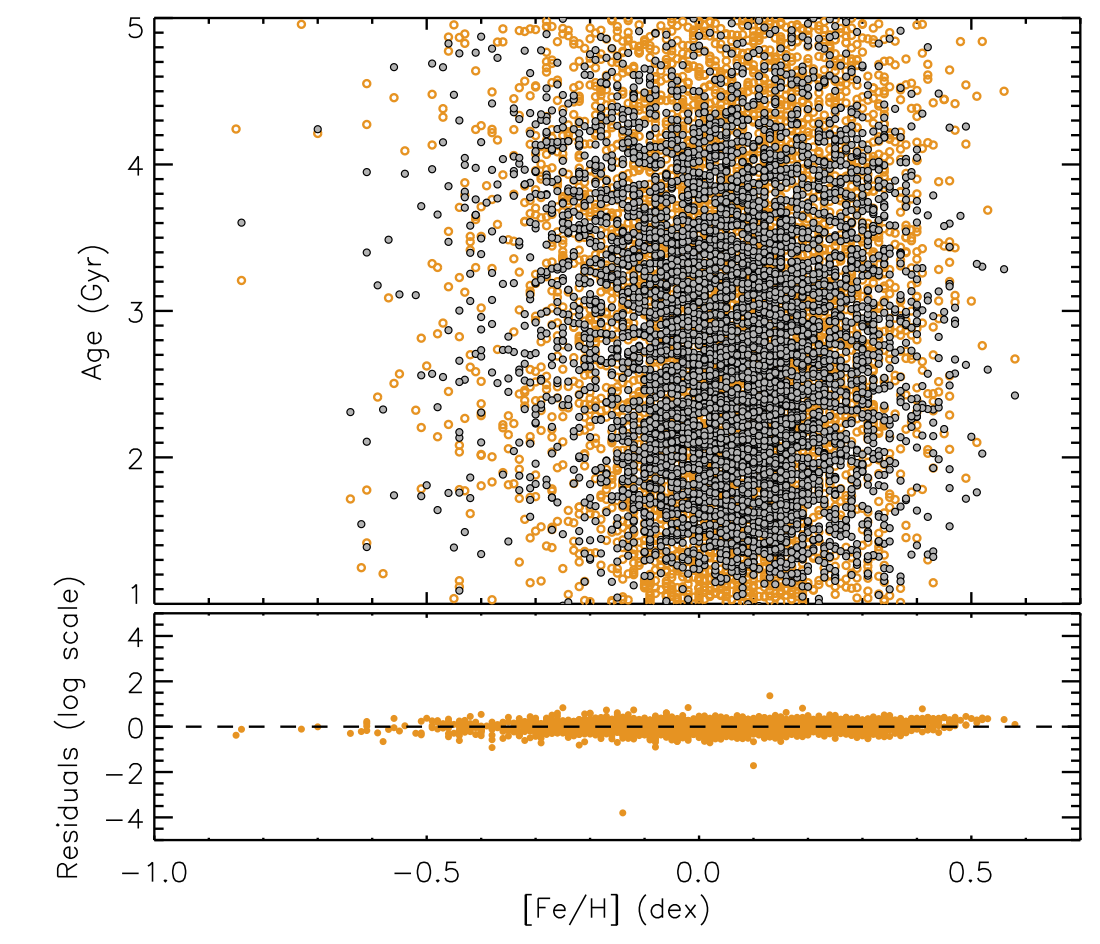}
    \includegraphics[width=5cm]{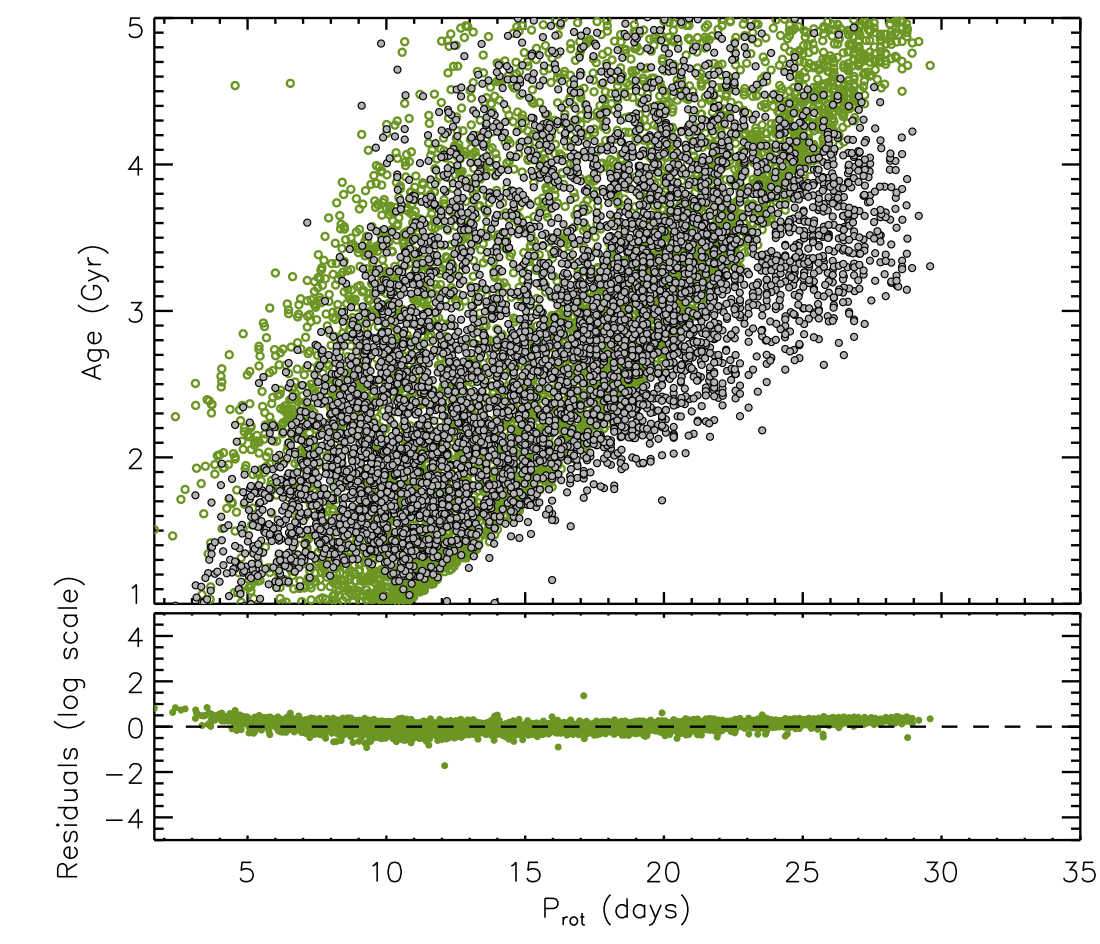}
    \caption{Residuals for late F and G dwarfs for the multivariate regression of Section~\ref{sec:bayesfit} with $\prot$. Same legend as in Figure~\ref{multi_G_noprot} but with also the residuals regarding $\prot$ (bottom right panel).}
    \label{multi_G_prot}
\end{figure}

We show here the residuals of the multivariate fits performed in Section~\ref{sec:multivar} for the late F, G and, K dwarfs. Figure~\ref{multi_G_noprot} shows the sample of late-F and G dwarfs and the residuals for the different parameters used in the fit: $\sph$, L, $\teff$, and [Fe/H] while Figure~\ref{multi_G_prot} also includes $\prot$. The residuals without including rotation  are smaller in general in particular for $\teff$ and L.

\begin{figure}[ht]
    \centering
    \includegraphics[width=5cm]{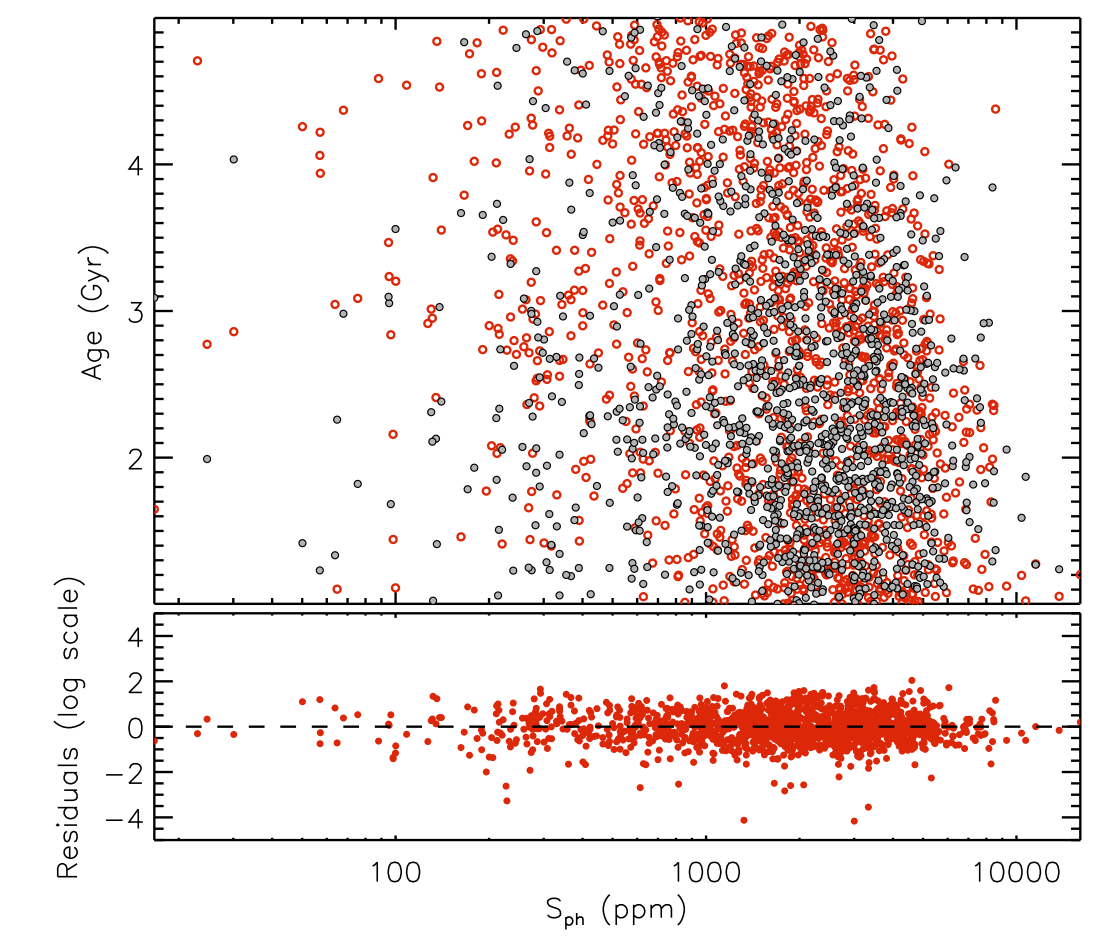}
    \includegraphics[width=5cm]{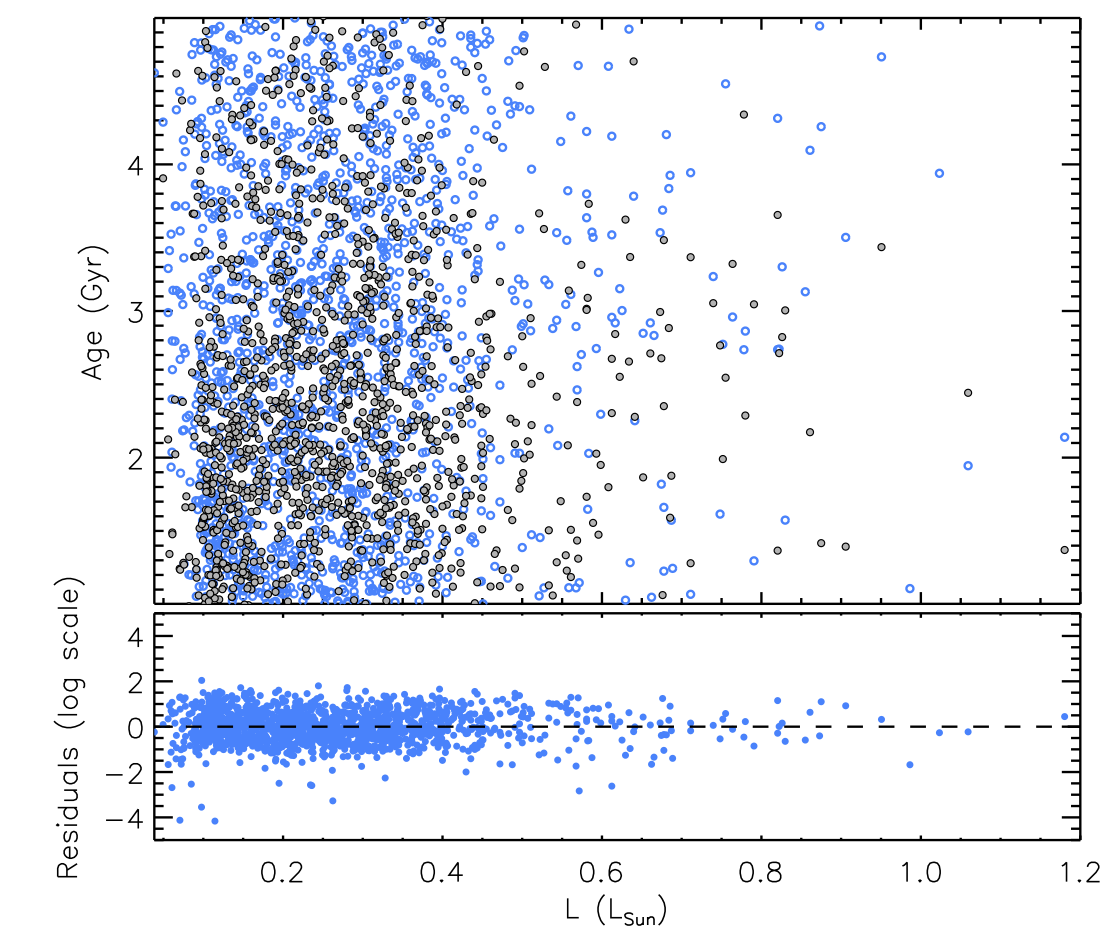}\\
    \includegraphics[width=5cm]{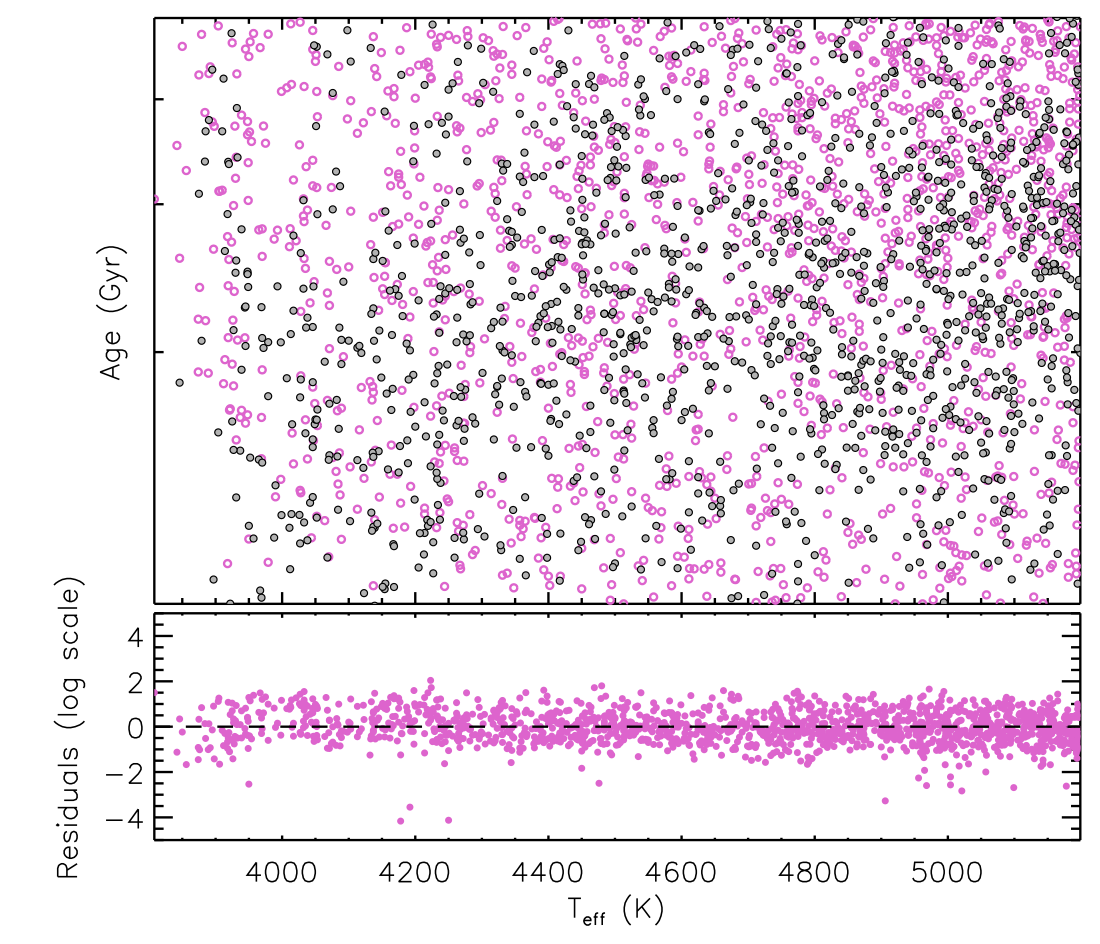}
    \includegraphics[width=5cm]{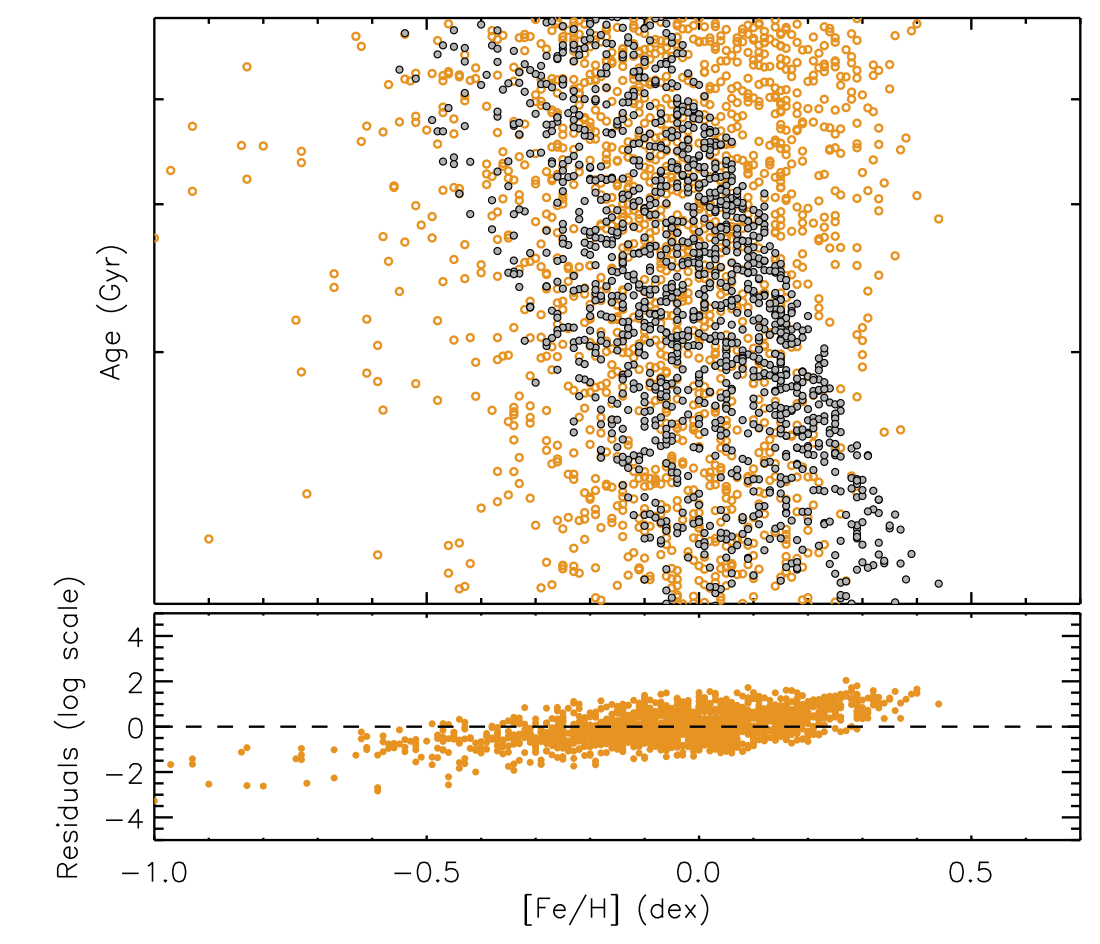}
    \caption{Residuals for K dwarfs for the multivariate regression of Section~\ref{sec:bayesfit} without $\prot$: Age (top left panel), L (top right panel), $\teff$ (bottom left panel), and [Fe/H] (bottom right panel). The colored symbols are the observables and the grey symbols correspond to the predicted values from the Bayesian fits.}
    \label{multi_K_noprot}
\end{figure}

\begin{figure}[ht]
    \centering
    \includegraphics[width=5cm]{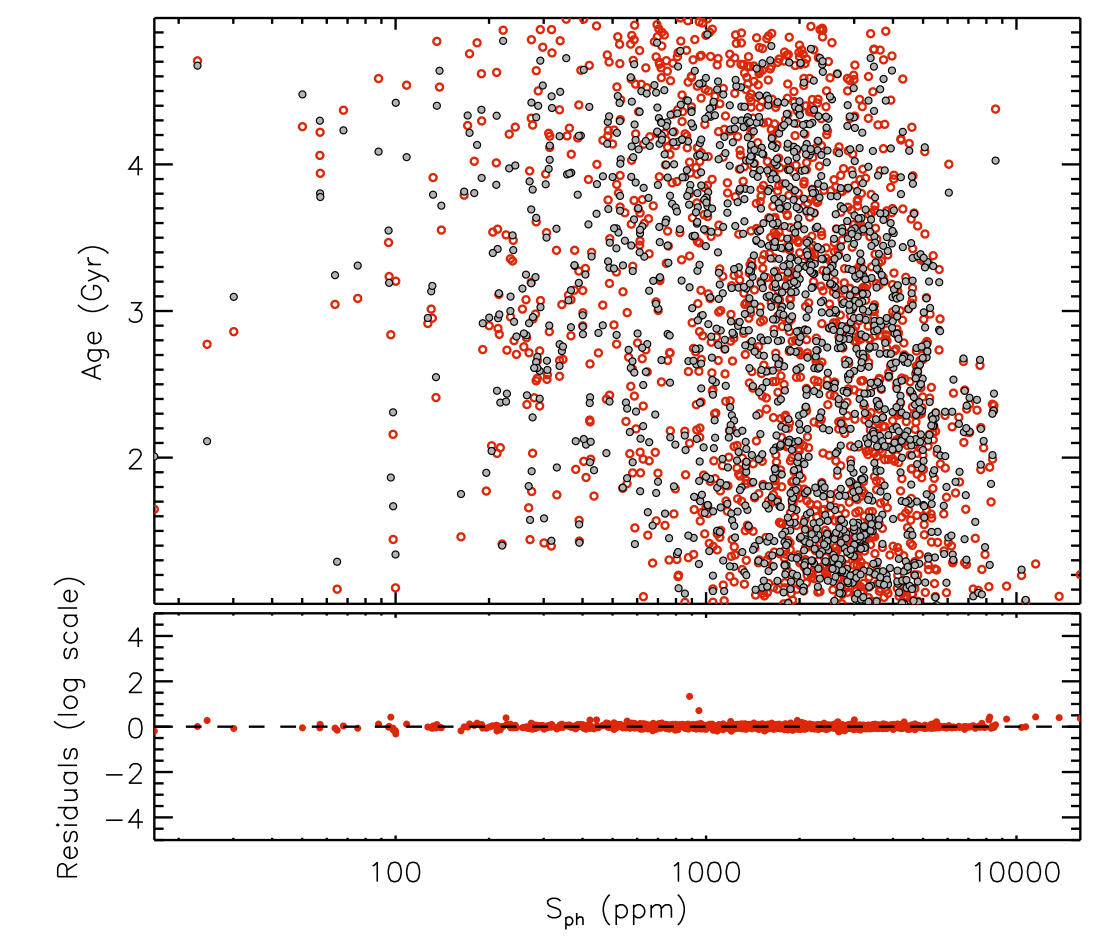}
    \includegraphics[width=5cm]{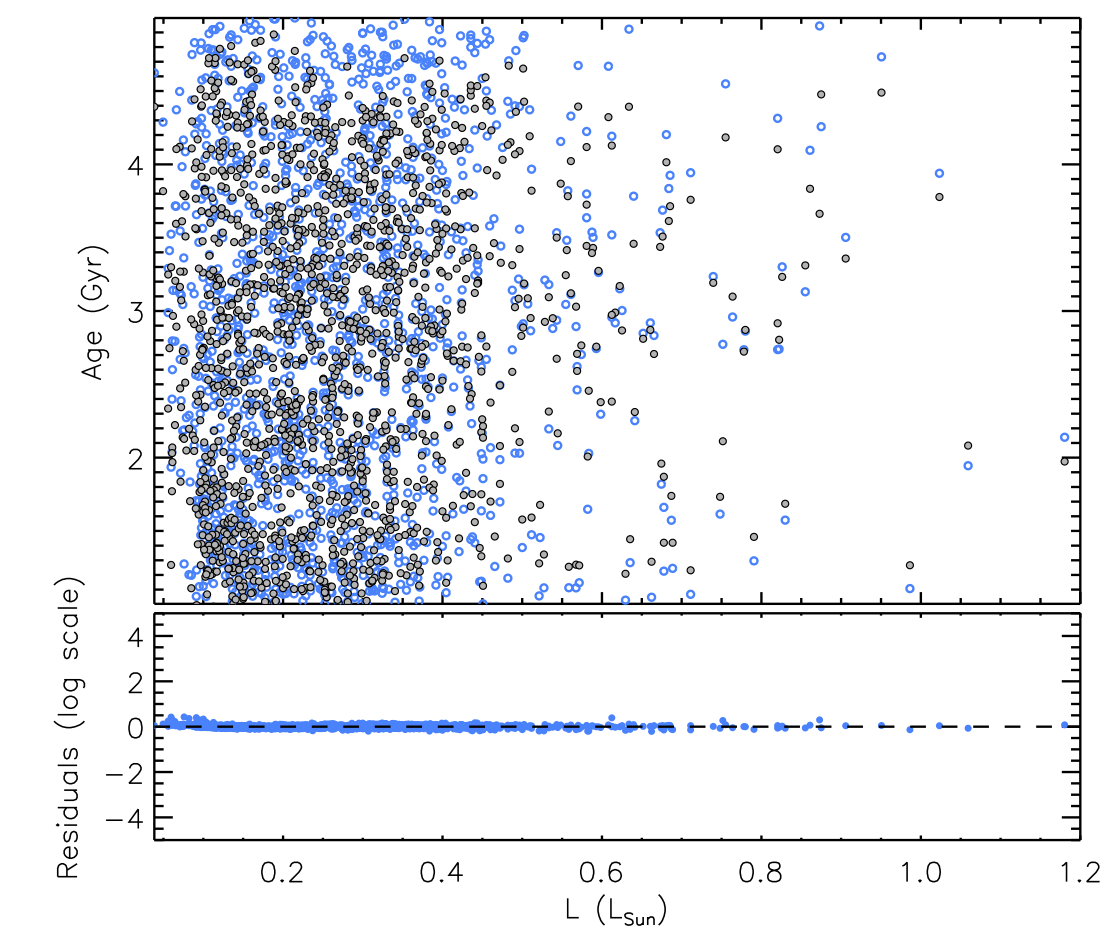}\\
    \includegraphics[width=5cm]{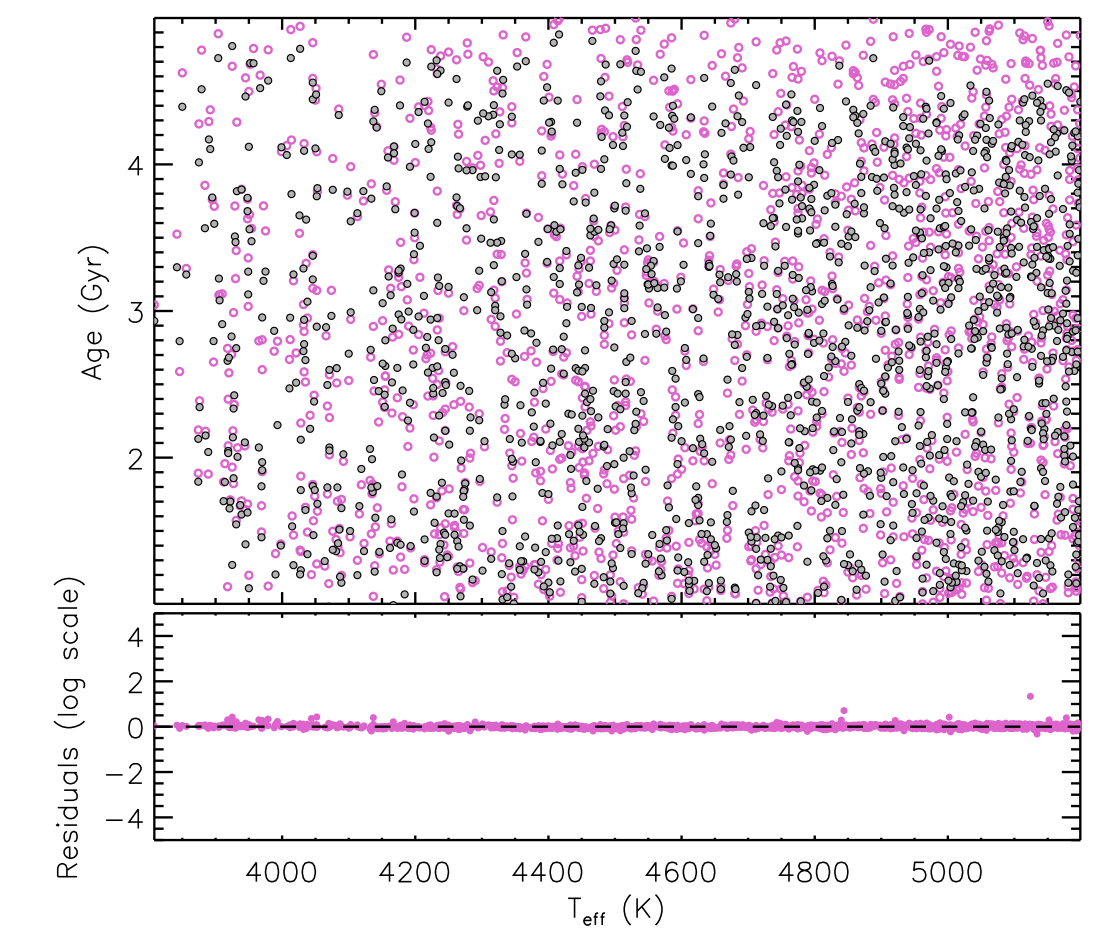}
    \includegraphics[width=5cm]{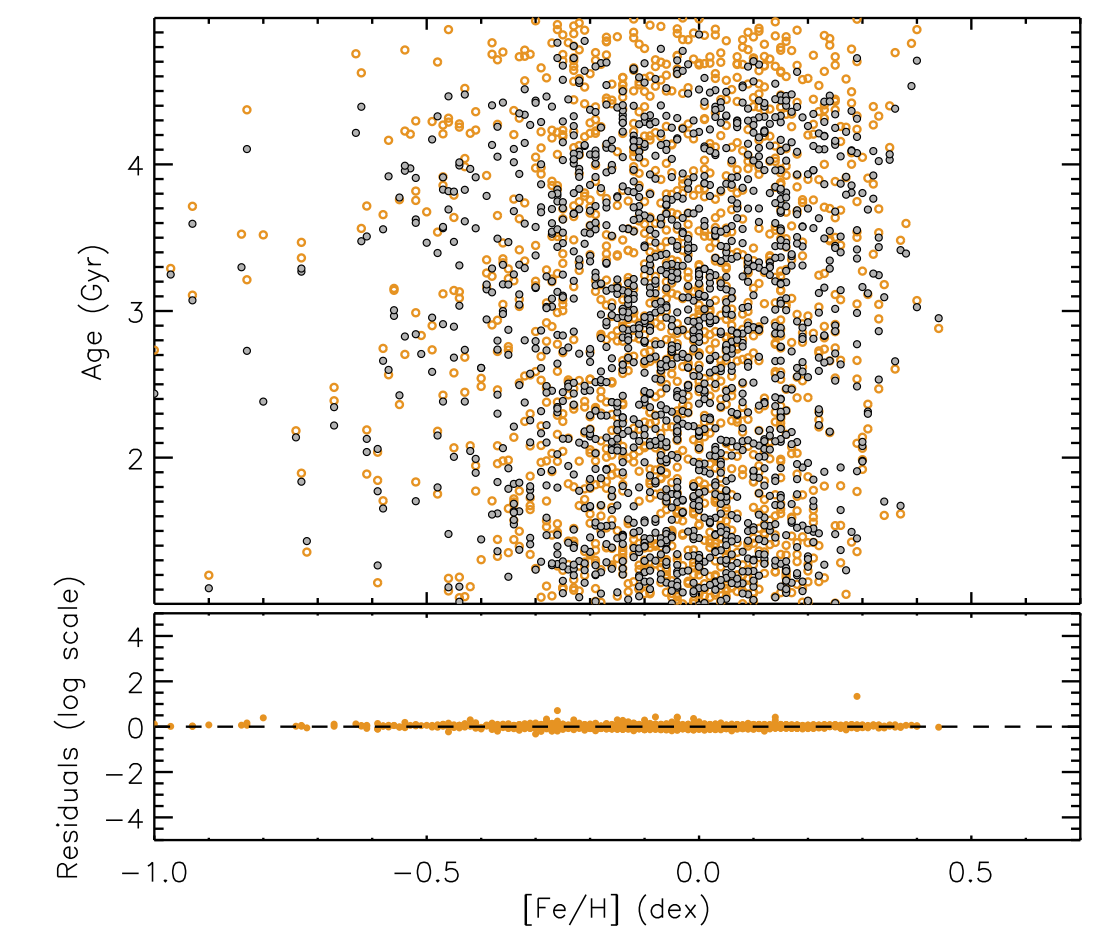}
    \includegraphics[width=5cm]{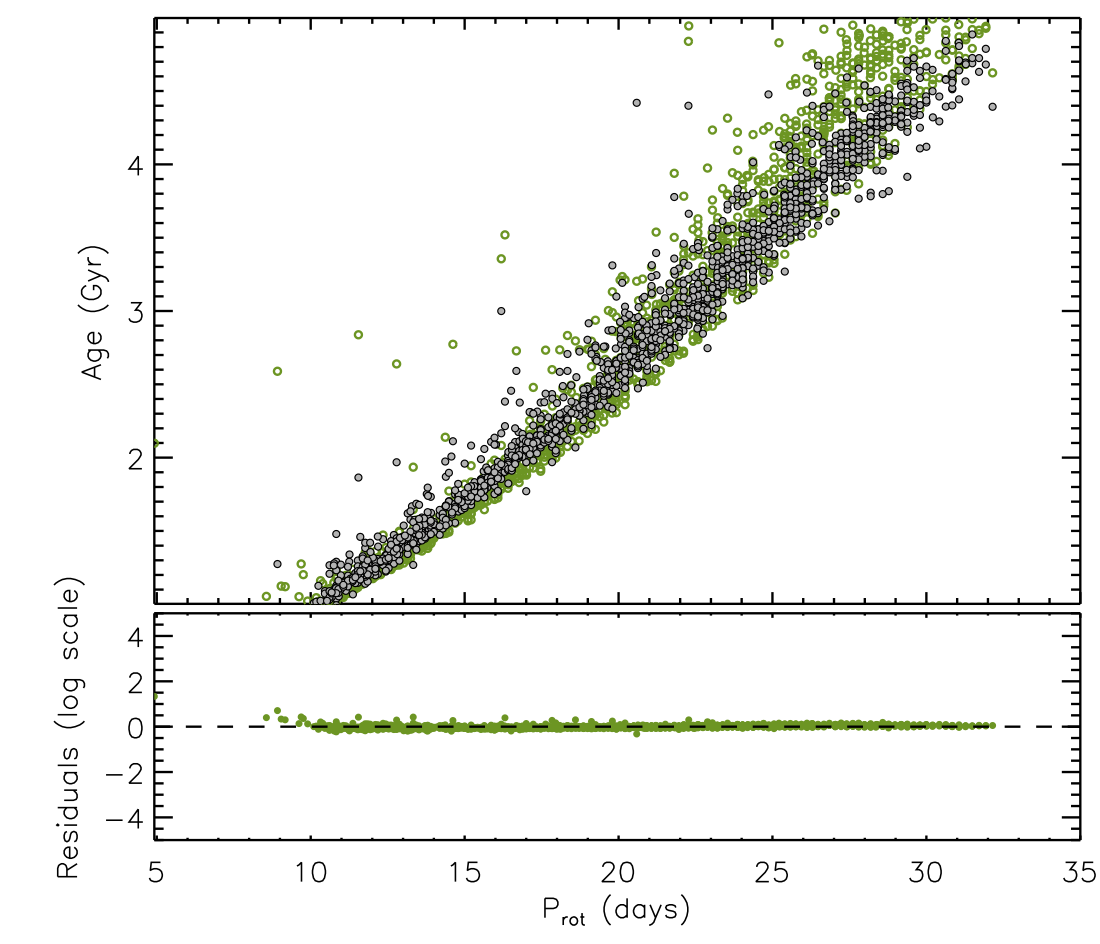}
    \caption{Residuals for K dwarfs for the multivariate regression of Section~\ref{sec:bayesfit} with $\prot$. Same legend as in Figure~\ref{multi_G_noprot} but with also the residuals regarding $\prot$ (bottom right panel).}
    \label{multi_K_prot}
\end{figure}

Figures~\ref{multi_K_noprot} and \ref{multi_K_prot} show the residuals for the K dwarfs without and with rotation, respectively. 

\clearpage

\section{Bayesian analysis to predict $\sph$}\label{appendix:Sph_pred}

While in the main paper, we focused on predicting ages from $\sph$ and other stellar parameters, we also performed the Bayesian analysis for the inverse relation compared to Section~\ref{sec:bayesfit}, which allows us to predict the level of magnetic activity expected for a star with some known stellar parameters. These relations can be useful for instance to assess whether solar-like oscillation modes can be detected for a given star. 

In Table~\ref{tab:multi-linear_Sph}, we only show the results of the favored analytical models for the different cases: the solar analogs, the late-F and G dwarfs, and K dwarfs.

For the late-F and G dwarfs, the Bayesian model comparison favors the model without including rotation (having $\alpha_5 = 0$) by a large extent ($\ln B > 893$, significantly over the condition of a strong Evidence of 5 as explained in \citet{Trotta08}), meaning that $\prot$ is introducing an additional level of complexity to the fit that is not justified by the improvement obtained in the residuals (which instead appear to worsen because they exhibit more structures in the model incorporating $\prot$).

For the K dwarfs, contrary to the case of late-F and G dwarfs, here the term incorporating the rotation ($\alpha_5 \ne 0$) is playing an important role in improving the overall matching between predictions and observations. 


\begin{table*}
\small
\centering
 \caption{Fitting coefficients from the multivariate fit for the favored analytical models similar to the ones described in Section~\ref{sec:multivar} but to compute $\sph$.}
\begin{tabular}{lrrrrrcc}
  \hline
  \hline
  \\[-8pt]
 Model & \multicolumn{1}{c}{$\alpha_1$} & \multicolumn{1}{c}{$\alpha_2$} & \multicolumn{1}{c}{$\alpha_3$} & \multicolumn{1}{c}{$\alpha_4$} & \multicolumn{1}{c}{$\alpha_5$} & \multicolumn{1}{c}{$\ln \beta$} & \multicolumn{1}{c} {$\ln{\mathcal{E}}$}\\[1pt]
  \hline
  \\[-8pt]
  Solar Analogs & $-5.221^{+0.076}_{-0.075}$ & -- & -- & -- & $1.449^{+0.066}_{-0.071}$ & $7.562^{+0.162}_{-0.132}$ & --\\[1pt]
 late-F and G dwarfs & $-2.801^{+0.025}_{-0.021}$ &  $1.254^{+0.022}_{-0.026}$ & $-2.601^{+0.063}_{-0.060}$ & $-37.31^{+0.32}_{-0.44}$ & -- & $331.8^{+3.8}_{-2.7}$ & $-10999$\\[1pt]
 K dwarfs & $4.289^{+0.063}_{-0.080}$ & $-1.141^{+0.071}_{-0.066}$ & $1.868^{+0.097}_{-0.098}$ & $2.883^{+0.578}_{-0.605}$ & $-7.060^{+0.128}_{-0.099}$ & $-1.397^{+6.302}_{-3.778}$ & $-2165$ \\[1pt]
  \hline
 \end{tabular}
 \flushleft {   Notes.} Median estimators and 68.3\,\% Bayesian credible limits are reported for each free parameter. 
\label{tab:multi-linear_Sph}
\end{table*}

\section{Age comparison for gyrochronology relations from the Bayesian fits}\label{appendix:Bayes_gyro}

\begin{figure}[ht]
    \centering
    \includegraphics[width=8.5cm]{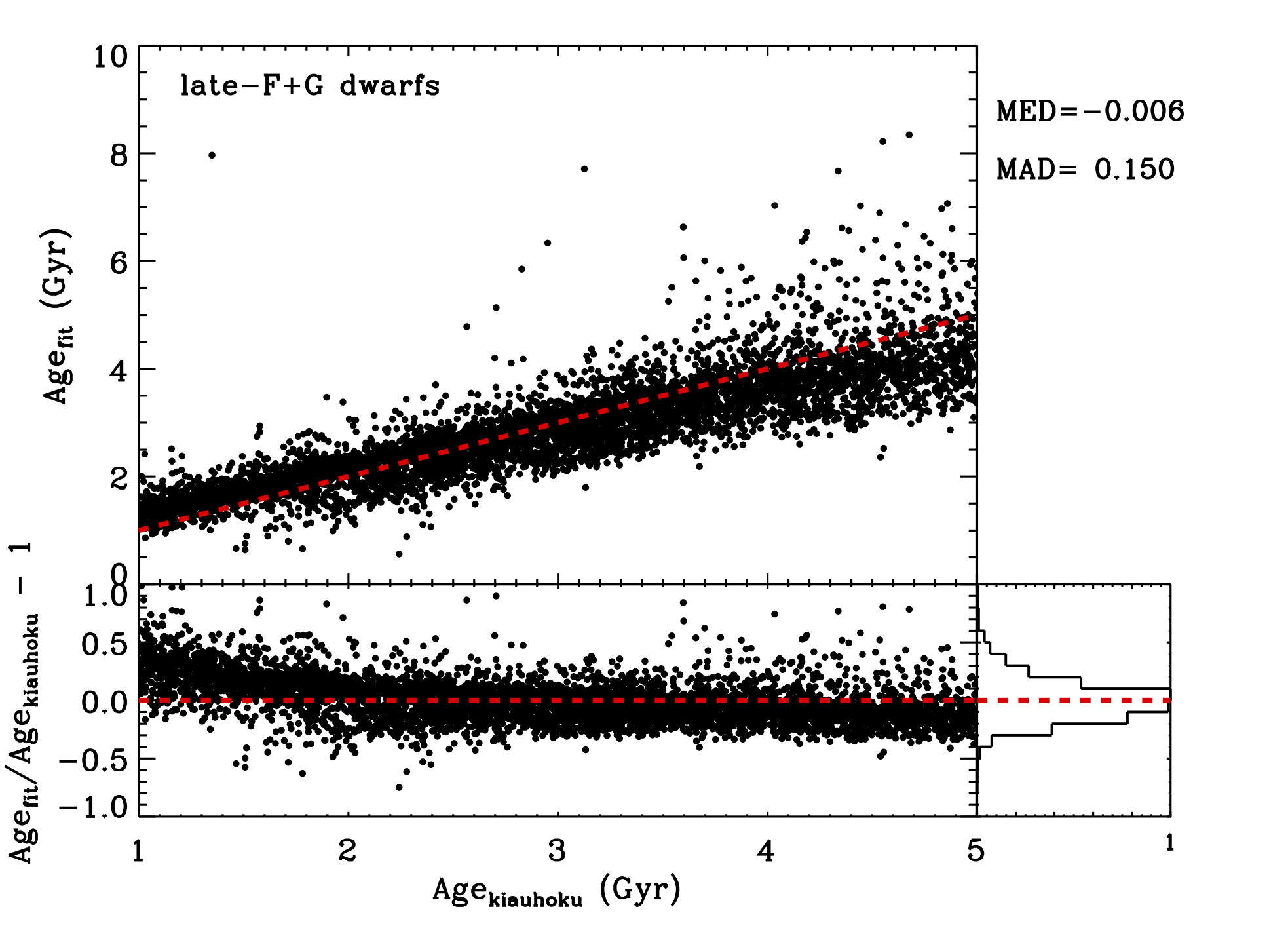}
    \includegraphics[width=8.5cm]{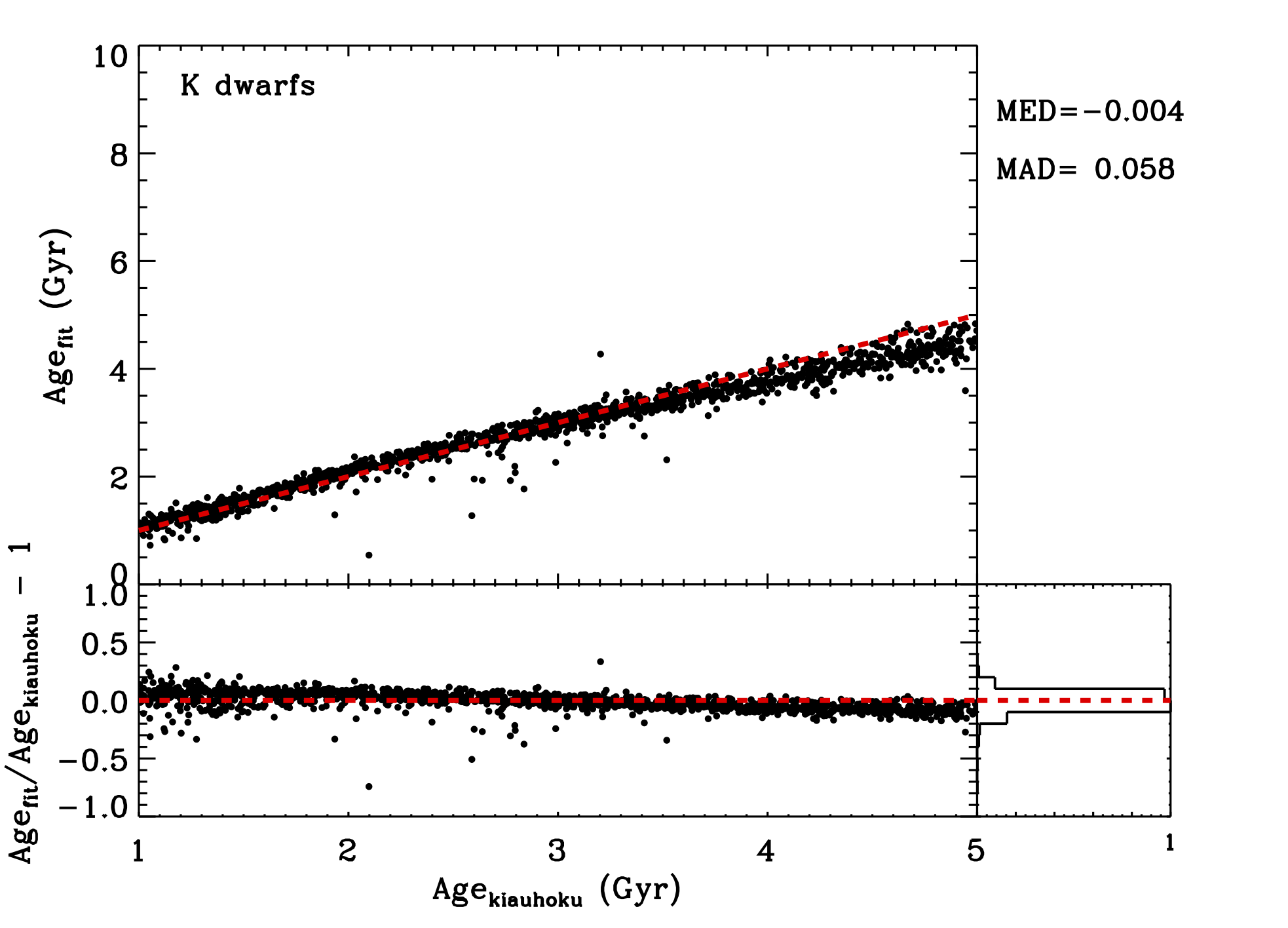}\\
    \caption{Age comparison for the predicted ages from Equation~(\ref{eq:multi-linear} without the $\sph$ term with the \texttt{kiauhoku} ones for the late-F and G dwarfs (left panel) and the K dwarfs (right panel). Same legend as in Figure~\ref{comp_age_fit_norot_rot}.}
    \label{comp_age_fit_gyro}
\end{figure}

We show in Figure~\ref{comp_age_fit_gyro} the comparison between the ages predicted the gyrochronology relation derived with Equation~(\ref{eq:multi-linear}) where the coefficient for $\sph$ is set to 0. For the late-F and G dwarfs, the median difference is of -0.006 with a MAD of 15\%, compared to 0.001 and 14\% when adding the magnetic activity proxy. The comparison for the K dwarfs is very similar to the case where $\sph$ is included in the relations as suggested by the Bayesian Evidence. 

\section{Comparison $S_{\rm ph}$ and $V_{\rm ph}$}\label{appendix:Vph}











We compare here the magnetic activity index $S_{\rm ph}$ with the variability index $V_{\rm ph}$ as described in Section~\ref{sec:detection_bias}, computed on subseries of length of 30 days for light curves filtered above 55 days. This comparison is done for stars with reliable rotation periods to see if there is any bias by using $V_{\rm ph}$ compared to the real index of magnetic activity.  We remind that the $\sph$ index is computed on light curves with a 20-d, 55-d or 80-d filter depending on the rotation period (see S19 for more details). To ensure that there is no bias due to the filter applied on the light curves, for this comparison we use the $\sph$ values computed on light curves filtered above 55 days. 



\begin{figure}[ht]
    \centering
    \includegraphics[width=12cm]{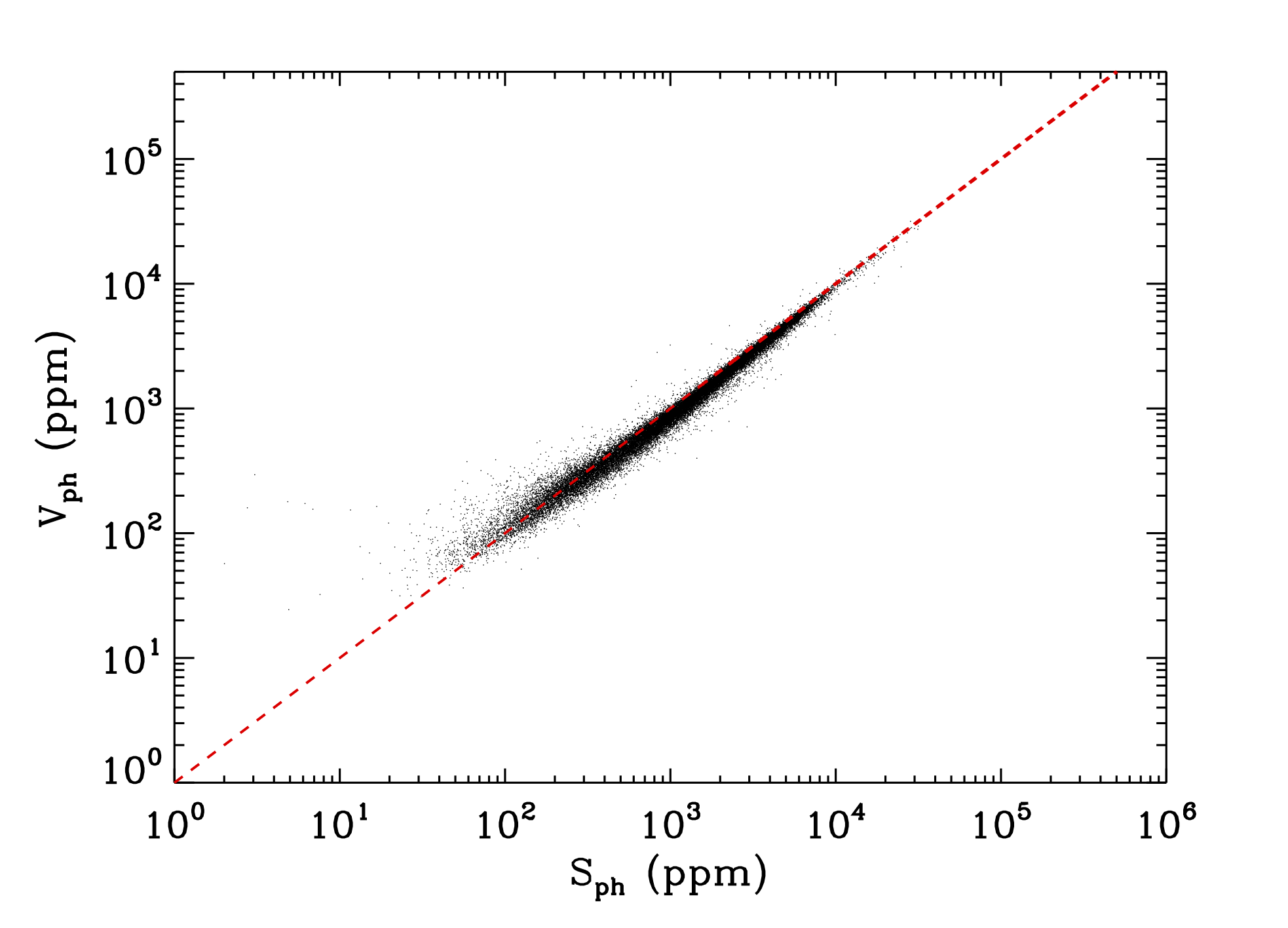}
    \caption{Comparison between the photometric magnetic activity index, $S_{\rm ph}$ computed on subseries of length 5\,$\times P_{\rm rot}$ and a newly defined variability index, $V_{\rm ph}$, computed on subseries of length 30\,days. Both indexes were computed with the 55-d filtered data.}
    \label{comp_Sph_Vph}
\end{figure}

From Figure~\ref{comp_Sph_Vph}, we see that $\sph$ and $V_{\rm ph}$ are close to the 1:1 line. Depending on the rotation period of the star, we find different small biases. For fast rotators with $\prot$ below 10 days, $V_{\rm ph}$ overestimates the magnetic variability for small $\sph$ values. These fast rotators are usually F stars that also show an additional modulation whose variability is captured with the 30-d subseries used for the $V_{\rm ph}$ calculation. In spite of the slight overestimation for the fast rotators, the median difference between $V_{\rm ph}$ and $S_{\rm ph, 55}$ is of -2.6\%. The median difference for periods between 10 and 30 days is of -12.6\% and -20.8\% for stars rotating slower than 30d. This underestimation of the magnetic variability for slower rotators is expected as with 30-d subseries, we have less than one rotation period and the $V_{\rm ph}$ does not capture the full modulation due to the active regions.

On average, the bias is not significant compared to the typical uncertainties on $\sph$ of around 10\% and we can say that $V_{\rm ph}$ is a reasonable measure of the variability of the star associated to magnetic activity, if rotational modulation is present.

\bibliography{BIBLIO_sav,biblio_zach}{}
\bibliographystyle{aasjournal}



\end{document}